\documentclass{aa}
\usepackage[english]{babel}
\usepackage[varg]{txfonts}

\usepackage[usenames,dvipsnames]{xcolor}

\graphicspath{{./}{figures/}}

\usepackage{epsfig}
\usepackage{graphicx,natbib}
\usepackage{amsmath,amssymb}
\PassOptionsToPackage{hyphens}{url}\usepackage{hyperref}
\usepackage{wasysym}
\usepackage{natbib}
\usepackage{ulem}
\usepackage{cancel}
\usepackage{subfigure}
\usepackage{gensymb}
\usepackage{lscape}
\usepackage{longtable}

\usepackage{caption}
\hypersetup{
    colorlinks=true,
    citecolor=blue,
    linkcolor=red,
    urlcolor=black
    }

\begin{document}

\def\nat{Nature }
\def\apj{Astrophys. J. }
\def\apjs{Astrophys. J., Suppl. Ser. }
\def\apjl{Astrophys. J., Lett. }
\def\apss{Astrophys. and Space Science}
\def\aap{A\&A}
\def\mnras{MNRAS}
\def\jgr{J. Geo. Reas.}
\def\zat{Zeitschrift f\"ur Astrophysik}

\def\arcsec{\hbox{$^{\prime\prime}$}}
\def\kms{\mathrm{km~s}^{-1}}
\def\tex{T_\mathrm{ex}}
\def\eup{E_\mathrm{up}}
\def\kb{k_\mathrm{B}}
\def\yesyes{\textbf{Y}}
\def\nono{\textbf{N}}

\title{Chemical inventory of the envelope of the Class~I protostar L1551~IRS~5}
\titlerunning{}

\author{ P. Marchand\inst{1}, A. Coutens\inst{1}, J. Scigliuto\inst{1,2}, F. Cruz-S{\'a}enz de Miera\inst{1,4}, A. Andreu\inst{1}, J.-C. Loison\inst{3}, \'A. K\'osp\'al\inst{4,5,6}, P. \'Abrah\'am\inst{4,5,7}\\
}

\institute{Institut de Recherche en Astrophysique et Planétologie, Université de Toulouse, UT3-PS, CNRS, CNES, 9 av. du Colonel Roche, 31028 Toulouse Cedex 4, France\\ \email{pierre.marchand@irap.omp.eu}
\and Université Côté d’Azur, Observatoire de la Côte d’Azur, CNRS, Laboratoire Lagrange, France
\and Institut des Sciences Mol\'eculaires (ISM), CNRS, Univ. Bordeaux, 351 cours de la Lib\'eration, F-33400 Talence, France
\and Konkoly Observatory, HUN-REN Research Centre for Astronomy and Earth Sciences, CSFK, MTA Centre of Excellence, Konkoly-Thege Mikl\'os \'ut 15-17, 1121 Budapest, Hungary
\and Institute of Physics and Astronomy, ELTE E\"otv\"os Lor\'and University, P\'azm\'any P\'eter s\'et\'any 1/A, 1117 Budapest, Hungary
\and Max Planck Institute for Astronomy, K\"onigstuhl 17, 69117 Heidelberg, Germany
\and University of Vienna, Dept. of Astrophysics, T\"urkenschanzstr. 17, 1180, Vienna, Austria}

\authorrunning{P. Marchand et~al.}

\date{}

\abstract{Episodic accretion in protostars leads to luminosity outbursts that end up heating their surroundings. This rise in temperature pushes  the snow lines back, enabling the desorption of chemical species from dust grain surfaces, which may significantly alter the chemical history of the accreting envelope. However, a limited number of extensive chemical surveys of eruptive young stars have been performed thus far.
In the present study, we carry out a large spectral survey of the binary Class~I protostar L1551~IRS~5,  known to be a FUor-like object, in the 3mm and 2mm bands with the IRAM-30m telescope. As a result, we detected more than 400 molecular lines. The source displays a great chemical richness with the detection of 75  species, including isotopologues. Among these species, there are 13 hydrocarbons, 25 N-bearing species, 30 O-bearing species, 15 S-bearing species, 12 deuterated molecules, and  a total of 10 complex organic molecules (l-C$_4$H$_2$, CH$_3$CCH, CH$_2$DCCH, CH$_3$CHO, CH$_3$CN, CH$_3$OCH$_3$, CH$_3$OCHO, CH$_3$OH, CH$_2$DOH, and HC$_5$N). 
With the help of local thermodynamic equilibrium (LTE) and non-LTE models, we determined the column densities of most molecules as well as excitation and kinetic temperatures. While most of those molecules trace the cold envelope ($\lesssim 20$ K), the OCS and CH$_3$OH emission arise from the warm ($> 100$ K) innermost ($< 2$\arcsec) regions. We compared the chemical inventory of L1551~IRS~5 and its column density ratios, including isotopic ratios, with other protostellar sources. A broad chemical  diversity is seen among Class~I objects. More observations with both single-dish telescopes and interferometers are needed to characterize the diversity in a larger sample of protostars, while more astrochemical models would help explain this diversity, in addition to the impact of luminosity outbursts on the chemistry of protostellar envelopes.
}

 \keywords{Astrochemistry, Stars: formation, ISM: molecules}

\maketitle

\section{Introduction}

In low-mass star formation environments, chemical species are present in the gas phase, but also condensed at the surface of grains or trapped in ice mantles. As material falls closer to the protostar during the collapse of the envelope, the rise in temperature experienced by dust grains causes a thermal desorption of their icy mantle and, consequently, the release of new molecules in the gas phase \citep{Ceccarelli2023}. For example, CO and CH$_4$ thermally desorb at T $\sim$ 20--25 K, while water sublimates above 100 K \citep{Minissale2022}. The sublimation of such species can trigger or impede gas phase reactions that affect the chemical content \citep[e.g., a less efficient deuteration due to the destruction of H$_3$$^+$ by CO,][]{Vastel2006}. Chemical species therefore show different spatial distributions in protostellar environments, and act as a proxy for the physical conditions of star formation (e.g., density, temperature, irradiation). The emission of relatively small molecules is usually detected at large scales, in cold protostellar envelopes. Conversely, water and complex organic molecules (COMs, i.e., molecules with carbon and at least six atoms) are very abundant in the gas phase in the warm inner regions ($\sim$ 100 au) of low-mass protostars \citep[e.g.,][]{Bottinelli2004}. To fully understand how molecules are formed and destroyed in protostars, it is thus essential to characterize the chemical content at both small and large scales.

Single-dish telescopes are particularly suited to probe protostellar envelopes at large scale. This has been the focus of several observing programs with the Institut de RadioAstronomie Millim\'etrique (IRAM)-30m telescope.
For example, the TIMASSS survey \citep{Caux2011} has focused on the well-known Class 0 protostar IRAS 16293--2422. The ASAI large program \citep{Lefloch2018} has surveyed several protostellar sources at different evolutionary stages, from prestellar cores to protoplanetary disks. \citet{Legal2020} also carried a spectral survey of seven Class~I protostars with various physical properties. These studies provide chemical inventories and help us improve our understanding of the chemical processes at work in protostellar envelopes.

However, these above-mentioned surveys were directed toward relatively quiescent protostars, yet FUor sources are protostars that are known to experience bursts of accretion \citep{Fischer2023}. Those lead to luminosity outbursts, which consequently warm up the protostellar envelope. The sudden increase in temperature can imply the thermal desorption of some key-species and potentially other changes in the chemical evolution of the system, even at the envelope scale \citep{Visser2015,Rab2017}.

In this work, we present a large spectral survey of the Class~I FUor-like protostar L1551~IRS~5, performed with the IRAM-30m telescope at 3mm and 2mm. The objective is to compare its chemical content with other protostars. The paper is organised as follows. Section \ref{Sec:obs} describes the source and its observations. Section \ref{Sec:results} presents the methodology of the analysis, the detected molecules, and their estimated column densities. We discuss those results and compare them with other surveys of quiescent protostars in Sect. \ref{Sec:discussion}. Section \ref{Sec:conclusion} presents our  conclusions.

\section{Observations}\label{Sec:obs}

\subsection{The source L1551~IRS~5}

L1551~IRS~5 is a Class~I proto-binary source \citep{Looney1997} located in Taurus, at a distance of $141 \pm 7$ pc \citep{Zucker2019}. The northern and southern source ($M\approx 0.8$ M$_\odot$ and $M\approx 0.3$ M$_\odot$, respectively) \citep{Liseau2005} are separated by $\sim 0.36$\arcsec ($\sim 50$ au) and have systemic velocities of $\sim 9$ $\kms$ and $6$ $\kms$, respectively \citep{Bianchi2020,Andreu2023}. They each possess a circumstellar disk and are embedded in a $\sim$ 300 au circumbinary disk \citep{Cruz2019}, which is itself embedded in a 2500 au to 8000 au scale collapsing envelope \citep{Ohashi1996,Osorio2003} with a systemic velocity of $\sim 6.4$ $\kms$ \citep{Chou2014,Mercimek2022}.
Based on its optical and near-infrared spectra, it has been classified as a FUor-like source \citep[][and references therein]{Connelley2018}, suggesting that it experienced a surge of accretion leading to a luminosity outburst. The temperature increase induced by the luminosity outburst could explain the water detection towards the northern component with NOEMA \citep{Andreu2023} and the reason why this low-mass protostar is one of the few Class~I sources with a rich complex organic chemistry in its inner regions \citep{Bianchi2020,Cruz2022}. However, the chemical composition of its envelope has been poorly known. Beside several dynamical studies of tracers such as $^{13}$CO and CS \citep{Kaifu1984,Ohashi1996,Fridlund2002,Chou2014,Takakuwa2020}, few chemical surveys have been carried out to date. L1551~IRS~5 was observed by \citet{Roberts2002} with the NRAO 12m radio-telescope, and by \citet{Jorgensen2004} using the Onsala 20m radio-telescope and the 15m James Clerk Maxwell Telescope (JCMT). Both studies focused only on a few of the brightest species. More recently, \citet{Mercimek2022} performed a comparative survey including this source, using the IRAM-30m telescope at frequencies above 200 GHz. IRAM-30m deep surveys at 3mm and 2mm of this source are still lacking. We find that by probing larger scales with lower frequency observations, such surveys ought to provide more clues on the characterization of its colder gas.

\subsection{Description of observations}

Observations were carried out with the IRAM-30m telescope (projects 047-22 and 115-22) with the Eight MIxer Receivers (EMIR) in band E0 (3mm) and E1 (2mm).
The telescope beam was centered on $\alpha_\mathrm{J2000}$= 04:31:34.161, $\delta_\mathrm{J2000}$= +18:08:04.722. The large extent of the beam at those frequencies, from $16$\arcsec ($\sim 2300$ au) in the E1 band, to $35$\arcsec ($\sim 5100$ au) in the E0 band, is ideal to study the cold envelope. We used the position-switching (PSW) mode with an off-source reference located at a (0,-900)\arcsec offset. 
The wide mode of the Fast Fourier Transform Spectrometer (FTS) provides a $2 \times 8$ GHz coverage for each setup with a resolution of 200 kHz (0.4 to 0.7 $\kms$). The E0 band has been entirely covered with the frequency range 71.75--115.75 GHz. In addition, we probed the ranges 135.25--143.25 GHz and 151.25--159.25 GHz in the E1 band. 

We used complementary observations of the source in the frequency range 143.389-147.029 GHz (project 080-16), at the same coordinates. The off-source reference for the PSW mode in this case is located at a (0,-300)\arcsec offset. For those observations, the fine mode of the FTS (50 kHz resolution) was used.

The observations were divided into six setups that are summarized in Table \ref{tab:setups} with the dates of observations.
Pointing and focusing were done on Mars, that was close to the source for all observations. Both vertical and horizontal polarisations were observed at the same time.

We reduced the data using \textsc{CLASS} from the \textsc{GILDAS}\footnote{https://www.iram.fr/IRAMFR/GILDAS} package \citep{Gildas2013}. The baselines were subtracted before the averaging of the spectra. We also averaged the vertical and horizontal polarizations. The data observed with a 50 kHz resolution were smoothed to match the spectral resolution of the other observations (200 kHz). The  achieved noise levels are indicated in Table \ref{tab:setups}. The line identification and some of the analysis have been done with the \textsc{CASSIS}\footnote{CASSIS (http://cassis.irap.omp.eu/) has been developed by IRAP-UPS/CNRS} software, using the CDMS \citep{Muller2001,Muller2005} and JPL \citep{JPL} spectroscopic databases.

\begin{table*}
  \caption{Setup summary. For setups 1 to 4, two frequency ranges were observed at the same time, with the EO: [LO+LI]+[UI+UO] receiver for setups 1 and 2; E0/E1: [LO+LI] for setup 3 and 4; and E1: [LO+LI] for setups 5 and 6.}
\label{tab:setups}
\centering
\begin{tabular}{lllllll}
\hline\hline
  Setup   &  Freq. range 1 (GHz) &  Freq. range 2 (GHz) &  On-source time (h) & Resolution (kHz)  &  rms (mK)$^a$ &   Date of observation  \\
\hline
1       & 83.75-91.75 & 99.75-107.75    &  2.5    & 200   &   3.8  &   2022-10-10 \\
2       & 91.75-99.75 & 107.75-115.75   &  3.33   & 200   &   4.0, 8.5$^b$  &  2022-09-01, 2022-10-10 \\
3       & 71.75-79.75 & 151.25-159.25   &  4.0    & 200   &   3.8, 5.0$^c$  &  2022-09-01 \\
4       & 79.75-87.75 & 135.25-143.25   &  2.0    & 200   &   6.0, 15.7$^c$  &  2023-02-14 \\
5       & 143.389-145.210 & --          &  0.7    & 50    &   17.0  &  2017-03-05 \\
6       & 145.210-147.029 & --          &  0.7    & 50    &   17.0  &  2017-10-20 \\
\hline
\end{tabular} 
\tablefoot{$^a$ for a 200 kHz spectral resolution. $^b$ For frequencies larger than $\sim 113$ GHz. $^c$ For frequency range 2.}
\end{table*}

\section{Results}\label{Sec:results}

\subsection{Line identification and spectral fitting}

\begin{figure}
\begin{center}
\includegraphics[trim=0cm 0cm 0cm 0cm, width=0.48\textwidth]{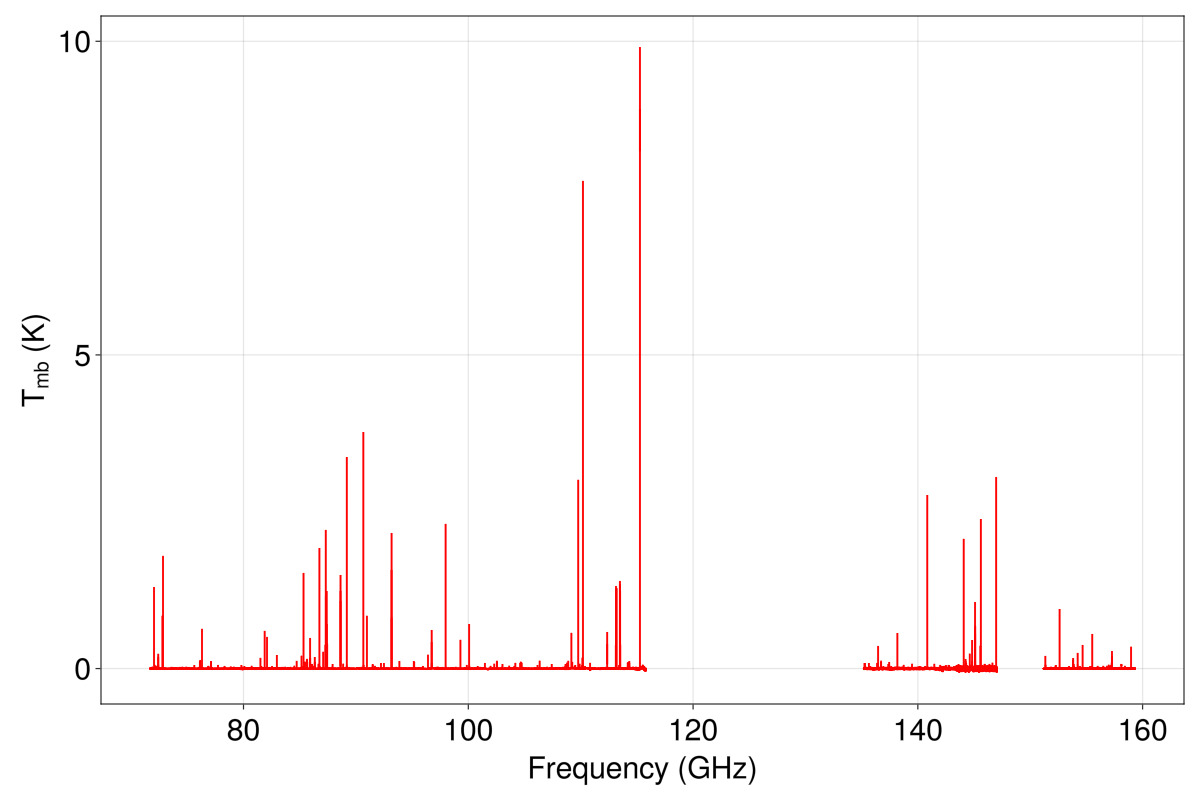}
  \caption{Spectrum observed towards L1551~IRS~5 by the IRAM-30m telescope as part of this study. The line intensity is displayed in main beam temperature.}
  \label{fig:spectrum}
\end{center}
\end{figure}

The full reduced spectrum is displayed in Fig. \ref{fig:spectrum}. With a signal-to-noise ratio (S/N) criterion of 3$rms$, we detected 403 lines. Hyperfine structure transitions were counted as only one line when blended, which accounts for 80 more transitions. We identified 75 species, 37 of which are main isotopologues and 38 are secondary isotopologues. We detect 68 organic species including 13 hydrocarbons, 25 nitrogen-bearing species, 30 oxygen-bearing species, 15 sulfur-bearing species, and 12 deuterated molecules. 10 species are COMs, namely l-C$_4$H$_2$, CH$_3$CCH, CH$_2$DCCH, CH$_3$CHO, CH$_3$CN, CH$_3$OCH$_3$, CH$_3$OCHO, CH$_3$OH, CH$_2$DOH, and HC$_5$N. The non-organic species are NH$_2$D, N$_2$H$^+$, N$_2$D$^+$, NS, SO, $^{34}$SO, and SO$_2$. We report two unidentified lines above 5$rms$, at 80.4805 GHz and 135.2495 GHz.

Each transition line has been fitted with a Gaussian curve using the Levenberg-Marquardt scheme of CASSIS, to obtain its full width at half maximum (FWHM), peak velocity, $v$, and peak intensity, $Int$. All lines, their properties, their fitting parameters, and their Gaussian fluxes with uncertainties are listed in Table \ref{tab:linelist}. The Gaussian flux, F, and its uncertainty, $\delta$F, are calculated as:
\begin{align}
    \mathrm{F} &= \sqrt{\frac{\pi}{4 \ln{2}}} Int \times FWHM,\\
    \delta_\mathrm{F} &= \mathrm{F}\sqrt{\left(\frac{\delta FWHM}{FWHM}\right)^2+\left(\frac{\delta Int}{Int}\right)^2},
\end{align}
where $\delta FWHM$ and $\delta Int$ are the one sigma uncertainties on the FWHM and the peak intensity.
If several transitions of the same species are blended into one apparent line, we used the same fit for all the transitions (indicated by `-' instead of numbers for the FWHM, intensity, and flux in the tables). Several OCS and CH$_3$OH lines clearly exhibit at least two separated components, each of which were fit with a Gaussian profile. Table \ref{tab:species} summarizes the number of lines detected for each species, with the minimum and maximum upper energy level, $E_\mathrm{up}$, of those transitions.

\begin{table*}
  \caption{Chemical species detected in our survey, with the number of detected lines and minimum and maximum upper energy levels, $E_\mathrm{up}$, of the transitions.}
\centering
\label{tab:species}
\begin{tabular}{llllcllll}
\hline\hline
Species & Lines  & $E_\mathrm{up,min}$ & $E_\mathrm{up,max}$ & \hspace{1cm} & Species  & Lines  & $E_\mathrm{up,min}$ & $E_\mathrm{up,max}$\\
\hline
CCH & 6 & 4.2 & 4.2 & & HC$^{15}$N & 1 & 4.1 & 4.1 \\ 
CCD & 10 & 3.5 & 10.4 & & HNC & 1 & 4.4 & 4.4 \\ 
C$^{13}$CH & 5 & 4.1 & 4.1 & & DNC & 2 & 3.7 & 11.0 \\ 
CCS & 16 & 15.4 & 57.2 & & HN$^{13}$C & 1 & 4.2 & 4.2 \\ 
C$_3$H & 8 & 7.8 & 12.5 & & H$^{15}$NC & 1 & 4.3 & 4.3 \\ 
c-C$_3$H & 3 & 4.4 & 4.4 & & HNCO & 3 & 10.6 & 29.5 \\ 
c-C$_3$H$_2$ & 14 & 6.4 & 82.6 & & HCO & 4 & 4.2 & 4.2 \\ 
c-C$_3$HD & 12 & 5.7 & 22.5 & & HCO$^+$ & 1 & 4.3 & 4.3 \\ 
c-CC$^{13}$CH$_2$ & 4 & 6.3 & 15.8 & & DCO$^+$ & 2 & 3.5 & 10.4 \\ 
l-C$_3$H$_2$ & 5 & 15.0 & 28.5 & & H$^{13}$CO$^+$ & 1 & 4.2 & 4.2 \\ 
C$_3$N $^{a}$ & 5 & 17.1 & 26.1 & & HC$^{18}$O$^+$ & 1 & 4.1 & 4.1 \\ 
C$_3$O & 4 & 16.6 & 30.5 & & HC$^{17}$O$^+$ & 1 & 4.2 & 4.2 \\ 
C$_3$S & 4 & 25.2 & 47.4 & & D$^{13}$CO$^+$ & 1 & 10.2 & 10.2 \\ 
C$_4$H $^{a}$ & 14 & 16.4 & 62.1 & & HC$_3$N & 8 & 15.7 & 66.8 \\ 
l-C$_4$H$_2$ & 7 & 23.6 & 46.8 & & DC$_3$N & 5 & 18.2 & 36.9 \\ 
CH$_3$CCH & 12 & 12.3 & 65.8 & & H$^{13}$CCCN & 2 & 19.0 & 23.3 \\ 
CH$_2$DCCH & 2 & 16.3 & 21.7 & & HC$^{13}$CCN & 2 & 23.9 & 33.9 \\ 
CH$_3$CHO & 23 & 5.0 & 42.5 & & HCC$^{13}$CN & 4 & 15.6 & 28.7 \\ 
CH$_3$CN & 8 & 8.8 & 132.8 & & HC$_5$N & 15 & 48.3 & 115.4 \\ 
CH$_3$OCH$_3$ $^{a}$ & 1 & 40.4 & 40.4 & & H$_2$CO & 3 & 3.5 & 21.9 \\ 
CH$_3$OCHO & 3 & 20.2 & 56.6 & & D$_2$CO & 1 & 13.4 & 13.4 \\ 
CH$_3$OH & 46 & 7.0 & 233.6 & & H$_2^{13}$CO & 4 & 4.3 & 22.4 \\ 
CH$_2$DOH & 7 & 6.4 & 25.8 & & H$_2$CCO & 8 & 9.7 & 40.5 \\ 
CN & 9 & 5.4 & 5.4 & & HCS$^+$ & 1 & 6.1 & 6.1 \\ 
$^{13}$CN & 15 & 5.2 & 5.2 & & H$_2$CS & 6 & 9.9 & 29.9 \\ 
C$^{15}$N & 3 & 5.3 & 5.3 & & H$_2$C$^{33}$S & 1 & 22.8 & 22.8 \\ 
CO & 1 & 5.5 & 5.5 & & HOCO$^+$ & 2 & 10.3 & 15.4 \\ 
$^{13}$CO & 1 & 5.3 & 5.3 & & HOCN & 2 & 10.1 & 15.1 \\ 
C$^{17}$O $^{a}$ & 2 & 5.4 & 5.4 & & NH$_2$D $^{a}$ & 10 & 20.7 & 21.3 \\ 
C$^{18}$O & 1 & 5.3 & 5.3 & & N$_2$H$^+$ $^{a}$ & 3 & 4.5 & 4.5 \\ 
$^{13}$C$^{18}$O & 1 & 5.0 & 5.0 & & N$_2$D$^+$ $^{a}$ & 5 & 3.7 & 11.1 \\ 
CS & 2 & 7.0 & 14.1 & & NS & 1 & 8.9 & 8.9 \\ 
$^{13}$CS & 2 & 6.7 & 13.3 & & OCS & 12 & 12.3 & 53.1 \\ 
C$^{33}$S & 1 & 7.0 & 7.0 & & OC$^{34}$S & 1 & 25.6 & 25.6 \\ 
C$^{34}$S & 2 & 6.9 & 13.9 & & SO & 6 & 9.2 & 38.6 \\ 
HCN & 3 & 4.2 & 4.2 & & $^{34}$SO & 2 & 9.1 & 15.6 \\ 
DCN $^{a}$ & 6 & 3.5 & 10.4 & & SO$_2$ & 5 & 7.7 & 54.7 \\ 
H$^{13}$CN & 3 & 4.1 & 4.1 \\
\hline
\end{tabular} 
\tablefoot{$^{a}$ Species with one or several lines that are made of several transitions.}
\end{table*}

\begin{figure}
\begin{center}
\includegraphics[trim=1cm 0cm 1cm 0cm, width=0.45\textwidth]{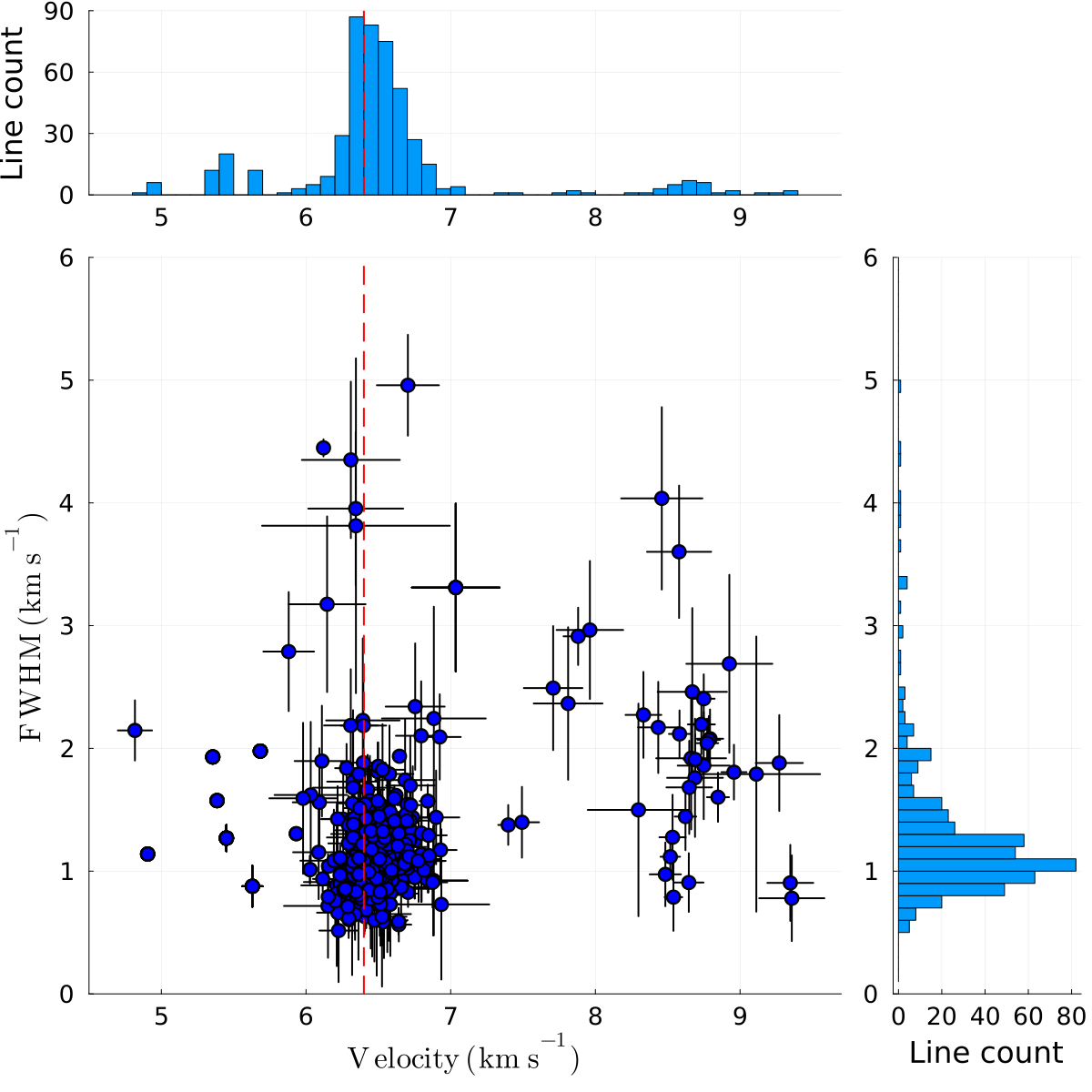}
  \caption{FWHM as a function of the Doppler shift velocity for all the detected lines, obtained through Gaussian fitting (bottom-left). The dashed red line indicates the envelope local standard of rest velocity of $6.4$ $\kms$. The top and right panels represent histograms of the line count for the Doppler shift velocity and the FWHM, respectively.}
  \label{fig:vfwhm}
\end{center}
\end{figure}

Figure \ref{fig:vfwhm} displays the FWHM and peak velocity of the fitted lines. Most lines are centered around $\sim$6.5 $\kms$, close to the estimated source systemic velocity of $\sim$6.4 $\kms$ \citep{Mercimek2022}, and have a width of $\sim$1 $\kms$. There is however a group centered around $\sim$ 9 $\kms$, composed of CH$_3$OH, CH$_2$DOH, OCS, and OC$^{34}$S. Several lines of these species actually exhibit doubly peaked profiles at $\sim 6.5$ $\kms$ and $\sim 9$ $\kms$, with the latter likely tracing the northern source as determined by ALMA and NOEMA observations \citep{Bianchi2020,Andreu2023,Cruz2022}. The two components are probably not always clearly separated, resulting in seemingly broad widths ($\geq 3$ $\kms$) for other lines of CH$_3$OH and OC$^{34}$S.

\begin{figure}
\begin{center}
\includegraphics[trim=1cm 0cm 1cm 0cm, width=0.45\textwidth]{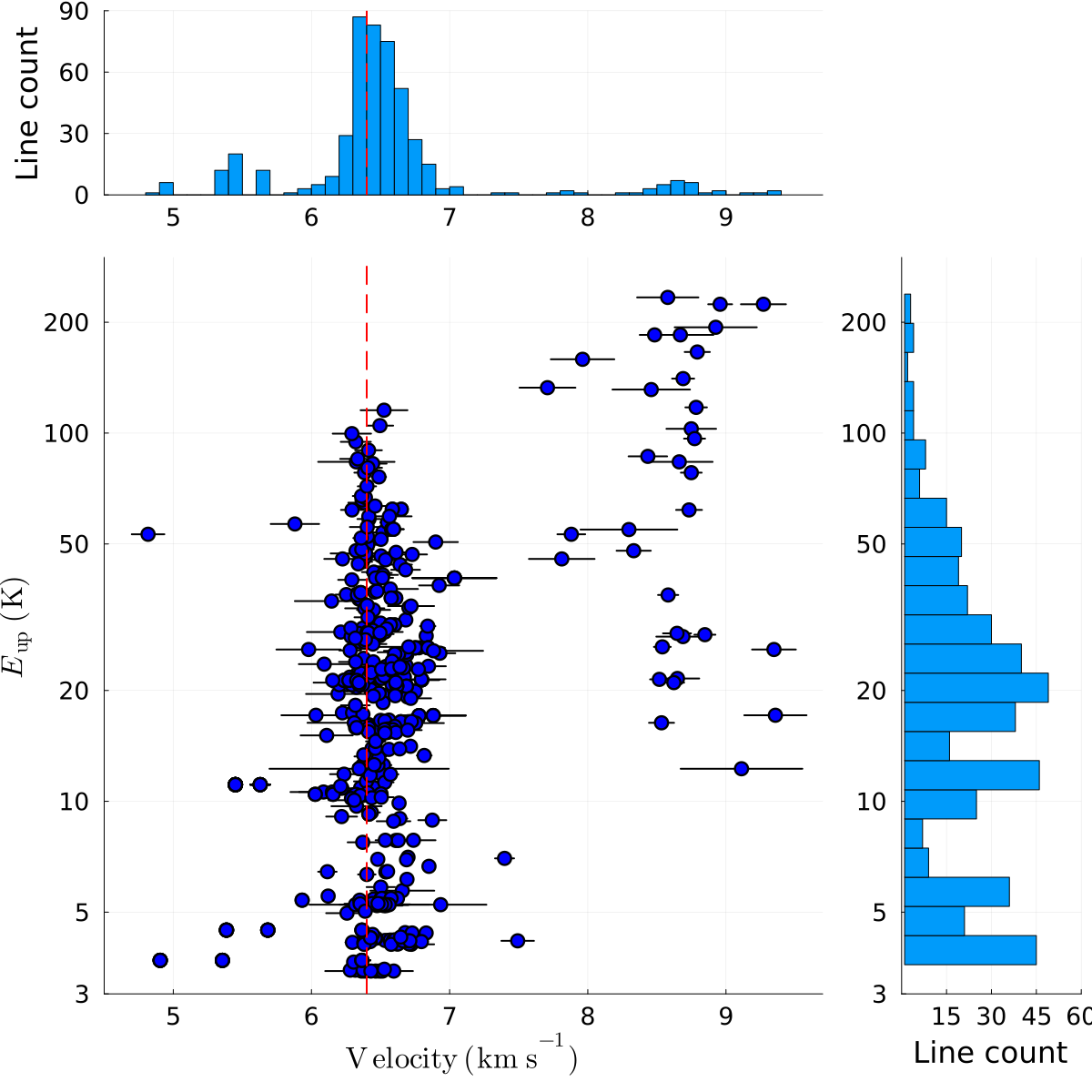}
  \caption{Same as Fig. \ref{fig:vfwhm}, for the $\eup$ as a function of the Doppler shift velocity.}
  \label{fig:veup}
\end{center}
\end{figure}

\begin{figure}
\begin{center}
\includegraphics[trim=1cm 0cm 1cm 0cm, width=0.45\textwidth]{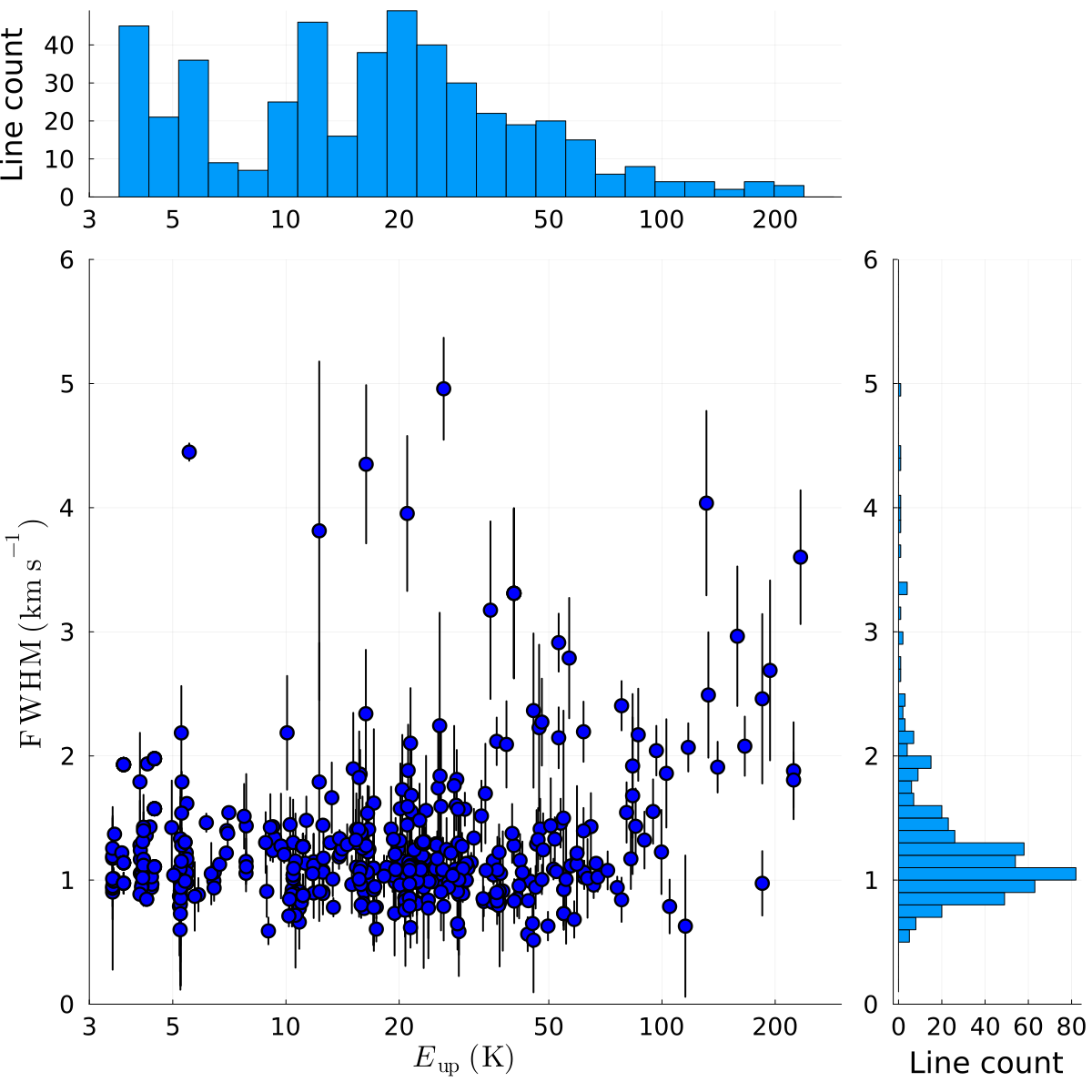}
  \caption{Same as Fig. \ref{fig:vfwhm}, for the FWHM as a function of the $\eup$.}
  \label{fig:eupfwhm}
\end{center}
\end{figure}

Likewise, we display the upper energy level $\eup$ of the transitions as a function of the velocity shift in Fig. \ref{fig:veup}. The number of detected lines smoothly declines above 20 K, but varies more randomly below this energy level. This is partly due to the numerous hyperfine components of CN, $^{13}$CN, N$_2$H$^+$, and N$_2$D$^+$ creating peaks around 4 K, 6 K, and 11 K.
At $6.4$ $\kms$, points are distributed at all $\eup$ values, from 3 K to $\sim$ 200 K. However, most lines at $\sim 9$ $\kms$ have an $\eup$ larger than 20 K. Above 100 K, there are more lines centered around 9 $\kms$ than 6.4 $\kms$. This suggests that this velocity regime traces hotter, and likely inner, emission, in agreement with interferometric studies \citep{Bianchi2020,Andreu2023}. 
Finally, we also display the FWHM as a function of $\eup$ in Fig. \ref{fig:eupfwhm}. The FWHM remains nearly constant around $\sim 1$ $\kms$ for $\eup < 100$ K. Nonetheless, nearly all $\eup > 100$ K have FWHM $\gtrsim 2$ $\kms$, which is broader than the vast majority of the other lines, and could be related to the presence of emission at both $\sim$6.4 and $\sim$9 $\kms$.

\subsection{Excitation temperatures and column densities}

We used radiative transfer modeling, assuming both local thermodynamic equilibrium (LTE) and non-LTE, to determine the column densities of the detected species, using the spectroscopic data of the CDMS and JPL databases.

\subsubsection{LTE modeling}\label{sec:lte}

\begin{figure*}
\begin{center}
\includegraphics[trim=2cm 0cm 2cm 0cm, width=0.99\textwidth]{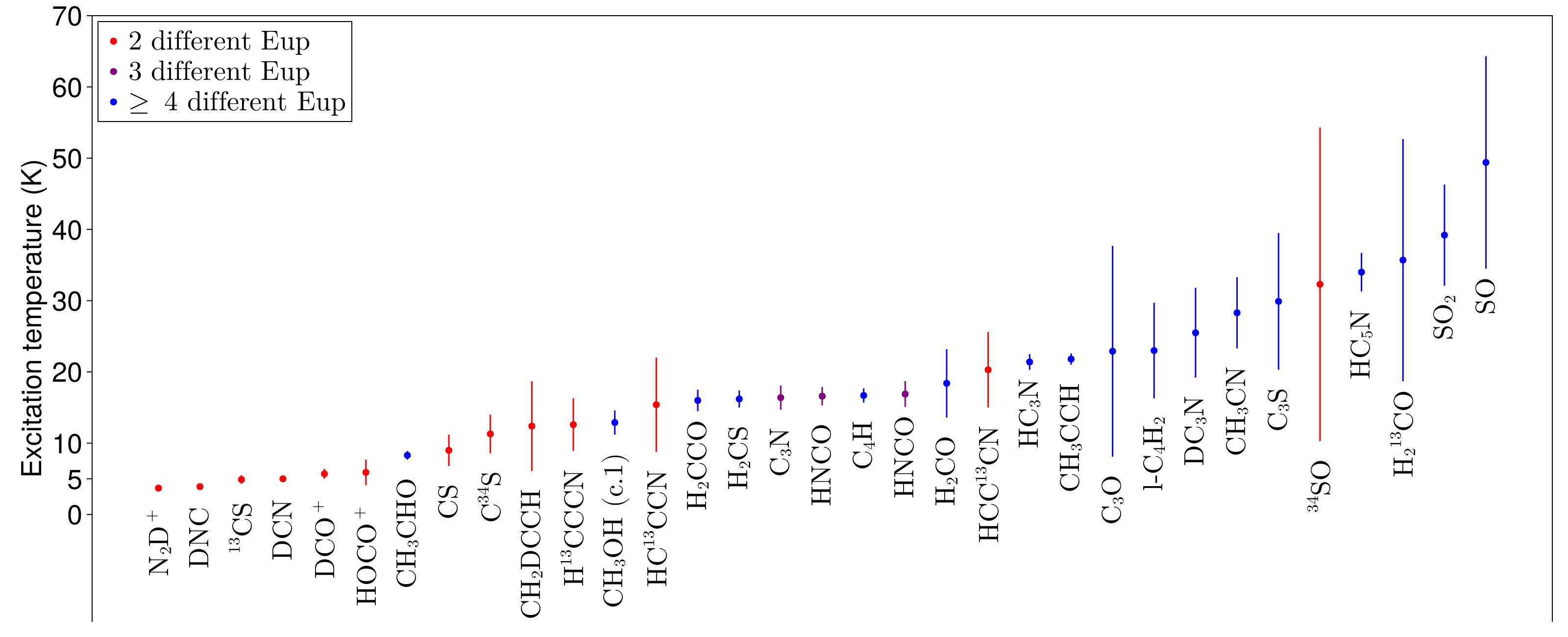}
  \caption{Excitation temperatures derived for each species with our LTE modeling, excluding OCS that has a very high excitation temperature (see Sect. \ref{sec:lte}). The species are sorted by increasing excitation temperatures. The colors represent the number of different $\eup$ available to derive the excitation temperatures: 2 (red), 3 (purple), and $\geq 4$ (blue). The vertical lines represent the error bars.}
  \label{fig:tex}
\end{center}
\end{figure*}

In the LTE case, the lines were fitted with a Python code \citep{Bottinelli2024}. It uses the Levenberg-Marquardt scheme of the Python package lmfit with a least-squares method, based on the line intensity equation:

\begin{equation}\label{eq:intensity}
  I(\nu) = \eta \left[ J(\nu,\tex)\left(1-e^{-\tau(\nu)}\right) - J(\nu,T_\mathrm{CMB})\left(1-e^{-\tau(\nu)}\right)\right],
\end{equation}
with 
\begin{align}
  J(\nu,T) & = \frac{h\nu}{\kb \left( e^{\frac{h\nu}{\kb T}}-1\right)},\\
  \tau(\nu) & = \tau_0 \exp\left[-\left(\frac{\nu-\nu_0}{\nu_0 (FWHM/c)}\right)^2 4 \ln(2) \right],\\
  \tau_0(\nu,\tex) & = \frac{c^3 A_\mathrm{ij} n_\mathrm{up}(\tex)\left(e^{\frac{h\nu}{\kb \tex}}-1\right)}{4\pi\nu_0^3 FWHM \sqrt{\pi/\ln(2)}},\\
  n_\mathrm{up}(\tex) & = \frac{{N_\mathrm{tot}} g_\mathrm{up}}{Q(\tex)e^{\frac{\eup}{\kb\tex}}}.   
\end{align}
Here, $\nu$ is the frequency, $\eta$ the filling factor, $n_\mathrm{up}$ the upper-level population, $g_\mathrm{up}$ the upper-level degeneracy, Q the partition function, $\eup$ the upper-level energy, $A_{ij}$ the spontaneous emission Einstein coefficient, $FWHM$ the full width  at half maximum of the line in the velocity space, $\nu_0$ the frequency of the transition, $\tex$ the excitation temperature, $T_\mathrm{CMB}=2.7$ K, $N_\mathrm{tot}$ the column density, and $\tau$ the optical depth. Also, $\eta$ is linked to the source size $\theta_\mathrm{s}$ by:
\begin{equation}\label{eq:eta}
    \eta = \frac{\theta_\mathrm{s}^2}{\theta_\mathrm{s}^2+\theta_\mathrm{B}^2},
\end{equation}
where $\theta_\mathrm{B}$ is the telescope beam size. Equation (\ref{eq:intensity}) is valid for a negligible continuum. $N_\mathrm{tot}$, $\tex$, $\theta_\mathrm{s}$, and $FWHM$ are unknowns and are used as parameters for the fit.
We first fit the species showing transitions spanning several $\eup$ values. When possible, for species with only one transition, we lift the degeneracy of the parameters by assuming the excitation temperature derived for an isotopologue. Indeed, we independently find similar excitation temperatures for c-C$_3$H$_2$ and c-C$_3$HD ($\sim$10 K), HC$_3$N, and its D and $^{13}$C isotopologues ($\sim$20 K), and CS and C$^{34}$S ($\sim$ 10 K). When taking the error intervals into account, the only exception is $^{13}$CS that seems to show a lower excitation temperature (4.9 K) than CS (9.0 K). Assuming a temperature of 9.0 K for $^{13}$CS results in a less accurate fit and a 40\% lower column density.
For several molecules, the partition function data is unavailable below 9.375 K. In that case, we extrapolated the partition function at lower temperatures, assuming a power law in log space (correlation coefficient $>0.99$ in every case). The resulting $\tex$, $N_\mathrm{tot}$, and $\theta_\mathrm{s}$ (with their uncertainties) are listed in columns 2 to 4 of Table \ref{tab:texncol}. All lines were fitted with a single component, except CH$_3$OH (see Sect. \ref{sec:methanol}) and OCS. Uncertainties (indicated in parenthesis) are given by lmfit and correspond to a 1 $\sigma$ confidence interval. We show all excitation temperatures derived using $\geq$ 2 $\eup$ values, sorted by increasing $T_\mathrm{ex}$, in Fig. \ref{fig:tex}. Fitting the emission of a species using a low number ($\lesssim 3$) of $\eup$ is easier than with a larger number of $\eup$, as the latter may mathematically result in smaller confidence intervals. Therefore, we used different colors in Fig. \ref{fig:tex} to illustrate the number of different $\eup$ used to derive each excitation temperature. We also display the radiative transfer models for all species in Appendix \ref{app:fit_lines}.

We also use CASSIS to perform rotational diagrams \citep{Goldsmith1999} for the species with more than two detected lines. 
All column densities found by fitting rotational diagrams agree within 20\% with the fitting of line intensities, while most excitation temperatures agree within 25\%. The exceptions are l-C$_3$H$_2$ (16.2$\pm$3.7 K with the line intensity fitting vs 25.3$\pm$3.8 K with the rotational diagram, respectively), CH$_3$CHO (8.3$\pm$0.6 K vs 11.4$\pm$0.9 K respectively), and H$_2$CO (18.4$\pm$4.8 K vs 9.7$\pm$3.6 K respectively). The values used in the rest of the paper are the ones obtained with the line intensity fitting method.

\subsubsection{The case of OCS}

\begin{figure}
\begin{center}
\includegraphics[trim=0cm 0cm 0cm 0cm, width=0.48\textwidth]{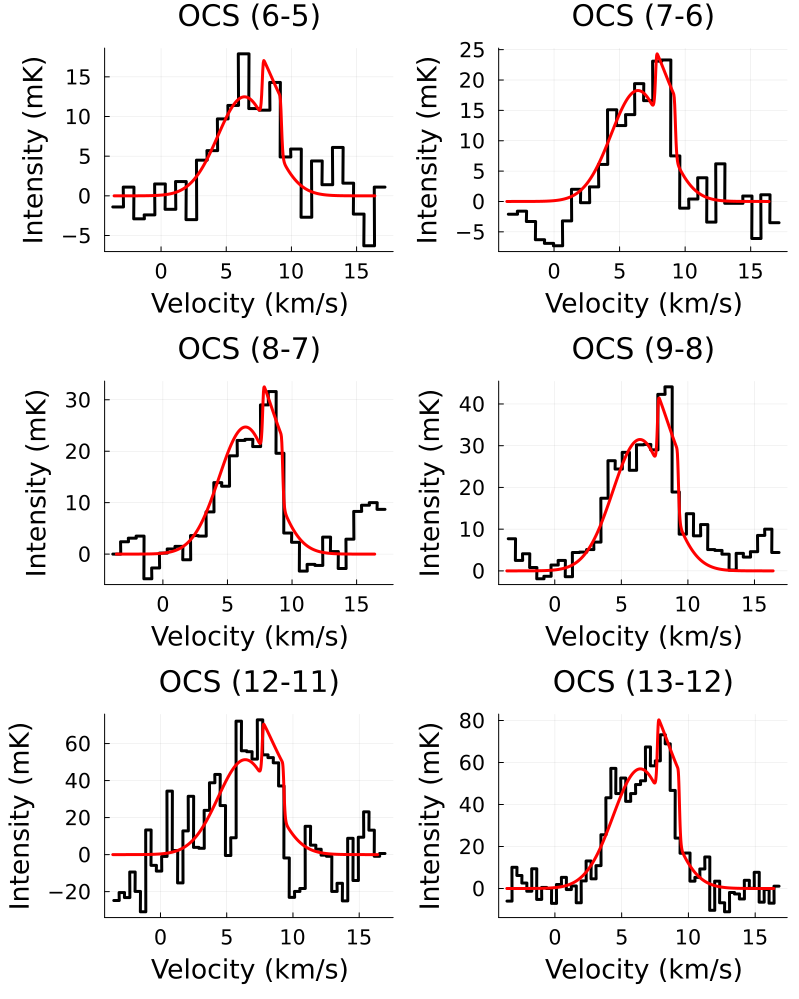}
 \caption{Detected OCS transition lines in black with the two-component LTE fit in red. The component at 9 $\kms$ is optically thick, which creates a sort of plateau.}
 \label{fig:OCS}
\end{center}
\end{figure}

Figure \ref{fig:OCS} shows the detected OCS transitions. The lines display broad ($\gtrsim$ 4 $\kms$) asymmetric profiles that can be fitted with two Gaussians, centered around $\sim$ 6.4 $\kms$ and $\sim$ 8.5 $\kms$. The narrow peak at 8.5 $\kms$ is close to the velocity of the northern source \citep{Bianchi2020,Andreu2023}. We therefore model this emission using two components, one corresponding to the envelope and centered on the system velocity of 6.4 $\kms$, and a second one at 8.5 $\kms$. ALMA observations of L1551~IRS~5 by \citet{Bianchi2020} suggest that the size of the methanol emission region (the hot corino) is about 0.15\arcsec. We therefore assume this source size for the second component. We fit the two components with the Levenberg-Marquardt scheme described above and display the model in red in Fig. \ref{fig:OCS}. The results of the fit are quite uncertain. We find a size of $42.0 \pm 30.3\arcsec$ for the first component, with an excitation temperature of $98.0 \pm 74.5$ K, a FWHM of $4.8 \pm 0.2$ $\kms$ and a column density of ($2.9 \pm 0.6) \times 10^{13}$ cm$^{-2}$. For the second component, we find a temperature of $400.9 \pm 39.2$ K and a column density of ($5.8 \pm 19.4) \times 10^{19}$ cm$^{-2}$. Although the model seems to reproduce the observation reasonably well, a nearly 100 K temperature at a scale of 40\arcsec ($>$5000 au) seems very unlikely. Moreover, the parameters have large uncertainties. Those results therefore cannot be considered as reliable even if they seem to suggest a warm emission for OCS. 
  With the IRAM-30m telescope, \citet{Mercimek2022} also observe broad OCS line profiles at higher frequencies ($\sim$220 GHz). They suggest that this emission originates from the $2\arcsec$ circumbinary disk, as observed by \citet{Takakuwa2020} with ALMA at $\sim$330 GHz.

\subsubsection{The case of methanol} \label{sec:methanol}
\begin{figure}
\begin{center}
\includegraphics[trim=0cm 0cm 0cm 0cm, width=0.48\textwidth]{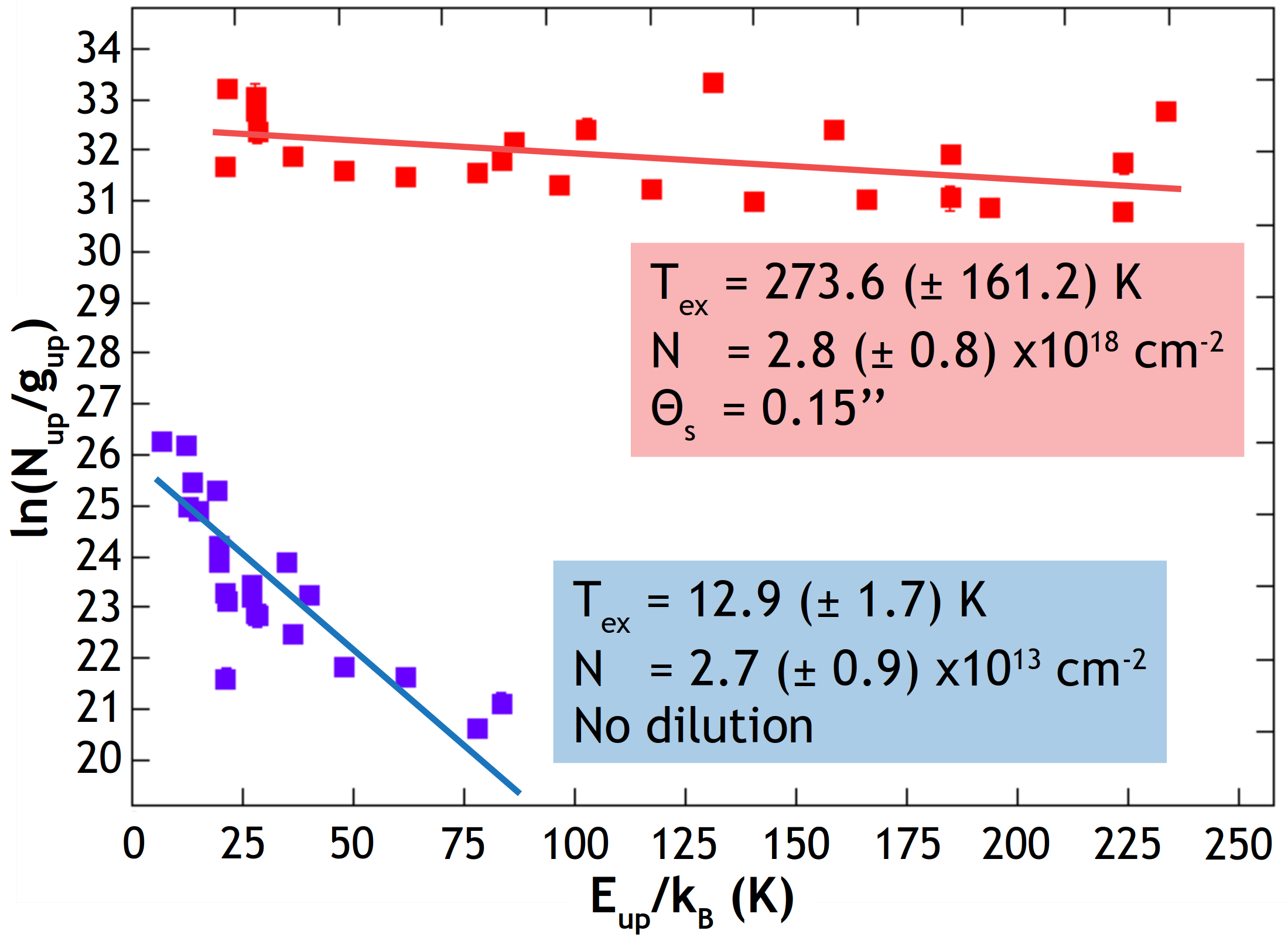}
  \caption{Rotational diagram for CH$_3$OH. Blue points represent components peaking at the system velocity $v_\mathrm{lsr} = 6.4$ $\kms$, for which we assume no beam dilution. Red points represent components peaking at $8.5$ km s$^{-1}$ for which we assume a $0.15$\arcsec emission region. The blue and red lines represent the linear regressions on those points.}
  \label{fig:ch3oh}
\end{center}
\end{figure}

We detected 37 lines of methanol (CH$_3$OH), distributed into 28 lines showing a single peak profile, and 9 lines with a double peak profile. Accordingly, we have fitted those lines with one or two Gaussians. We were not able to model all lines using the LTE and non-LTE methods described above. Instead, we used CASSIS to plot and fit the rotational diagram, displayed in Fig. \ref{fig:ch3oh}. We separated the lines in two groups. The blue points represent the component centered around the system velocity $v_\mathrm{lsr}$ of 6.4 $\kms$, while the red points are for the component centered around 8.5 $\kms$. The points in the first group all have $\eup$ $<100$ K, and forms a steep slope in the rotational diagram, indicative of a low excitation temperature. The linear regression on those points gives $T_\mathrm{ex}=12.9 \pm 1.7$ K and $N=(2.7\pm0.9)\times10^{13}$ cm$^{-2}$. This emission likely originates from the cold envelope at large scale.
The second group of points forms a very shallow slope and extends to $\eup>200$ K, suggesting a hot emission. The velocity of 8.5 $\kms$ matches the northern source velocity. Following \citet{Bianchi2020}, we therefore consider a $\theta_\mathrm{s}=0.15\arcsec$ emission region for the methanol in this source. The linear regression gives $T_\mathrm{ex}=273.6 \pm 161.2$ K and $N=(2.8 \pm 0.8) \times 10^{18}$ cm$^{-2}$, which is half an order of magnitude lower than the lower limit of $10^{19}$ cm$^{-2}$ derived by \citet{Bianchi2020} for CH$_3$OH in the northern source. However, the opacity of several of these excited lines is larger than 1, and the column density is consequently very probably underestimated.
These two fits are listed in Table \ref{tab:texncol}, with the labels "c.1" and "c.2." We displayed the radiative transfer models of the lines combining those two fits in Fig. \ref{fig:page_meth_1}. Overall, those parameters  reproduced the emission poorly, with several lines either dimmer or significantly brighter than what is observed. Assuming different values for $\theta_\mathrm{s}$ does not improve the results. Consequently, although it is fairly certain that the methanol emission can be decomposed into at least two components, namely, cold and hot, we did not use those fits in our further analysis.

\subsubsection{Non-LTE modeling}
We also carried out non-LTE models for the species with available collisional rate coefficients. We used the radiative transfer code Radex \citep{Vandertak2007}, assuming an expanding sphere geometry. The collision coefficients with H$_2$ (ortho, para or scaled from He) were taken from the Lamda, Basecol and EMAA databases (the references for each species are listed in the notes of Table \ref{tab:texncol}). When both ortho and para-H$_2$ coefficients were available, we assumed an ortho/para ratio of $10^{-3}$, as long as the kinetic temperature remained below 20 K; otherwise we assumed the standard ratio at LTE \citep{Faure2019}. For each species, we run Radex for a grid of kinetic temperatures $T_\mathrm{kin}$, column densities of the species $N_\mathrm{tot}$, and number densities of the collider n(H$_2$). Each model gives an opacity and an excitation temperature for each transition that we use to reproduce the line profile using Equation (\ref{eq:intensity}). We choose parameters that reproduce the best our measured intensity on all lines for a given species, by minimizing the L2 norm of the weighted residuals:
\begin{equation}
    \chi^2 = \sum_i \frac{(I_{\mathrm{obs},i} - I_\nu(\mathbf{\theta}))^2}{\sigma_i^2}.
\end{equation}
Here, $I_{\mathrm{obs},i}$ is the intensity in the spectral channel $i$ of the transition lines, $I_\nu(\mathbf{\theta})$ is the modeled intensity calculated with the set of parameters $\mathbf{\theta}$ following Equation (\ref{eq:intensity}), and $\sigma_i^2$ is the error on the intensity, calculated as
\begin{equation}
    \sigma_i^2 = rms_i^2 + (cal\times I_{\mathrm{obs},i})^2,
\end{equation}
where $rms_i$ is the noise at the channel frequency and $cal=0.1$, the estimated calibration error of our observations (10\%). We define confidence intervals by calculating parameter ranges, where $\chi^2$ is less than 1.5 than the minimum value, to evaluate how  constrained those parameters are. The excitation temperatures, column densities, kinetic temperatures, and H$_2$ densities are listed in columns 5 to 8 of Table \ref{tab:texncol}.

\begin{table*}
  \caption{Best-fit parameters for each species, with both LTE and non-LTE modeling when possible. The number in parenthesis indicate the uncertainties. Bracketed values for the non-LTE $T_\mathrm{ex}$ indicate the range of excitation temperatures for the lines of the species.}
\label{tab:texncol}
\centering
\begin{tabular}{lllllllll}
\hline\hline
Species & $T_\mathrm{ex}$ (K) & N (cm$^{-2}$)  & $\Theta_\mathrm{s}$(\arcsec) & $T_\mathrm{ex}$ (K)  & N (cm$^{-2}$)  & $T_\mathrm{kin}$ (K) & n(H$_2$) (cm$^{-3}$)\\
 &  (LTE) & (LTE) & (LTE) & (non-LTE) &  (non-LTE)  & (non-LTE) & (non-LTE)\\
\hline
CCH $^{b,g}$ & 8.9  & $3.6 (0.2)\times 10^{14}$ & $>100$ &  - & - & - & - \\
CCD $^{e,g}$ &  8.9 (0.7)   &  $9.4(0.3)\times 10^{12}$ &  - & - & - & - \\
C$^{13}$CH $^{b,g}$ & 8.9  & $4.0 (0.3)\times 10^{12}$ &  - & - & - & - \\ 
CCS $^e$ &  12.1 (1.1)   &  $2.8(1.0)\times 10^{12}$ & 39.2 (18.1)  & - & - & - & - \\
C$_3$H $^e$ &  9.0 (0.7)   &  $3.5(0.5)\times 10^{11}$  & $>100$ & - & - & - & - \\
c-C$_3$H$_2$ $^e$ &  10.4 (0.3)   &  $1.3(0.1)\times 10^{13}$  & 64.0 (16.8) & - & - & - & - \\
c-C$_3$HD $^e$ &  9.8 (0.8)   &  $9.3(1.4)\times 10^{11}$ & 52.2 (22.9)  & - & - & - & -\\
c-CC$^{13}$CH$_2$ $^e$ &  7.8 (2.7)   &  $3.6(0.5)\times 10^{11}$ & $>100$  & - & - & - & - \\
l-C$_3$H$_2$ $^e$ &  16.2 (3.7)   &  $1.7(0.2)\times 10^{11}$ & $>100$  & - & - & - & -\\
C$_3$N $^e$ &  16.4 (1.7)   &  $2.3(0.3)\times 10^{11}$ & $>100$  & - & - & - & -\\
C$_3$O $^e$ &  22.9 (14.8)   &  $1.2(0.2)\times 10^{11}$ & $>100$  & - & - & - & - \\ 
C$_3$S  &  29.9 (9.6)   &  $2.0(0.3)\times 10^{11}$ & $>100$  & - & - & - & -\\ 
C$_4$H $^e$ &  16.7 (1.0)   &  $4.3(0.7)\times 10^{12}$ & $>100$  &  - & - & - & - \\ 
l-C$_4$H$_2$ $^e$ &  23.0 (6.7)   &  $2.6(1.6)\times 10^{11}$ & $>100$  & - & - & - & -\\ 
CH$_3$CCH $^e$ &  21.8 (0.8)   &  $1.8(0.0)\times 10^{13}$ & $>100$  & - & - & - & -\\ 
CH$_2$DCCH  &  12.4 (6.3)   &  $3.3(0.8)\times 10^{12}$ & $>100$  & - & - & - & -\\ 
CH$_3$CHO $^{a,e}$  &  8.3 (0.6)   &  $2.3(0.4)\times 10^{12}$ & $>100$  & - & - & - & -\\ 
CH$_3$CN  &  28.3 (5.0)   &  $2.1(0.3)\times 10^{11}$  & $>100$ & - & - & - & -\\ 
CH$_3$OH (c.1) $^d$ &  12.9 (1.7)   &  $2.7(0.9)\times 10^{13}$ & $>100$  & - & - & - & -\\ 
CH$_3$OH (c.2) $^d$ &  273.6 (161.2)   &  $2.8(0.8)\times 10^{18}$ & $0.15$  & - & - & - & -\\ 
CS $^e$ &  9.0 (2.2)   &  $1.2(0.2)\times 10^{13}$ & $>100$  & [10.2 -- 12.5] & $1.5(0.1)\times 10^{13}$ & 14.0 (3.7) & $6.7(93)\times 10^{5}$\\ 
$^{13}$CS $^{a,e}$  &  4.9 (0.6)   &  $7.3(1.5)\times 10^{11}$ & $>100$  & [4.8 -- 5.5] & $5.0(0.1)\times 10^{11}$ & 42.0 (7.4) & $4.0(0.1)\times 10^{4}$\\ 
C$^{33}$S $^{a,b}$ &  9.0   &  $2.1(0.2)\times 10^{11}$ & $>100$  & - & - & - & -\\ 
C$^{34}$S  &  11.3 (2.7)   &  $1.4(0.1)\times 10^{12}$  & $>100$ & 13.0 & $1.2(0.1)\times 10^{12}$ & 13.0 (10.8) & $1.0(1.0)\times 10^{8}$\\ 
HCN $^b$ &  5.0   &  $3.0(0.7)\times 10^{13}$ & $>100$  & 10.0 & $6.8(2.3)\times 10^{12}$ & 11.0 & $2.0\times 10^{6}$\\ 
DCN $^{a,e}$ &  5.0 (0.3)   &  $1.3(0.1)\times 10^{12}$ & $>100$  & [6.1 -- 12.3] & $1.1(0.3)\times 10^{12}$ & 11.0 (4.9) & $2.0(3.4)\times 10^{6}$\\ 
H$^{13}$CN $^{a,b}$ &  5.0   &  $6.1(0.3)\times 10^{11}$ & $>100$  & - & - & - & -\\ 
HC$^{15}$N $^{a,b}$ &  5.0   &  $1.9(0.2)\times 10^{11}$ & $>100$  & - & - & - & -\\ 
HNC $^b$ &  -$^f$   &  - & - & 8.7 & $1.2(0.1)\times 10^{13}$ & 9.0 & $1.2\times 10^{6}$\\
DNC  &  3.9 (0.3)   &  $5.2(2.4)\times 10^{12}$  & $>100$ & [7.2 -- 9.4] & $1.4(0.4)\times 10^{12}$ & 9.0 (4.7) & $1.2(8.8)\times 10^{6}$\\ 
HN$^{13}$C $^b$  &  3.9   &  $1.3(0.1)\times 10^{12}$ & $>100$  & 9.2 & $5.6(0.5)\times 10^{11}$ & 9.0 & $1.2\times 10^{6}$\\ 
H$^{15}$NC $^{a,b}$ &  3.9   &  $2.4(0.2)\times 10^{11}$ & $>100$  & - & - & - & -\\ 
HNCO $^e$ &  16.9 (1.8)   &  $1.1(0.2)\times 10^{12}$ & $>100$  & - & - & - & -\\ 
DCO$^+$  &  5.7 (0.7)   &  $3.0(0.7)\times 10^{12}$ & $>100$  & [9.1 -- 10.9] & $2.0(0.1)\times 10^{12}$ & 10 $^c$ (2.4) & $1.4(98)\times 10^{6}$\\ 
H$^{13}$CO$^+$ $^b$  &  5.7   &  $4.1(0.7)\times 10^{12}$ & $>100$  & - & - & - & -\\ 
HC$^{18}$O$^+$ $^b$  &  5.7   &  $2.7(0.2)\times 10^{11}$ & $>100$  & - & - & - & -\\ 
HC$_3$N $^e$ &  21.4 (1.1)   &  $3.5(0.2)\times 10^{12}$ & $>100$  & 21.0 & $3.6(0.5)\times 10^{12}$ & 21.0 (4.0) & $6.7(6.6)\times 10^{7}$\\ 
DC$_3$N $^e$ &  25.5 (6.3)   &  $1.2(0.1)\times 10^{11}$ & $>100$  & - & - & - & -\\ 
H$^{13}$CCCN $^e$ &  12.6 (3.7)   &  $1.1(0.5)\times 10^{11}$ & $>100$  & - & - & - & -\\ 
HC$^{13}$CCN  &  15.4 (6.6)   &  $6.2(3.6)\times 10^{10}$ & $>100$  & - & - & - & -\\ 
HCC$^{13}$CN $^e$ &  20.3 (5.3)   &  $7.2(1.4)\times 10^{10}$ & $>100$  & - & - & - & -\\ 
HC$_5$N $^e$ &  34.0 (2.7)   &  $5.2(0.6)\times 10^{11}$ & $>100$  & - & - & - & -\\ 
H$_2$CO $^e$ &  18.4 (4.8)   &  $2.2(0.5)\times 10^{13}$ & $>100$  & [7.9 -- 11.6] & $2.6(0.7)\times 10^{12}$ & 10.0 $^c$ (4.1) & $6.7(93)\times 10^{5}$\\ 
D$_2$CO $^b$ &  18.4    &  $2.7(0.2)\times 10^{12}$ & $>100$  & - & - & - & -\\ 
H$_2$$^{13}$CO $^e$ &  35.7 (17.0)   &  $1.1(0.5)\times 10^{12}$ & $>100$  & - & - & - & -\\ 
H$_2$CCO $^e$ &  16.0 (1.5)   &  $2.2(0.1)\times 10^{12}$ & $>100$  & - & - & - & -\\ 
H$_2$CS $^e$ &  16.2 (1.2)   &  $2.4(0.1)\times 10^{12}$ & $>100$  & - & - & - & -\\ 
\hline
\end{tabular}
\end{table*}

\begin{table*}
  \caption*{\textbf{Table \ref*{tab:texncol}} continued}
\centering
\begin{tabular}{llllllll}
\hline\hline
Species & $T_\mathrm{ex}$ (K) & N (cm$^{-2}$)  & $\Theta_\mathrm{s}$(\arcsec) & $T_\mathrm{ex}$ (K)  & N (cm$^{-2}$)  & $T_\mathrm{kin}$ (K) & n(H$_2$) (cm$^{-3}$) \\
 &  (LTE) & (LTE) & (LTE) & (non-LTE) &  (non-LTE)  & (non-LTE) & (non-LTE)\\
\hline
HNCO  &  16.6 (1.3)   &  $1.1(0.1)\times 10^{12}$ & $>100$  & - & - & - & -\\ 
HOCO$^+$  &  5.9 (1.8)   &  $2.2(1.1)\times 10^{11}$  & $>100$ & - & - & - & -\\ 
N$_2$D$^{+}$ $^a$ &  3.7 (0.0)   &  $1.2(0.0)\times 10^{12}$  & $>100$ & - & - & - & - \\ 
SO $^e$ &  49.4 (14.9)   &  $2.3(0.5)\times 10^{13}$  & $>100$ & - & - & - & -\\ 
$^{34}$SO  &  32.3 (22)   &  $9.4(2.9)\times 10^{11}$ & $33.1 (33.3)$  & - & - & - & -\\ 
SO$_2$ $^e$ &  39.2 (7.1)   &  $7.3(4.2)\times 10^{12}$  & $17.3(7.8)$ & - & - & - & -\\ 
\hline
\end{tabular}
  \tablefoot{$^a$ The partition functions of those species have been extrapolated below 9.375 K. $^b$ The column densities of those species have been calculated using the excitation temperature of their isotopologue. $^c$ Minimum kinetic temperature below which collision coefficients are not available. $^d$ See Sect. \ref{sec:methanol}. $^e$ Species for which a rotational diagram has been fitted (see Sect. \ref{sec:lte}). The following is the list of references for collision coefficients; CS: \citet{DenisAlpizar2013}, $^{13}$CS and C$^{34}$S: \citet{Lique2006}, HCN: \citet{Goicoechea2022}, DCN: \citet{Magalhaes2018}, HNC and DNC: \citet{HernandezVera2017}, HN$^{13}$C: \citet{Dumouchel2010}, DCO$^+$: \citet{Flower1999}, HC$_3$N: \citet{Faure2016}, H$_2$CO: \citet{Wiesenfeld2013}. $^f$ We could not find a good fit of HNC in LTE, so we did not include it in the Table. $^g$ We do not report our non-LTE results for CCH, CCD, and C$^{13}$CH due to very poor constraints with the fitting method.}
\end{table*}

\subsubsection{Comparison between the LTE and non-LTE models}

The excitation temperatures given by both methods are similar for CS and $^{13}$CS. On the other hand, the LTE modeling of DCN, DNC and DCO$^+$ converges toward very low $\tex \leq 5$ K, while the non-LTE models are closer to 10 K. The kinetic temperature of the gas is not well constrained since we only have access to two different $\eup$ values for most of the modeled species here. The uncertainties are also large for the H$_2$ density. 

The column densities are mostly consistent between LTE and non-LTE models. They are very close to each other for C$^{34}$S, DCN, and HC$_3$N, with a difference of less than 20\%, while the difference is less than a factor of 2 for CS, $^{13}$CS, DCO$^+$, and H$_2$CO. However, DNC and HN$^{13}$C differ by a factor 2 to 3, and HCN by a factor 4.5. Those discrepancies do not seem correlated to the difference in excitation temperature between LTE and non-LTE.

The non-LTE modeling of $^{13}$CS results in a large kinetic temperature (42 K) but an H$_2$ density of only a few $10^4$ cm$^{-3}$. A temperature of 42 K is consistent with the inner envelope or a disk, while densities of $10^4$ cm$^{-3}$ are typical of outer envelopes at $T \lesssim 10$ K. In addition, for CS and C$^{34}$S, we find kinetic temperatures and excitation temperatures, both in LTE and non-LTE, close to 10 K. Those are indications that the $T_\mathrm{kin}=42$ K value for $^{13}$CS needs to be taken with caution. However, the excitation temperature and column density match the LTE modeling with only a $\sim$30\% difference.

\subsubsection{Upper limits}
Here, we provide upper limits on the column densities of several non-detected isotopologues. For each species, we calculated the maximum line flux of all transitions covered by our spectral survey:
\begin{equation}
  F_\mathrm{max} = 3 rms \sqrt{2\mathrm{d}v\times FWHM},
\end{equation}
where $rms$ is the noise level at the current frequency, $\mathrm{d}v$ is the channel width in km s$^{-1}$ (200 kHz in frequency space), and $FWHM$ is the line width in km s$^{-1}$.
The corresponding column densities are calculated based on $F_\mathrm{max}$ assuming LTE with the same excitation temperature, FWHM, and source size as the main isotopologue. The upper limit is taken as the lowest value among all transitions. The results are reported in Table \ref{tab:upperlimits} with the superscript $^a$. Those upper limits range from $ 4.5\times 10^{10}$ to $1.7 \times 10^{12}$ cm$^{-2}$. $^{13}$CCH and c-C$_3$D$_2$ are more than a hundred times less abundant than their main isotopologue. The upper limits on l-C$_3$HD and l-C$_3$D$_2$ are only three times lower than the column density of l-C$_3$H$_2$, while the upper limits for HDCS and D$_2$CS are ten times lower compared to H$_2$CS. 

We also calculated the upper limits of several species detected in other Class~I protostars (see Sect. \ref{sec:classI}), listed with the superscript $^b$ in Table \ref{tab:upperlimits}. We employed the same method and assumed no beam dilution and a FWHM of 1 $\kms$. We ran the calculations for excitation temperatures of 10, 20, and 30 K and took the largest of the three upper limits. Apart from C$_6$H, which shows an upper limit of $2.0 \times 10^{12}$ cm$^{-2}$, all the other species have column densities lower than $10^{12}$ cm$^{-2}$.

\begin{table}
  \caption{Upper limit on column densities for undetected isotopologues.}
\label{tab:upperlimits}
\centering
\begin{tabular}{lll}
\hline\hline
Species  & Column density (cm$^{-2}$)\\
\hline
$^{13}$CCH $^a$ & $\leq 1.7\times 10^{12}$ \\ 
c-C$_3$D$_2$ $^a$ & $\leq 1.1\times 10^{11}$ \\ 
l-C$_3$HD $^a$ & $\leq 5.6\times 10^{10}$ \\ 
l-C$_3$D$_2$ $^a$ & $\leq 4.5\times 10^{10}$ \\ 
CH$_3$CCD $^a$ & $\leq 1.4\times 10^{12}$ \\ 
H$_2$C$^{18}$O $^a$ & $\leq 3.4\times 10^{11}$ \\
HDCS $^a$ & $\leq 2.3\times 10^{11}$ \\
D$_2$CS $^a$ & $\leq 3.2\times 10^{11}$ \\
C$_2$H$_3$CN $^b$ & $\leq 2.8\times 10^{11}$ \\
C$_5$H $^b$ & $\leq 6.8\times 10^{11}$ \\
C$_6$H $^b$ & $\leq 2.0\times 10^{12}$ \\
HNCCC $^b$ & $\leq 8.9\times 10^{10}$ \\
HCCNC $^b$ & $\leq 3.1\times 10^{11}$ \\
HCCCHO $^b$ & $\leq 6.0\times 10^{11}$ \\
H$_2$CCN $^b$ & $\leq 1.1\times 10^{11}$ \\
HCNO $^b$ & $\leq 5.0\times 10^{10}$ \\
HSCN $^b$ & $\leq 1.3\times 10^{11}$ \\
NH$_2$CHO $^b$ & $\leq 3.8\times 10^{11}$ \\
\hline
\end{tabular}
\tablefoot{$^a$ We assumed the same excitation temperature, FWHM and beam dilution than the main isotopologue in our LTE models (see Table \ref{tab:texncol}). $^b$ We assumed no beam dilution, a FHWM of 1 $\kms$, and excitation temperatures of 10, 20, and 30 K. The displayed upper limit is the largest among the three.}
\end{table}

\subsection{Isotopic fractionation}

The calculation of column densities allows us to determine the isotopic fractionation ratios of several molecules, for the isotopes D, $^{13}$C, $^{17}$O, $^{18}$O, $^{15}$N, $^{33}$S, and $^{34}$S. The results are listed in Table \ref{tab:isotop}. The statistically corrected ratios, that account for the probability that a D atom replaces a H atom in equivalent sites in the molecule are calculated as:
\begin{align}
    \frac{\mathrm{XD}}{\mathrm{XH}}&=\frac{\mathrm{XHD}}{\mathrm{XH}_2} = \frac{1}{2}\frac{N(\mathrm{XHD})}{N(\mathrm{XH}_2)}, \\
    \frac{\mathrm{XD}}{\mathrm{XH}}&=\frac{\mathrm{XH}_2\mathrm{D}}{\mathrm{XH}_3} = \frac{1}{3}\frac{N(\mathrm{XH}_2\mathrm{D})}{N(\mathrm{XH}_3)}, \\
    \frac{\mathrm{XD}}{\mathrm{XH}}&=\frac{\mathrm{XD}_2}{\mathrm{XH}_2} = \sqrt{\frac{N(\mathrm{XHD})}{N(\mathrm{XH}_2)}}, 
\end{align}
where $N$(X) is the column density of molecule X. Similarly to the deuterium ratios, we have divided the c-C$_3$H$_2$/c-HCC$^{13}$CH column density ratio by 2 to account for the statistical weight of the substituted atom. We also added upper and lower limits based on the estimates of Table \ref{tab:upperlimits}.

The $^{12}$C/$^{13}$C ratios lie between $\sim$16 and $\sim$60, except for CCH, which exhibits high column density ratios with both isotopologues: C$^{13}$CH (90) and $^{13}$CCH ($\geq 211$) in LTE. The $^{13}$CCH/C$^{13}$CH ratio is comparable to previous measurements in TMC1 and L1527 and is well explained by the H + $^{13}$CCH $\rightarrow$ H + C$^{13}$CH reaction, which favors C$^{13}$CH \citep{Loison2020}. 
Comparing with the ISM standard value of $\sim$68 \citep{Milam2005}, we can infer that the lines of the main isotopologues of CS, HCN, HNC, and H$_2$CO are most likely optically thick; we consider the ratios involving those species as upper or lower limits. For HC$_3$N, there seems to be a slight enrichment in $^{13}$C, contrary to previous measurements in dense clouds (Table 1 of \citealt{Loison2020}). This apparent enrichment may be due to the $^{13}$C + HC$_3$N reaction \citep{Loison2020} or (more likely) to a slightly optically thick emission of the main isotopologue.

For the mono-deuterated molecules, the deuterium fractionation ratios vary from 2\% to 6\% using the LTE column densities. 
Those values are higher than the cosmic D/H ratio of $\sim$1.4 $\times$ 10$^{-5}$ \citep{Linsky2003}, but consistent with other protostellar environments \citep{Ceccarelli2014,Riaz2022,Giers2023}.
In Table \ref{tab:isotop}, we also display several ratios of a secondary isotopologue with respect to the $^{13}$C isotopologue, for example, DCN/H$^{13}$CN 
that we renormalize by the standard factor of $^{12}$C/$^{13}$C=68. 
This allows us to estimate the deuterium fractionation ratio of HCN, HNC, and HCO$^+$. 
The DCN/(H$^{13}$CN $\times$ 68) and DNC/(HN$^{13}$C $\times$ 68)  are about 3.1\% and 5.8\%, respectively, which are lower than the upper limits derived with the main isotopologues ($<$4.3\% and $<$16\%, respectively). The DCO$^{+}$/ (H$^{13}$CO$^+$ $\times$ 68) is found to be $\sim$1.1\%.
For the doubly deuterated molecules, we only provide upper limits on the statistically corrected ratios, due to their non-detection or the optical thickness of the lines of the main isotopologue. The lowest of those upper limits is $\sim$9\% for c-C$_3$D$_2$/c-C$_3$H$_2$, while the upper limits for l-C$_3$D$_2$/l-C$_3$H$_2$, D$_2$CO/H$_2$CO, and D$_2$CS/H$_2$CS stand above 35\%. 

 We detect two species bearing the $^{15}$N isotope, HC$^{15}$N, and H$^{15}$NC. The direct ratio of HCN/HC$^{15}$N only gives a lower limit of $>$ 158, since the HCN emission is optically thick. Using the $^{13}$C isotopologue, we find a $^{14}$N/$^{15}$N ratio of 217.0 $\pm$ 54.4 for HCN and 367.2 $\pm$ 50.8 for HNC. Those values are in the range of what is found in other protostars  using indirect measurements with H$^{13}$CN and and HN$^{13}$C \citep[$\sim$160--370 for HCN and $\sim$240--460 for HNC,][]{Wampfler2014,Yoshida2019} as well as a direct measurement for HCN in the Class 0 protostar L483 \citep[321~$\pm$~96,][]{Agundez2019}. 
  However, indirect measurements using the $^{13}$C isotopologues could lead to erroneous results if the $^{12}$C/$^{13}$C differs from 68, as shown by observations toward molecular clouds and prestellar cores \citep[e.g.,][]{Ikeda2002,Magalhaes2018,Jensen2024} and theoretical studies \citep[e.g.,][]{Roueff2015,Loison2020,Colzi2020}. The actual HCN/HC$^{15}$N and HNC/H$^{15}$NC ratios in L1551~IRS~5 may therefore be different from the values we obtain through their $^{13}$C isotopologues. Despite the many uncertainties surrounding measurements of the $^{14}$N/$^{15}$N ratio in HCN and HNC, it seems that these measurements tend to give a $^{14}$N/$^{15}$N ratio notably smaller than the solar value of 441.

H$_2$C$^{18}$O and HC$^{18}$O$^+$ are the only two $^{18}$O-bearing species that we detected. The local measurements of H$_2$CO/H$_2$C$^{18}$O ratios typically gives values of $\gtrsim$ 500 \citep[e.g.,][]{Lucas1998,Wilson1999}. 
However, we only derived a very low lower limit of 65 for this ratio, due to the probably optically thick emission of H$_2$CO and the non-detection of H$_2$C$^{18}$O. 
We also obtain an (H$^{13}$CO$^+$ $\times$ 68)/HC$^{18}$O$^+$ ratio of 1033 $\pm$ 190, which is 50\% higher than what is measured in diffuse molecular clouds for HCO$^+$/HC$^{18}$O$^+$ (672 $\pm$ 110, \citealt{Lucas1998}) and in the protostar L1527 based on H$^{13}$CO$^+$ \citep{Yoshida2019}.

\citet{Yan2023} measured the sulfur isotopologue ratios in the Milky Way, and find $^{32}$S/$^{34}$S $\sim 15-25$, $^{32}$S/$^{33}$S $\sim 50-100$, and $^{34}$S/$^{33}$S $\sim 3-6$. We find consistent values here for L1551~IRS~5, in agreement with \citet{Esplugues2023} for the Class 0 protostar B335. The ratios are however on the lower end of the intervals for CS, which supports the idea of an optically thick CS emission.

\begin{table}
  \caption{Statistical isotopic fractionation ratios for molecules whose column density has been estimated, for the LTE and non-LTE cases.}
\label{tab:isotop}
\centering
\begin{tabular}{lcc}
\hline\hline
Species   &   LTE ratio   &  non-LTE ratio\\
\hline
& D / H & \\
\hline
CCD / CCH & 0.026 $\pm$ 0.002 & -- \\
c-C$_3$HD / c-C$_3$H$_2$ & 0.036 $\pm$ 0.006 & -- \\
c-C$_3$D$_2$ / c-C$_3$H$_2$ $^a$ & $\leq 0.09$ & -- \\
l-C$_3$HD / l-C$_3$H$_2$ $^a$ & $\leq 0.33$ & -- \\
l-C$_3$D$_2$ / l-C$_3$H$_2$ $^a$ & $\leq 0.51$ & -- \\
CH$_2$DCCH / CH$_3$CCH & 0.061 $\pm$ 0.015 & -- \\
CH$_3$CCD / CH$_3$CCH $^a$ & $\leq 0.078$ & -- \\
DCN / HCN & $<0.043$ & $<0.162$ \\
DCN / (H$^{13}$CN $\times 68$) & 0.031 $\pm$ 0.003 & --\\
DNC / HNC & -- & $<0.117$ \\
DNC / (HN$^{13}$C $\times 68$) & 0.058 $\pm$ 0.028 & 0.037 $\pm$ 0.011\\
DC$_3$N / HC$_3$N & 0.034 $\pm$ 0.003 & -- \\
D$_2$CO / H$_2$CO & $<0.35$ & -- \\
HDCS / H$_2$CS $^a$ & $\leq 0.095$ & -- \\
D$_2$CS / H$_2$CS $^a$ & $\leq 0.36$ & -- \\
DCO$^+$ / (H$^{13}$CO$^+$ $\times 68$) & 0.011 $\pm$ 0.003 & -- \\
\hline
& $^{12}$C / $^{13}$C & \\
\hline
CCH / C$^{13}$CH & 90 $\pm$ 8.4 & -- \\
CCH / $^{13}$CCH $^a$ & $\geq 211$ & -- \\
c-C$_3$H$_2$ / c-HCC$^{13}$CH & 72.2 $\pm$ 10.4 & -- \\
CS / $^{13}$CS & $>16.4$ & $ > 10.0$ \\
HCN / H$^{13}$CN & $> 49.2$ & -- \\
HNC / HN$^{13}$C & -- & $> 21.4$ \\
HC$_3$N / H$^{13}$CCCN & 31.8 $\pm$ 14.6 & -- \\
HC$_3$N / HC$^{13}$CCN & 56.5 $\pm$ 32.9 & -- \\
HC$_3$N / HCC$^{13}$CN & 48.6 $\pm$ 9.9 & -- \\
H$_2$CO / H$_2^{13}$CO & $>20.0$ & -- \\
\hline
& $^{14}$N / $^{15}$N & \\
\hline
HCN / HC$^{15}$N & $>157.9$ & -- \\
(H$^{13}$CN $\times 68$) / HC$^{15}$N & 217.0 $\pm$ 54.4 & -- \\
(HN$^{13}$C $\times 68$) / H$^{15}$NC & 367.2 $\pm$ 50.8 & -- \\
\hline
& $^{16}$O / $^{18}$O & \\
\hline
H$_2$CO /H$_2$C$^{18}$O $^a$ & $\geq 65$ & -- \\
(H$^{13}$CO$^+$ $\times 68$) / HC$^{18}$O$^+$ & 1033 $\pm$ 190 & -- \\
\hline
& $^{32}$S / $^{34}$S, $^{32}$S / $^{33}$S & \\
\hline
CS / C$^{34}$S & $>8.6$ & $>12.5$ \\
CS / C$^{33}$S & $>57.1$ & -- \\
C$^{34}$S / C$^{33}$S & 6.7 $\pm$ 0.8 & -- \\
SO / $^{34}$SO & 24.5 $\pm$ 9.2 & -- \\
\hline
\end{tabular}
\tablefoot{$^a$ We provide upper and lower limits for non-detected species (see Table \ref{tab:upperlimits}).}
\end{table}

\section{Discussion}\label{Sec:discussion}

\subsection{Comparison with previous studies of L1551~IRS~5}

In this section, we compare the estimated physical parameters with what has been found by other observations of the same source. \citet{Roberts2002} observed a similar frequency range, focusing on HCN, H$_2$CO, and their isotopologues between $\sim$72 GHz and $\sim$150 GHz\footnote{They also observed the (4,0,4--3,0,3) transition of HDCO at 256.585 GHz.}. The observations were done with the NRAO 12m radio telescope, resulting in a beam twice larger than ours (40\arcsec-88\arcsec or $\sim$5750-13000 au). They derived column densities in LTE assuming several excitation temperatures from 5 K to 40 K, except for HCN and DCN; in those cases, they found excitation temperatures of 5 K and 6 K, respectively, in accordance with our values. The column densities for DCN, H$^{13}$CN, and H$_2^{13}$CO only have a 20\% difference with ours, sign of an extended emission from those species. We however find lower column densities on H$_2$CO (by a factor of 2) and HCN (by a factor of 3). This may be a bias due to the optical thickness of the detected lines, that \citet{Roberts2002} derive to be 2.4 for H$_2$CO and 16.9 for HCN. Our upper limit of 4.3\% for the DCN /HCN ratio (in LTE) is therefore higher than their 1.9\% estimate.

\citet{Jorgensen2004} performed observations in the 3mm, 2mm, and 1mm bands using the 15m James Clerk Maxwell Telescope and the Onsala 20m telescope. Comparing the integrated fluxes of the lines detected in both surveys, we obtained similar values for SO (2,3--1,2), with a 15\% higher value in our case, and for CN (1--0), for which we have $\leq 30$\% larger fluxes on each hyperfine component. We  found a flux of $\approx$ 0.33 $\pm$ 0.02 K km s$^{-1}$ on the H$^{13}$CN (1 -- 0) line when integrating the three hyperfine components, whereas they estimated an upper limit of 0.12 K km s$^{-1}$. Conversely, they measured a flux that is twice as high for N$_2$H$^+$ (1 -- 0) despite their larger beam. This is probably a consequence of a strong large-scale N$_2$H$^{+}$ emission, as observed by \citet{Tatematsu2004}.

\citet{Mercimek2022} surveyed the frequency range 214.5--238.0 GHz with the IRAM-30m telescope. They therefore probed the source with a smaller beam (11\arcsec or $\sim$ 1500 au). They derive excitation temperatures and column densities using a rotational diagram analysis. We estimated higher column densities for DCN (3 times higher), N$_2$D$^{+}$ (10 times in LTE, 4 times in non-LTE), $^{13}$CS (3 times in LTE, 2 times in non-LTE), and DCO$^{+}$ (9 times). Those species with excitation temperatures $<10$ K are usually associated with the colder envelope. However, \citet{Mercimek2022}  assumed a 20-35 K rotational temperature in their analysis. The differences may therefore be due to the excitation temperature used to derive the column density, or to a decreasing column density of those species in the inner regions.
We also find lower excitation temperatures than the 20-35 K assumed in their analysis for CCD (8.9 K in our case), CCS (12.1 K) and c-C$_3$H$_2$ \citep[24.4 K in][10.4 K in our case]{Mercimek2022}, but our column densities are similar within 20\% for those three species. There are also a few species for which our estimated column density is much lower than \citet{Mercimek2022}, namely H$_2$CCO (10 times lower), CH$_3$CN (5 times), CH$_3$CCH (3 times), OCS (2 times), and CH$_3$CHO (7 times). Except for CH$_3$CHO, those species have excitation temperatures $>15$ K, which suggests that their emission region is not as extended (and therefore not as cold) as species with $T_\mathrm{ex}<10$ K, resulting in different dilution factors ($\eta$ in Equations (\ref{eq:intensity}) and (\ref{eq:eta})). We however find excitation temperatures and column densities in agreement within 10\% for SO, SO$_2$, and o-D$_2$CO, despite their excitation temperatures of $\gtrsim 20$ K.

\subsection{Isotopic ratios}

Excluding the upper limits, most deuterium fractionation ratios range from 1 to 6\%, and are consistent with values found in other Class 0/I sources; namely DC$_3$N/HC$_3$N \citep{Agundez2019,Bianchi2019}, c-C$_3$HD/c-C$_3$H$_2$ \citep{Agundez2019,Giers2022}, HDCS/H$_2$CS \citep{Agundez2019} as well as DNC/HNC, DCO$^{+}$/H$^{13}$CO$^{+}$, DCN/H$^{13}$CN,  and DNC/HN$^{13}$C \citep{Riaz2022}. These values are higher than the HDO/H$_2$O ratio of $\sim$0.1\% found by \citet{Andreu2023} in L1551~IRS~5, but this difference between water and organic species is observed in other sources \citep{Parise2005,Coutens2012,Jorgensen2018}.
Our D$_2$CO/H$_2$CO upper limit ($<$35\%) is lower than the high ratio found by \citet{Mercimek2022} in L1551~IRS~5 (45-84\%), which is likely due to the optical thickness of the H$_2$CO lines. Our upper limit of $<$9.5\% for the HDCS/H$_2$CS ratio is consistent with their estimate of 9\% to 14\%.

The highest (non-upper limit) ratio is CH$_2$DCCH/CH$_3$CCH with 6.1\%, which is almost twice the level of deuteration of the other species in our study. This is also consistent with other measurements, as \citet{Markwick2002} found a ratio of $\geq 10$\% in TMC-1 with the Onsala 20m telescope, while \citet{Agundez2019} measured 6.5\% in L483 with the IRAM-30m. Our upper limit of 7.8\% on the CH$_3$CCD/CH$_3$CCH ratio is also high, while \citet{Agundez2019} measured a 5.9\% ratio, close to their CH$_2$DCCH/CH$_3$CCH ratio of 6.5\%. The proximity of both ratios is also supported by chemical models \citep{Agundez2021}. We could therefore expect a CH$_3$CCD/CH$_3$CCH also close to 6\%. CH$_3$CCH is a relatively complex molecule with seven atoms. ALMA observations of IRAS~16293--2422 show that the deuteration of the largest COMs is higher by a factor of about 2-4 than the simplest COMs \citep{Jorgensen2018}. A similar trend is seen in our data between the simplest species and CH$_3$CCH. \citet{Jorgensen2018} suggested that for the COMs, it could be due to different timescales with the most complex ones forming at a later stage with denser and colder conditions. However, this explanation could be different in the case of the present study as CH$_3$CCH is predicted to form very early on by chemical models \citep[e.g.,][]{Coutens2020} and it is also detected in various molecular clouds \citep{Turner2000}. D-H exchanges could potentially occur at later stages and increases the D/H ratio of CH$_3$CCH.
Conversely, DCO$^+$ seems to display a much lower deuteration ratio of $\sim$ 1\%, reflecting its formation in the diffuse ISM.

We found a high CCH/C$^{13}$CH ratio of 90 with the LTE models, at the same level as in the prestellar core L1544 and the Class 0 protostars HH211 and L483 \citep{Agundez2019,Giers2023}. This may indicate that this ratio experiences little variation during the evolution of a protostar, at least in the envelope. Similarly, an even higher CCH/$^{13}$CCH ratio is found in those sources ($>150$), which may be consistent with our lower limit of 211. 
However, single-dish observations of L1527 by \citet{Yoshida2019} and TMC1 by \citet{Sakai2010} show ratios almost twice larger for both isotopologues, which suggests that the carbon fractionation in CCH may be sensitive to the environment in which it occurs. 

The three HC$_3$N isotopologues containing single $^{13}$C substitutions show similar fractionation ratios. However, their relative abundances H$^{13}$CCCN : HC$^{13}$CCN : HCC$^{13}$CN of $\sim$ 1:0.6$\pm$0.4:0.7$\pm$ 0.3 are quite different from other protostellar sources, which display a 1:1:x ratio, with x$\approx$ 1.2-2.1 \citep[see][and references therein]{Agundez2019}. Despite relatively large error bars on our measurements, HCC$^{13}$CN does not seem to be more abundant than the two others in L1551~IRS~5.

\subsection{Abundance ratios as evolutionary tracers}\label{sec:evol}

\begin{figure}
\begin{center}
\includegraphics[trim=0cm 0cm 0cm 0cm, width=0.48\textwidth]{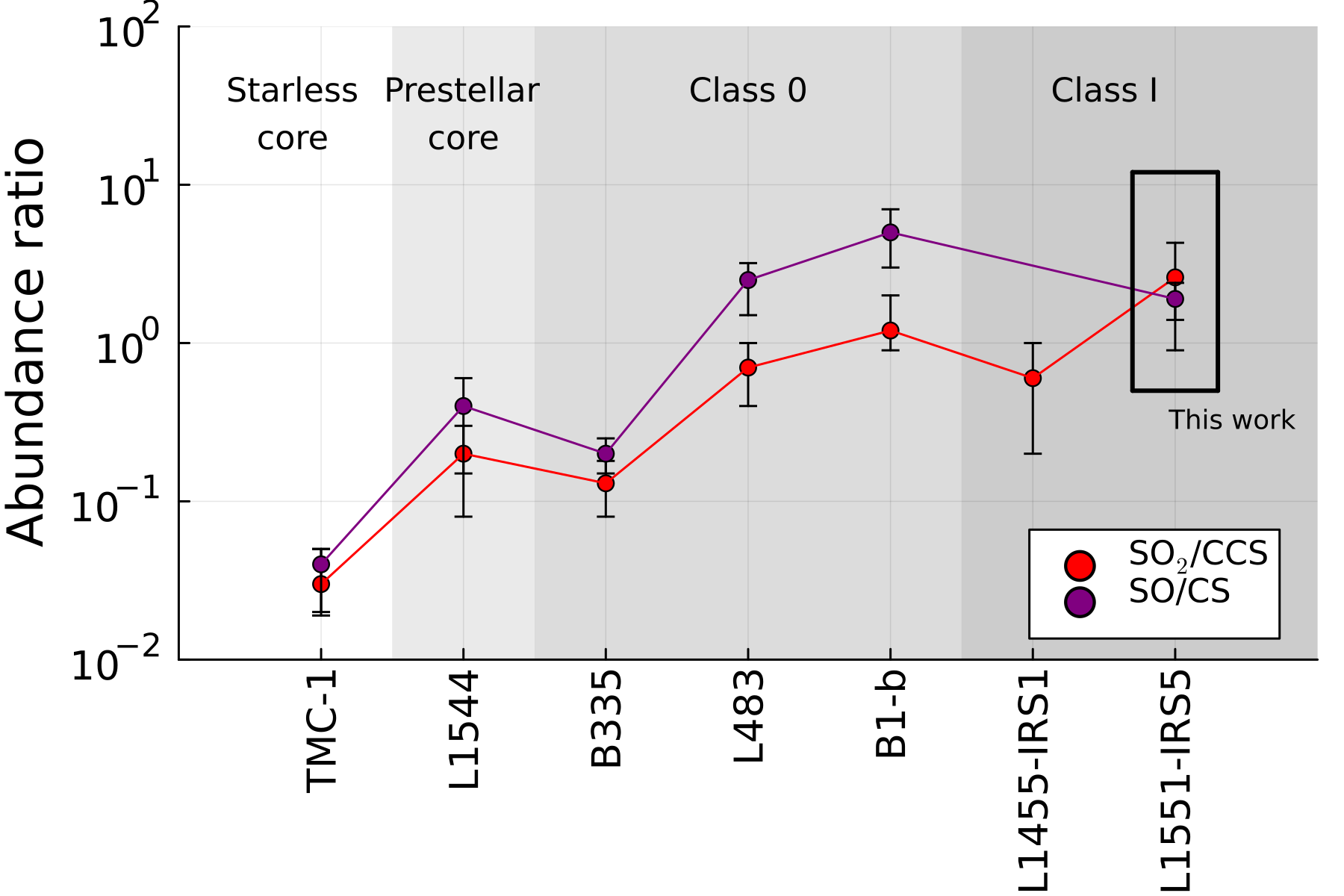}
  \caption{Abundance ratios of SO$_2$/CCS (red) and SO/CS (purple) in several sources and in L1551~IRS~5. The ratios for TMC-1, L1544, L483 and B1-b are taken from \citet{Agundez2019}, B335 from \citet{Esplugues2023}, and the value for L1455 IRS 1 is taken from \citet{Mercimek2022}.}
  \label{fig:evol_ratio}
\end{center}
\end{figure}

Abundance ratios are often proposed as evolution tracers in protostellar environments. \citet{Agundez2019} and \citet{Esplugues2023} propose that the SO$_2$/CCS and SO/CS abundance ratios should increase with the evolutionary stage (from starless core to Class~I). 
They find overall increasing ratios going from the starless core TMC1, then to the prestellar core L1544, to the Class 0 protostars B335, L483, and B1-b. As a Class~I protostar, L1551~IRS~5 represents the next evolutionary stage, and we plot those ratios alongside our measurements in Fig. \ref{fig:evol_ratio}. Our points lie in the same ranges as the Class 0 sources and so is the SO$_2$/CCS ratio for the Class~I protostar L1455 IRS 1. These results need to be taken with caution, as we measure very different excitation temperatures for the two species in each ratio (39.2 K and 12.1 K for SO$_2$ and CCS respectively, and 49.4 K and 9.0 K for SO and CS), indicating that their emission originates from different regions. That is not apparent in \citet{Agundez2019} that assume a $\sim 10$ K rotational temperature for deriving the column densities, while \citet{Esplugues2023} assume a kinetic temperature of $15$ K for their non-LTE calculations. Assuming a 10 K excitation temperature for those species results in a SO/CS ratio that is three times greater and a SO$_2$/CCS ratio three times lower, which does not change the trend. We therefore conclude that the increasing trends may be good indicators of evolutionary stages, but only in the early phases of the star formation process.

\subsection{Comparison with other Class~I protostars}\label{sec:classI}

Here, we compare the chemical composition of L1551~IRS~5 with other Class~I protostars for which large spectral surveys with the IRAM-30m telescope are available. L1527 and SVS~13A have been observed as part of the ASAI large program \citep{Lefloch2018}. The survey spans several frequency ranges from 85 GHz to 272 GHz with a spectral resolution of 200 kHz, and achieves a rms of 2 to 7 mK. L1489~IRS, B5~IRS~1 and L1455~IRS~1 were observed by \citet{Mercimek2022} at 2mm, also with a spectral resolution of 200 kHz and a rms of 10 mK. The distance and luminosity of each source is listed in Table \ref{tab:othersources}, alongside L1551~IRS~5. Except for SVS~13A (another an eruptive young star), those sources are dimmer than the 22 L$_\odot$ of L1551~IRS~5 \citep{Froebrich2005}. We do not include the sources surveyed by \citet{Legal2020} because they only reported the molecules with $\leq 3$ atoms so far.

Table \ref{tab:detections} lists the species detected in each source. Comparing the detections (\yesyes) with non-detections (\nono), L1527 and L1551~IRS~5 show the largest numbers of species. Particularly, L1527 is known for its rich carbon-chain chemistry \citep{Sakai2008}. Comparatively, we do not detect several species in L1551~IRS~5 that are present in L1527; namely C$_2$H$_3$CN, C$_5$H, C$_6$H, HCCCHO, HCCNC, H$_2$CCN, HNCCC, HCNO, and HSCN. We provide upper limits on the column densities of those species in Table \ref{tab:upperlimits}. 
\citet{Yoshida2019} performed a survey towards L1527 with the Nobeyama 45m telescope at 3mm and calculated a column density for C$_5$H of ($5.3 \pm 1.1)\times 10^{11}$ cm$^{-2}$, which is about 400 times less abundant than C$_4$H. A similar ratio in L1551~IRS~5 would yield a C$_5$H column density of $\sim$10$^{10}$ cm$^{-2}$, which is consistent with the upper limit of $6.8\times 10^{11}$ cm$^{-2}$ we derived. In the same manner, \citet{Araki2017} determined a C$_4$H/C$_6$H abundance ratio of 153 in L1527. This would mean for L1551~IRS~5 a C$_6$H column density of $2.8\times 10^{10}$ cm$^{-2}$, which is lower than the upper limit of $2.0\times 10^{12}$ cm$^{-2}$.
\citet{Marcelino2010} calculated column densities for HNCO and its isomers in several protostellar sources including L1527. They found HCNO/HNCO ratios of 40 in L1527 and ranging from $\sim$20 to $\sim$ 80 in the other sources. Similar ratios in L1551~IRS~5 result in an HCNO column density between $1.4$ and $5.5 \times 10^{10}$ cm$^{-2}$, also consistent with our upper limit of $5.0 \times 10^{10}$ cm$^{-2}$. HCCNC and HNCCC are both isomers of HC$_3$N. Their relative abundance have been measured in TMC-1 by \citet{Cernicharo2020}, with ratios of $77 \pm 8$ for HC$_3$N/HCCNC, and $392 \pm 22$ for HC$_3$N/HNCCC. Applying those ratios to HC$_3$N, we find column densities of $4.5 \times 10^{10}$ cm$^{-2}$ for HCCNC and $8.9 \times 10^{9}$ cm$^{-2}$ for HNCCC, both lower than the upper limits. 
In summary, C$_5$H, C$_6$H, HCCNC, HNCCC, and HCNO may still be present in L1551~IRS~5 with the same abundances relative to C$_4$H (for C$_5$H and C$_6$H) or their isomer as L1527, but the lower abundances of C$_4$H and the main isomers in L1551~IRS~5 make their detection more difficult.

Except HCCCHO, all O-bearing COMs detected in L1527 are also present in L1551~IRS~5 as well as in SVS~13A. On the other hand, a fewer number of COMs are detected in the three sources L1489~IRS, B5~IRS~1, and L1455~IRS~1 by the survey of \citet{Mercimek2022}. Notably, CH$_3$CHO and CH$_3$OCHO are missing despite being detected in L1551~IRS~5 within the same study. 
Those three sources and SVS~13A also show a lower number of hydrocarbons than L1527 and L1551~IRS~5, with the lack of detection of c-C$_3$H, l-C$_3$H$_2$, and l-C$_4$H$_2$. The survey of \citet{Mercimek2022} only covers transitions with $\eup > 50$ K for the last two species though. More generally, all species detected in L1489~IRS, B5~IRS~1, and L1455~IRS~1 are also detected in L1551~IRS~5, but not the opposite. H$_2$CCO and CH$_3$CN are not detected by the survey of \citet{Mercimek2022} in any of those three sources, while CCS, H$_2$S, HNCO, and OCS are only seen in L1455~IRS~1 (in addition to L1551~IRS~5). The most notable exception is SiO, that is detected in L1455~IRS~1 and SVS~13A, but not in L1527 and L1551~IRS~5 despite their closer distance. The lack of SiO may suggest the absence of shocks in those systems, although shocked gas have been reported in L1551~IRS~5 using other tracers \citep{Yang2022}. 
NH$_2$CHO is also a species detected in SVS~13~A but not seen in L1551~IRS~5.

\begin{table}
  \caption{Properties of the selected sources from the ASAI program \citep{Lefloch2018} and from \citet{Mercimek2022}.}
\label{tab:othersources}
\centering
\begin{tabular}{llll}
\hline\hline
Source   &  Lum. (L$_\odot$) & Dist. (pc) & Location\\
\hline
L1551 IRS 5 & 22 & 141 & Taurus\\
L1527 & 2.75 & 141 & Taurus\\
L1489 IRS & 3.5 & 141 & Taurus\\
SVS 13A & 34 & 260 & Perseus\\
B5 IRS 1 & 5.0 & 294 & Perseus\\
L1455 IRS 1 & 3.6 & 294 & Perseus\\
\hline
\end{tabular}
\end{table}

\begin{table*}
  \caption{List of detected species (only the main isotopologues) in Class~I protostars with IRAM-30m surveys. }
\label{tab:detections}
\centering
\begin{tabular}{llllllll}
\hline\hline
Species & L1527 & SVS 13A & L1489 IRS & B5 IRS 1 & L1455 IRS 1 & L1551~IRS~5 $^a$ & L1551~IRS~5 (This work)\\
\hline
CCH &\yesyes&\yesyes&\yesyes&\yesyes&\yesyes&\yesyes&\yesyes\\
C$_2$H$_3$CN &\yesyes&\nono& - & - & - & - &\nono\\
CCS &\yesyes&\yesyes&\nono&\nono&\yesyes&\yesyes&\yesyes\\
C$_3$H &\yesyes&\nono& h & h & h & h &\yesyes\\
c-C$_3$H &\yesyes&\nono&\nono&\nono&\nono&\yesyes&\yesyes\\
c-C$_3$H$_2$ &\yesyes&\yesyes&\yesyes&\yesyes&\yesyes&\yesyes&\yesyes\\
l-C$_3$H$_2$ &\yesyes&\nono& h & h & h & h &\yesyes\\
C$_3$N &\yesyes&\nono& h & h & h & h &\yesyes\\
C$_3$O &\yesyes&\nono& h & h & h & h &\yesyes\\
C$_3$S &\yesyes&\yesyes& r & r & r & r &\yesyes\\
C$_4$H &\yesyes&\yesyes& h & h & h & h &\yesyes\\
l-C$_4$H$_2$ &\yesyes&\nono& h & h & h & h &\yesyes\\
C$_5$H &\yesyes&\nono& r & r & r & r &\nono\\
C$_6$H &\yesyes&\nono& r & r & r & r &\nono\\
CH$_3$CCH &\yesyes&\yesyes&\nono&\nono&\nono&\yesyes&\yesyes\\
CH$_3$CHO &\yesyes&\yesyes&\nono&\nono&\nono&\yesyes&\yesyes\\
CH$_3$CN &\yesyes&\yesyes&\nono&\nono&\nono&\yesyes&\yesyes\\
CH$_3$OCH$_3$ &\yesyes&\yesyes& - & - & - & - &\yesyes\\
CH$_3$OCHO &\yesyes&\yesyes&\nono&\nono&\nono&\yesyes&\yesyes\\
CH$_3$OH &\yesyes&\yesyes&\nono&\yesyes&\yesyes&\yesyes&\yesyes\\
CN &\yesyes&\yesyes&\yesyes&\yesyes&\yesyes&\yesyes&\yesyes\\
CO &\yesyes&\yesyes&\yesyes&\yesyes&\yesyes&\yesyes&\yesyes\\
CS &\yesyes&\yesyes& - &\yesyes&\yesyes&\yesyes&\yesyes\\
HCCCHO &\yesyes&\nono&\nono&\nono&\nono&\nono&\nono\\
HC$_3$N &\yesyes&\yesyes& h & h & h & h &\yesyes\\
HC$_5$N &\yesyes&\yesyes& r & r & r & r &\yesyes\\
HCCNC &\yesyes&\yesyes& h & h & h & h &\nono\\
HCN &\yesyes&\yesyes&\yesyes&\yesyes&\yesyes&\yesyes&\yesyes\\
HCNO &\yesyes&\nono& h & h & h & h &\nono\\
HCO &\yesyes&\yesyes& - & - & - & - &\yesyes\\
HCO$^{+}$ &\yesyes&\yesyes&\yesyes&\yesyes&\yesyes&\yesyes&\yesyes\\
HCS$^+$ &\yesyes&\yesyes& r & r & r & r &\yesyes\\
H$_2$CO &\yesyes&\yesyes&\yesyes&\yesyes&\yesyes&\yesyes&\yesyes\\
H$_2$CCN &\yesyes&\yesyes& h & h & h & h &\nono\\
H$_2$CCO &\yesyes&\nono&\nono&\nono&\nono&\yesyes&\yesyes\\
H$_2$CS &\yesyes&\yesyes&\nono&\yesyes&\yesyes&\yesyes&\yesyes\\
H$_2$S & h & h &\nono&\nono&\yesyes&\yesyes& r \\
HNC &\yesyes&\yesyes& r & r & r & r &\yesyes\\
HNCCC &\yesyes&\nono& h & h & h & h &\nono\\
HNCO &\yesyes&\yesyes&\nono&\nono&\yesyes&\yesyes&\yesyes\\
HNO &\yesyes&\nono& r & r & r & r &\yesyes\\
HOCN &\yesyes&\nono& h & h & h & h &\yesyes\\
HOCO$^{+}$ &\yesyes&\nono& h & h & h & h &\yesyes\\
HSCN &\yesyes&\nono& h & h & h & h &\nono\\
NH$_2$CHO &\nono&\yesyes& - & - & - & - &\nono\\
NH$_3$ / NH$_2$D &\yesyes&\yesyes& h & h & h & h &\yesyes\\
N$_2$H$^+$ / N$_2$D$^+$ &\yesyes&\yesyes&\nono&\yesyes&\yesyes&\yesyes&\yesyes\\
NS & - & - & r & r & r & r &\yesyes\\
OCS &\yesyes&\yesyes&\nono&\nono&\yesyes&\yesyes&\yesyes\\
SiO &\nono&\yesyes&\nono&\nono&\yesyes&\nono&\nono\\
SO &\yesyes&\yesyes&\yesyes&\yesyes&\yesyes&\yesyes&\yesyes\\
SO$_2$ &\yesyes&\yesyes&\nono&\yesyes&\yesyes&\yesyes&\yesyes\\
\hline
Total & 48/50 & 33/50 & 7/23 & 12/24 & 17/24 & 23/25 & 41/52 \\
Fraction & 96\% & 66\% & 30\% & 50\% & 71\% & 92\% & 72\% \\
\hline
\end{tabular}
\tablefoot{Data for L1527 and SVS 13A are taken from \citet{Lefloch2018} in the [80 -- 116] GHz frequency range, and data for L1489~IRS, B5~IRS~1, L1455~IRS~1 and L1551~IRS~1 (with the superscript $^a$) are taken from \citet{Mercimek2022} in the [214.5 -- 238] GHz range. Bold-font letters stand for confirmed detections (\yesyes) and non-detections (\nono). Dashes (-) stand for species that have not been reported. The letter r (as in "range") indicates that the species that do not have transitions in the frequency ranges of the studies. The letter h (as in "hot") indicates that the species have transitions within the frequency range of the studies but only with high E$_\mathrm{up}$ values ($\geq 50$ K) and they were not reported. The "total" line counts the number of \yesyes$~$over the number of \yesyes + \nono$~$ for each source.}
\end{table*}

\subsection{Consequences of a luminosity outburst}

As a FUor-like protostar, L1551~IRS~5 probably experienced a luminosity outburst that could have impacted the chemical history of its envelope. The starting date, duration, and intensity of this burst are however unconstrained, and its influence on the envelope composition unknown.

\begin{figure}
\begin{center}
\includegraphics[trim=0cm 0cm 0cm 0cm, width=0.48\textwidth]{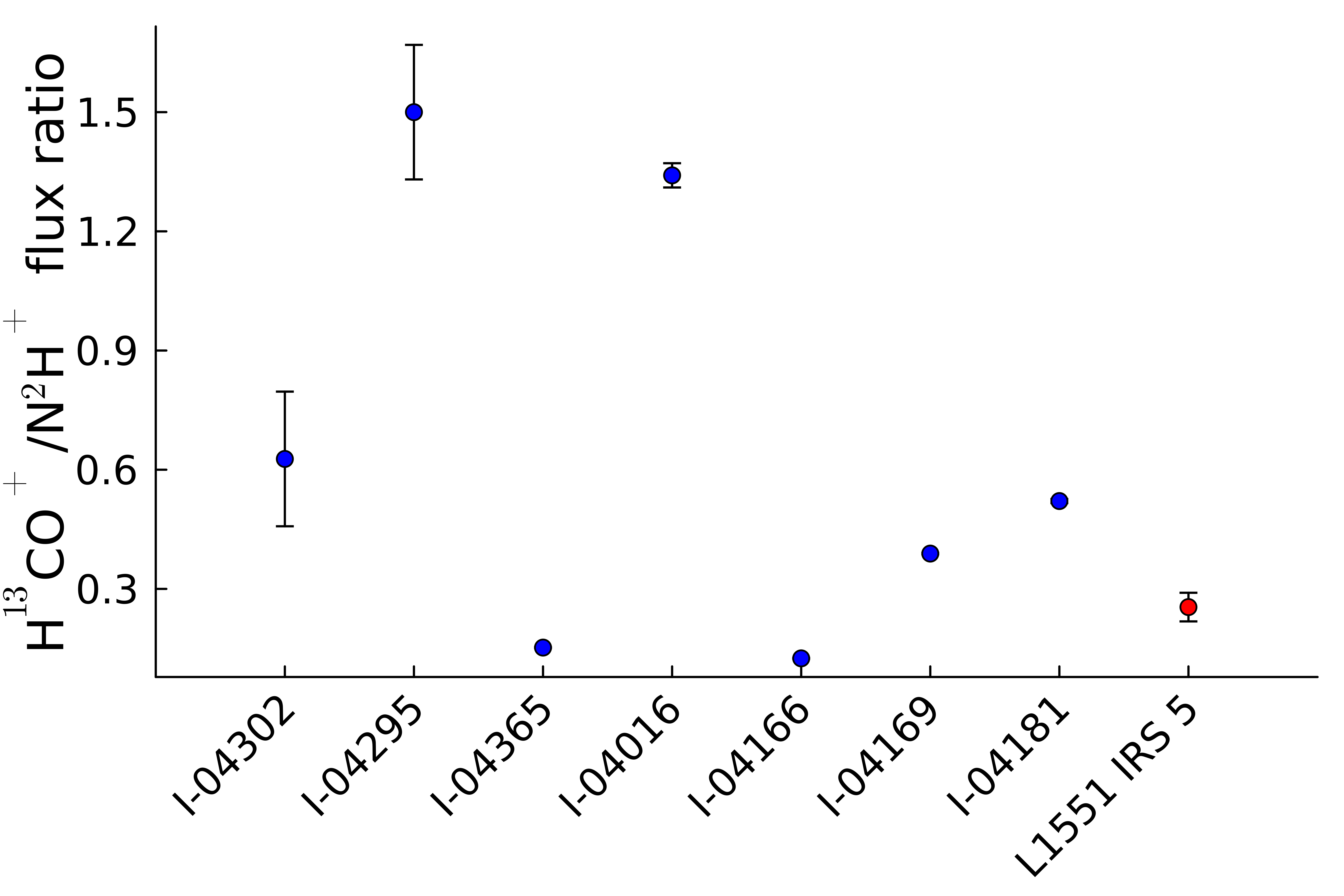}
  \caption{H$^{13}$CO+/N$_2$H+ 1-0 line flux ratios for L1551~IRS~5 (red point) compared to the Class~I sources in \citet{Legal2020} (blue points). Black lines represent error bars. The N$_2$H$^+$ line flux for I-04302 provided by \citet{Legal2020} is an upper limit.}
  \label{fig:line_ratio_legal}
\end{center}
\end{figure}

\citet{Visser2015} suggest that the line flux ratio H$^{13}$CO$^+$ (1 -- 0)/N$_2$H$^+$ (1 -- 0) (integrated over the hyperfine components) is an outburst tracer, due to the desorption of CO from grain mantles that promotes the formation of HCO$^+$ and that is anti-correlated with N$_2$H$^+$. A high ratio ($\gtrsim 1$) would therefore be indicative of a recent ($<10^3$ years) outburst. Figure \ref{fig:line_ratio_legal} displays this ratio for L1551~IRS~5, compared to other Class~I protostars measured in the spectral survey of \citet{Legal2020}. We find a rather low ratio, although in the same range of values as the other protostars, which are not known to experience outbursts. The results of this method to characterize the outburst are therefore inconclusive in our case. Using this principle, \citet{Hsieh2019} measured the location of the peak of N$_2$H$^+$ emission in Class 0/I sources using ALMA. They compare it with the theoretical CO snowline derived from the luminosity of the protostar to determine whether the source experienced a recent outburst. Similar observations of L1551~IRS~5 may be necessary to characterize its luminosity history.

\section{Conclusion}\label{Sec:conclusion}

We have performed a spectral survey at 3mm and 2mm of the Class~I protostar L1551~IRS~5 with the IRAM-30m telescope.  Our main results are as follows:

\begin{itemize}
    \item[$\bullet$] We detect 403 molecular lines from 75 chemical species including 38 secondary isotopologues. Among these species, we find 13 hydrocarbons, 25 N-bearing species, 30 O-bearing species, 15 S-bearing species, and 12 deuterated molecules. 
Ten complex organic molecules (l-C$_4$H$_2$, CH$_3$CCH, CH$_2$DCCH, CH$_3$CHO, CH$_3$CN, CH$_3$OCH$_3$, CH$_3$OCHO, CH$_3$OH, CH$_2$DOH, and HC$_5$N) are also detected. 
 Most lines are centered around the velocity of the envelope $v_\mathrm{lsr}=6.4$ $\kms$, but a few lines are also found around $8.5$ $\kms$, the velocity of the northern source.

    \item[$\bullet$] Column densities, source sizes and excitation temperatures were derived for most detected species with LTE models. When possible, non-LTE models were also used. The models reveal that most species trace the cold envelope with temperatures $\lesssim10$ K: CCH, CCS, C$_3$H, c-C$_3$H$_2$, CH$_3$CHO, CS, HCN, HNC, DCO$^+$, HOCO$^+$, N$_2$D$^+$, and their respective isotopologues. We also see several high-temperature ($\gtrsim 30$ K) species tracing warmer regions, typically molecules with $\geq 5$ atoms such as CH$_3$CN and HC$_5$N, and the S-bearing species C$_3$S, SO, and SO$_2$. The lines of CH$_3$OH and OCS exhibit a component whose emission seems to originate from the warm innermost regions ($<2$\arcsec).

    \item[$\bullet$] Using the derived column densities, we calculated isotopic fractionation ratios for C, H, N, O, and S. They are found to be in agreement with other protostellar sources, in particular the $^{33}$S/$^{32}$S and $^{34}$S/$^{32}$S ratios in CS and SO. The $^{12}$C/$^{13}$C ratios are very low for CS, HCN, HNC, and H$_2$CO, which suggests that their emission is optically thick at those frequencies. We detect both HC$^{15}$N and H$^{15}$NC with $^{14}$N/$^{15}$N isotopic ratios of $\sim$200 and $\sim$370 (using the indirect double isotope method with $^{13}$C). The $^{16}$O/$^{18}$O ratio of HCO$^+$ when using H$^{13}$CO$^+$ and a $^{12}$C/$^{13}$C of 68 is $\gtrsim$ 1000, namely, significantly higher than the values measured in diffuse molecular clouds and L1527. The D/H ratios range from 1\% for HCO$^+$ to 6\% for CH$_3$CCH with in between CCH,  
    c-C$_3$H$_2$, HCN, and HC$_3$N that all show D/H ratios close to 3\%.

    \item[$\bullet$]  Comparisons of our results with spectral surveys of other Class I protostars show that there exists a large diversity of chemistry between those sources. The origin of these differences still needs to be explained.
\end{itemize}

In conclusion, this work is a first step toward improving our understanding of the impact of  protostellar outbursts on protostellar envelopes. Additional studies, both theoretical and observational, are needed to further constrain this effect and to provide statistically significant results.

\begin{acknowledgements}
    This study is part of a project that has received funding from the European Research Council (ERC) under the European Union’s Horizon 2020 research and innovation program (Grant agreement No. 949278, Chemtrip). This work is based on observations carried out under project numbers 047-22,  115-22, and 080-16 with the IRAM 30m telescope. IRAM is supported by INSU/CNRS (France), MPG (Germany) and IGN (Spain). We thank the people at the IRAM-30m facility that were very kind and made these observations possible. This work was also supported by the NKFIH excellence grant TKP2021-NKTA-64.
\end{acknowledgements}

\bibliographystyle{aa}
\bibliography{MaBiblio}

\newpage

\appendix

\section{Detected lines}\label{app:tables}

\begin{landscape}
\begin{table}
  \caption{Identified lines in the survey. Line characteristics (columns 1 to 6) : frequency in GHz, species name, molecular tag, transition, upper energy level in K, Einstein A$_{ij}$ coefficient in s$^{-1}$. Gaussian fitting parameters (columns 7 to 14): FWHM in km s$^{-1}$, error on the FWHM, velocity shift in km s$^{-1}$, error on the velocity shift, intensity in K, error on the intensity, Gaussian flux in K km s$^{-1}$, and error on the Gaussian flux. An empty fitting parameter indicates that the line is blended with the line just above.}
\label{tab:linelist}
\centering
\begin{tabular}{lllcllllllllll}
\hline\hline
  Freq.  &  Species  & Tag & Transition   & E$_{\mathrm{up}}$  & A$_{\mathrm{ij}}$   &  FWHM  & $\delta$FWHM  &  v &   $\delta$v  &  Int &   $\delta$Int & Flux  & $\delta$Flux\\
  (GHz)        &       &      &              &  (K)  & (s$^{-1}$)  & (km s$^{-1}$) &  (km s$^{-1}$)  &  (km s$^{-1}$) &  (km s$^{-1}$)   &  (K)  &   (K)  & (K km s$^{-1}$)  & (K km s$^{-1}$)\\
\hline
87.2841 & CCH & 25501 & 1 1.5 1 -- 0 0.5 1 & 4.19 & $2.60\times 10^{-7}$ & $1.035$ & $0.0176$ & $6.5652$ & $0.0076$ & $0.3988$ & $0.0056$ & $0.4394$ & $0.0097$ \\
87.3169 & CCH & 25501 & 1 1.5 2 -- 0 0.5 1 & 4.19 & $1.53\times 10^{-6}$ & $1.0501$ & $0.0236$ & $6.5589$ & $0.0078$ & $2.434$ & $0.0405$ & $2.7208$ & $0.0761$ \\
87.3286 & CCH & 25501 & 1 1.5 1 -- 0 0.5 0 & 4.19 & $1.27\times 10^{-6}$ & $1.0708$ & $0.0171$ & $6.5329$ & $0.0046$ & $1.3954$ & $0.0165$ & $1.5905$ & $0.0316$ \\
87.4020 & CCH & 25501 & 1 0.5 1 -- 0 0.5 1 & 4.2 & $1.27\times 10^{-6}$ & $1.1387$ & $0.0119$ & $6.5709$ & $0.0041$ & $1.3772$ & $0.0111$ & $1.6693$ & $0.022$ \\
87.4072 & CCH & 25501 & 1 0.5 0 -- 0 0.5 1 & 4.2 & $1.54\times 10^{-6}$ & $1.0094$ & $0.0091$ & $6.5947$ & $0.0044$ & $0.7321$ & $0.0056$ & $0.7866$ & $0.0093$ \\
87.4465 & CCH & 25501 & 1 0.5 1 -- 0 0.5 0 & 4.2 & $2.61\times 10^{-7}$ & $1.0434$ & $0.0074$ & $6.6026$ & $0.0042$ & $0.3963$ & $0.0026$ & $0.4402$ & $0.0043$ \\
\hline
\end{tabular}
\end{table}
\end{landscape}

\begin{landscape}
\begin{table}
  \caption*{\textbf{Table \ref*{tab:linelist}} continued }
\label{}
\centering
\begin{tabular}{lllcllllllllll}
\hline\hline
  Freq.  &  Species  & Tag & Transition   & E$_{\mathrm{up}}$  & A$_{\mathrm{ij}}$   &  FWHM  & $\delta$FWHM  &  v &   $\delta$v  &  Int &   $\delta$Int & Flux  & $\delta$Flux\\
  (GHz)        &       &      &              &  (K)  & (s$^{-1}$)  & (km s$^{-1}$) &  (km s$^{-1}$)  &  (km s$^{-1}$) &  (km s$^{-1}$)   &  (K)  &   (K)  & (K km s$^{-1}$)  & (K km s$^{-1}$)\\
\hline
156.9817 & CCS & 56502 & 12 13 -- 11 12 & $51.47$ & $1.80\times 10^{-4}$ & $1.0882$ & $0.0833$ & $6.5033$ & $0.0351$ & $0.0608$ & $0.004$ & $0.0704$ & $0.0071$ \\
76.1987 & C$_3$H & 37501 & 4 1 0 3.5 4 -- 3 -1 0 2.5 3 & 7.83 & $2.78\times 10^{-5}$ & $1.055$ & $0.1462$ & $6.6106$ & $0.0869$ & $0.0322$ & $0.0038$ & $0.0362$ & $0.0066$ \\
76.1999 & C$_3$H & 37501 & 4 1 0 3.5 3 -- 3 -1 0 2.5 2 & 7.83 & $2.64\times 10^{-5}$ & $1.1507$ & $0.1834$ & $6.6282$ & $0.1122$ & $0.0238$ & $0.0036$ & $0.0292$ & $0.0064$ \\
76.2042 & C$_3$H & 37501 & 4 -1 0 3.5 4 -- 3 1 0 2.5 3 & 7.84 & $2.78\times 10^{-5}$ & $1.1077$ & $0.1179$ & $6.5347$ & $0.0706$ & $0.0328$ & $0.0031$ & $0.0387$ & $0.0055$ \\
76.2051 & C$_3$H & 37501 & 4 -1 0 3.5 3 -- 3 1 0 2.5 2 & 7.84 & $2.64\times 10^{-5}$ & $1.4364$ & $0.4178$ & $6.7378$ & $0.1601$ & $0.0137$ & $0.0033$ & $0.021$ & $0.0079$ \\
97.9952 & C$_3$H & 37501 & 4 1 0 4.5 5 -- 4 -1 0 3.5 4 & 12.54 & $6.12\times 10^{-5}$ & $1.0765$ & $0.1314$ & $6.3909$ & $0.0625$ & $0.0269$ & $0.003$ & $0.0308$ & $0.0051$ \\
\hline
\end{tabular}
\end{table}
\end{landscape}

\begin{landscape}
\begin{table}
  \caption*{\textbf{Table \ref*{tab:linelist}} continued }
\label{}
\centering
\begin{tabular}{lllcllllllllll}
\hline\hline
  Freq.  &  Species  & Tag & Transition   & E$_{\mathrm{up}}$  & A$_{\mathrm{ij}}$   &  FWHM  & $\delta$FWHM  &  v &   $\delta$v  &  Int &   $\delta$Int & Flux  & $\delta$Flux\\
  (GHz)        &       &      &              &  (K)  & (s$^{-1}$)  & (km s$^{-1}$) &  (km s$^{-1}$)  &  (km s$^{-1}$) &  (km s$^{-1}$)   &  (K)  &   (K)  & (K km s$^{-1}$)  & (K km s$^{-1}$)\\
\hline
115.5244 & c-CC$^{13}$CH$_2$ & 39510 & 3 1 3 -- 2 0 2 & $11.85$ & $6.50\times 10^{-5}$ & $0.8954$ & $0.2012$ & $6.2358$ & $0.1013$ & $0.0359$ & $0.0075$ & $0.0342$ & $0.0105$ \\
153.8946 & c-CC$^{13}$CH$_2$ & 39510 & 3 2 2 -- 2 1 1 & $15.85$ & $1.05\times 10^{-4}$ & $0.7995$ & $0.1047$ & $6.3192$ & $0.0437$ & $0.0404$ & $0.0045$ & $0.0344$ & $0.0059$ \\
82.3951 & l-C$_3$H$_2$ & 38501 & 4 1 4 -- 3 1 3 & $23.24$ & $4.56\times 10^{-5}$ & $0.9402$ & $0.6489$ & $6.5388$ & $0.1494$ & $0.0276$ & $0.0157$ & $0.0277$ & $0.0248$ \\
83.9337 & l-C$_3$H$_2$ & 38501 & 4 1 3 -- 3 1 2 & $23.43$ & $4.82\times 10^{-5}$ & $1.174$ & $0.2486$ & $6.6104$ & $0.0876$ & $0.0268$ & $0.0046$ & $0.0335$ & $0.0091$ \\
102.9924 & l-C$_3$H$_2$ & 38501 & 5 1 5 -- 4 1 4 & $28.19$ & $9.33\times 10^{-5}$ & $1.0872$ & $0.1422$ & $6.4957$ & $0.0534$ & $0.0343$ & $0.0037$ & $0.0397$ & $0.0067$ \\
103.9529 & l-C$_3$H$_2$ & 38501 & 5 0 5 -- 4 0 4 & $14.97$ & $9.99\times 10^{-5}$ & $0.9642$ & $0.2311$ & $6.4898$ & $0.0784$ & $0.0231$ & $0.0044$ & $0.0237$ & $0.0073$ \\
\hline
\end{tabular}
\end{table}
\end{landscape}

\begin{landscape}
\begin{table}
  \caption*{\textbf{Table \ref*{tab:linelist}} continued }
\label{}
\centering
\begin{tabular}{lllcllllllllll}
\hline\hline
  Freq.  &  Species  & Tag & Transition   & E$_{\mathrm{up}}$  & A$_{\mathrm{ij}}$   &  FWHM  & $\delta$FWHM  &  v &   $\delta$v  &  Int &   $\delta$Int & Flux  & $\delta$Flux\\
  (GHz)        &       &      &              &  (K)  & (s$^{-1}$)  & (km s$^{-1}$) &  (km s$^{-1}$)  &  (km s$^{-1}$) &  (km s$^{-1}$)   &  (K)  &   (K)  & (K km s$^{-1}$)  & (K km s$^{-1}$)\\
\hline
95.1889 & C$_4$H & 49503 & 10 9.5 9 -- 9 8.5 8 & 25.13 & $2.08\times 10^{-5}$ & $1.2818$ & $0.0348$ & $6.6027$ & $0.0146$ & $0.0993$ & $0.0023$ & $0.1355$ & $0.0048$ \\
95.1889 & C$_4$H & 49503 & 10 9.5 10 -- 9 8.5 9 & 25.13 & $2.10\times 10^{-5}$ & - & - & - & - & - & - & - & -\\
104.6666 & C$_4$H & 49503 & 11 11.5 11 -- 10 10.5 10 & 30.14 & $2.80\times 10^{-5}$ & $0.9986$ & $0.0364$ & $6.6017$ & $0.0136$ & $0.1205$ & $0.0035$ & $0.1281$ & $0.006$ \\
104.6666 & C$_4$H & 49503 & 11 11.5 12 -- 10 10.5 11 & 30.14 & $2.81\times 10^{-5}$ & - & - & - & - & - & - & - & -\\
104.7051 & C$_4$H & 49503 & 11 10.5 10 -- 10 9.5 9 & 30.16 & $2.79\times 10^{-5}$ & $1.133$ & $0.0455$ & $6.571$ & $0.0187$ & $0.1023$ & $0.0035$ & $0.1234$ & $0.0065$ \\
104.7051 & C$_4$H & 49503 & 11 10.5 11 -- 10 9.5 10 & 30.16 & $2.81\times 10^{-5}$ & - & - & - & - & - & - & - & -\\
\hline
\end{tabular}
\end{table}
\end{landscape}

\begin{landscape}
\begin{table}
  \caption*{\textbf{Table \ref*{tab:linelist}} continued }
\label{}
\centering
\begin{tabular}{lllcllllllllll}
\hline\hline
  Freq.  &  Species  & Tag & Transition   & E$_{\mathrm{up}}$  & A$_{\mathrm{ij}}$   &  FWHM  & $\delta$FWHM  &  v &   $\delta$v  &  Int &   $\delta$Int & Flux  & $\delta$Flux\\
  (GHz)        &       &      &              &  (K)  & (s$^{-1}$)  & (km s$^{-1}$) &  (km s$^{-1}$)  &  (km s$^{-1}$) &  (km s$^{-1}$)   &  (K)  &   (K)  & (K km s$^{-1}$)  & (K km s$^{-1}$)\\
\hline
74.9241 & CH$_3$CHO & 44003 & 4 1 4 1 -- 3 1 3 1 & 11.33 & $1.24\times 10^{-5}$ & $1.4812$ & $0.1917$ & $6.3976$ & $0.0898$ & $0.0272$ & $0.0032$ & $0.0428$ & $0.0075$ \\
76.8664 & CH$_3$CHO & 44003 & 4 0 4 2 -- 3 0 3 2 & 9.33 & $1.43\times 10^{-5}$ & $1.3371$ & $0.1279$ & $6.4333$ & $0.0627$ & $0.0362$ & $0.0032$ & $0.0515$ & $0.0067$ \\
76.8790 & CH$_3$CHO & 44003 & 4 0 4 0 -- 3 0 3 0 & 9.23 & $1.43\times 10^{-5}$ & $1.2377$ & $0.0904$ & $6.4153$ & $0.0501$ & $0.0424$ & $0.0029$ & $0.0558$ & $0.0056$ \\
79.0993 & CH$_3$CHO & 44003 & 4 1 3 2 -- 3 1 2 2 & 11.84 & $1.46\times 10^{-5}$ & $1.1188$ & $0.2526$ & $6.5718$ & $0.0575$ & $0.0332$ & $0.0057$ & $0.0396$ & $0.0112$ \\
79.1502 & CH$_3$CHO & 44003 & 4 1 3 0 -- 3 1 2 0 & 11.77 & $1.46\times 10^{-5}$ & $1.0519$ & $0.1278$ & $6.4254$ & $0.0829$ & $0.0316$ & $0.0035$ & $0.0354$ & $0.0058$ \\
84.2197 & CH$_3$CHO & 44003 & 2 1 2 0 -- 1 0 1 0 & 4.97 & $2.29\times 10^{-6}$ & $1.423$ & $0.3788$ & $6.2571$ & $0.1506$ & $0.0158$ & $0.0035$ & $0.0239$ & $0.0083$ \\
\hline
\end{tabular}
\end{table}
\end{landscape}

\begin{landscape}
\begin{table}
  \caption*{\textbf{Table \ref*{tab:linelist}} continued }
\label{}
\centering
\begin{tabular}{lllcllllllllll}
\hline\hline
  Freq.  &  Species  & Tag & Transition   & E$_{\mathrm{up}}$  & A$_{\mathrm{ij}}$   &  FWHM  & $\delta$FWHM  &  v &   $\delta$v  &  Int &   $\delta$Int & Flux  & $\delta$Flux\\
  (GHz)        &       &      &              &  (K)  & (s$^{-1}$)  & (km s$^{-1}$) &  (km s$^{-1}$)  &  (km s$^{-1}$) &  (km s$^{-1}$)   &  (K)  &   (K)  & (K km s$^{-1}$)  & (K km s$^{-1}$)\\
\hline
95.1694 & CH$_3$OH & 32504 & 8 0 8 0 -- 7 1 7 0 & $83.54$ & $4.26\times 10^{-6}$ & $1.9195$ & $0.5794$ & $8.6619$ & $0.2414$ & $0.0166$ & $0.0029$ & $0.0338$ & $0.0118$\\
95.9143 & CH$_3$OH & 32504 & 2 1 2 0 -- 1 1 1 0 & $21.44$ & $2.49\times 10^{-6}$ & $2.1029$ & $0.4435$ & $6.796$ & $0.1746$ & $0.016$ & $0.0027$ & $0.0358$ & $0.0097$\\
96.7394 & CH$_3$OH & 32504 & 2 1 2 2 -- 1 1 1 2 & $12.54$ & $2.56\times 10^{-6}$ & $1.4417$ & $0.0176$ & $6.4812$ & $0.0074$ & $0.4258$ & $0.0045$ & $0.6534$ & $0.0106$\\
96.7414 & CH$_3$OH & 32504 & 2 0 2 0 -- 1 0 1 0 & $6.96$ & $3.41\times 10^{-6}$ & $1.3988$ & $0.0129$ & $6.4799$ & $0.0054$ & $0.6276$ & $0.0049$ & $0.9345$ & $0.0113$\\
96.7445 & CH$_3$OH & 32504 & 2 0 2 1 -- 1 0 1 1 & $20.09$ & $3.41\times 10^{-6}$ & $1.317$ & $0.0563$ & $6.4526$ & $0.0229$ & $0.0864$ & $0.0031$ & $0.1211$ & $0.0068$\\
96.7555 & CH$_3$OH & 32504 & 2 1 1 1 -- 1 1 0 1 & $28.01$ & $2.62\times 10^{-6}$ & $1.046$ & $0.1709$ & $6.287$ & $0.0839$ & $0.0218$ & $0.0031$ & $0.0243$ & $0.0053$\\
\hline
\end{tabular}
\end{table}
\end{landscape}

\begin{landscape}
\begin{table}
  \caption*{\textbf{Table \ref*{tab:linelist}} continued }
\label{}
\centering
\begin{tabular}{lllcllllllllll}
\hline\hline
  Freq.  &  Species  & Tag & Transition   & E$_{\mathrm{up}}$  & A$_{\mathrm{ij}}$   &  FWHM  & $\delta$FWHM  &  v &   $\delta$v  &  Int &   $\delta$Int & Flux  & $\delta$Flux\\
  (GHz)        &       &      &              &  (K)  & (s$^{-1}$)  & (km s$^{-1}$) &  (km s$^{-1}$)  &  (km s$^{-1}$) &  (km s$^{-1}$)   &  (K)  &   (K)  & (K km s$^{-1}$)  & (K km s$^{-1}$)\\
\hline
157.2708 & CH$_3$OH & 32504 & 1 0 1 1 -- 1 1 1 2 & $15.45$ & $2.20\times 10^{-5}$ & $1.2991$ & $0.0292$ & $6.4125$ & $0.0122$ & $0.2547$ & $0.0049$ & $0.3523$ & $0.0104$\\
157.2723 & CH$_3$OH & 32504 & 3 0 3 1 -- 3 1 3 2 & $27.05$ & $2.15\times 10^{-5}$ & $0.9431$ & $0.0377$ & $6.3858$ & $0.0158$ & $0.148$ & $0.0051$ & $0.1485$ & $0.0078$\\
157.276 & CH$_3$OH & 32504 & 2 0 2 1 -- 2 1 2 2 & $20.09$ & $2.18\times 10^{-5}$ & $0.9602$ & $0.0261$ & $6.4298$ & $0.0109$ & $0.2121$ & $0.0049$ & $0.2168$ & $0.0077$\\
86.6688 & CH$_2$DOH & 33004 & 2 1 1 0 -- 2 0 2 0 & $10.6$ & $4.65\times 10^{-6}$ & $1.1537$ & $0.382$ & $6.0873$ & $0.1765$ & $0.0113$ & $0.0033$ & $0.0138$ & $0.0061$ \\
88.7545 & CH$_2$DOH & 33004 & 3 1 2 0 -- 3 0 3 0 & $17.13$ & $4.92\times 10^{-6}$ & $1.6206$ & $0.5949$ & $6.0325$ & $0.25$ & $0.0087$ & $0.0027$ & $0.015$ & $0.0072$ \\
88.7545 & CH$_2$DOH & 33004 & 3 1 2 0 -- 3 0 3 0 & $17.13$ & $4.92\times 10^{-6}$ & $0.779$ & $0.349$ & $9.3577$ & $0.2253$ & $0.0125$ & $0.0042$ & $0.0103$ & $0.0058$ \\
\hline
\end{tabular}
\end{table}
\end{landscape}

\begin{landscape}
\begin{table}
  \caption*{\textbf{Table \ref*{tab:linelist}} continued }
\label{}
\centering
\begin{tabular}{lllcllllllllll}
\hline\hline
  Freq.  &  Species  & Tag & Transition   & E$_{\mathrm{up}}$  & A$_{\mathrm{ij}}$   &  FWHM  & $\delta$FWHM  &  v &   $\delta$v  &  Int &   $\delta$Int & Flux  & $\delta$Flux\\
  (GHz)        &       &      &              &  (K)  & (s$^{-1}$)  & (km s$^{-1}$) &  (km s$^{-1}$)  &  (km s$^{-1}$) &  (km s$^{-1}$)   &  (K)  &   (K)  & (K km s$^{-1}$)  & (K km s$^{-1}$)\\
\hline
112.359 & C$^{17}$O & 29503 & 1 3.5 -- 0 2.5 & $5.39$ & $6.70\times 10^{-8}$ & - & - & $6.4795$ & $0.0093$ & - & - & - & -\\
112.36 & C$^{17}$O & 29503 & 1 2.5 -- 0 2.5 & $5.39$ & $6.70\times 10^{-8}$ & $0.9913$ & $0.0312$ & $6.3496$ & $0.0145$ & $0.3744$ & $0.0106$ & $0.3951$ & $0.0167$ \\
109.7822 & C$^{18}$O & 30502 & 1 -- 0 & $5.27$ & $6.27\times 10^{-8}$ & $1.1511$ & $0.0143$ & $6.3523$ & $0.0059$ & $3.1462$ & $0.0333$ & $3.8552$ & $0.0629$ \\
104.7114 & $^{13}$C$^{18}$O & 31502 & 1 0.5 -- 0 0.5 & $5.03$ & $5.45\times 10^{-8}$ & $1.0402$ & $0.0713$ & $6.3889$ & $0.03$ & $0.0571$ & $0.0034$ & $0.0632$ & $0.0057$ \\
97.981 & CS & 44501 & 2 0 -- 1 0 & 7.05 & $1.68\times 10^{-5}$ & $1.5431$ & $0.0265$ & $6.6987$ & $0.0111$ & $2.4843$ & $0.0366$ & $4.0806$ & $0.0923$ \\
146.969 & CS & 44501 & 3 0 -- 2 0 & 14.11 & $6.07\times 10^{-5}$ & $1.2504$ & $0.0237$ & $6.7168$ & $0.01$ & $2.99$ & $0.0488$ & $3.9797$ & $0.0995$ \\
\hline
\end{tabular}
\end{table}
\end{landscape}

\begin{landscape}
\begin{table}
  \caption*{\textbf{Table \ref*{tab:linelist}} continued }
\label{}
\centering
\begin{tabular}{lllcllllllllll}
\hline\hline
  Freq.  &  Species  & Tag & Transition   & E$_{\mathrm{up}}$  & A$_{\mathrm{ij}}$   &  FWHM  & $\delta$FWHM  &  v &   $\delta$v  &  Int &   $\delta$Int & Flux  & $\delta$Flux\\
  (GHz)        &       &      &              &  (K)  & (s$^{-1}$)  & (km s$^{-1}$) &  (km s$^{-1}$)  &  (km s$^{-1}$) &  (km s$^{-1}$)   &  (K)  &   (K)  & (K km s$^{-1}$)  & (K km s$^{-1}$)\\
\hline
144.0773 & DCO$^+$ & 30510 & 2 -- 1 & $10.37$ & $2.12\times 10^{-4}$ & $0.8829$ & $0.0082$ & $6.3531$ & $0.0036$ & $2.0437$ & $0.0166$ & $1.9206$ & $0.0237$ \\
86.7543 & H$^{13}$CO$^+$ & 30504 & 1 -- 0 & $4.16$ & $3.85\times 10^{-5}$ & $1.0178$ & $0.0029$ & $6.4251$ & $0.0011$ & $2.1066$ & $0.0048$ & $2.2824$ & $0.0083$ \\
85.1622 & HC$^{18}$O$^+$ & 31506 & 1 -- 0 & $4.09$ & $3.64\times 10^{-5}$ & $0.8862$ & $0.0192$ & $6.3814$ & $0.011$ & $0.2181$ & $0.0038$ & $0.2058$ & $0.0057$ \\
87.0575 & HC$^{17}$O$^+$ & 30505 & 1 -- 0 & $4.18$ & $3.89\times 10^{-5}$ & $1.398$ & $0.2885$ & $7.4925$ & $0.1179$ & $0.0201$ & $0.0035$ & $0.03$ & $0.0081$ \\
141.4651 & D$^{13}$CO$^+$ & 31508 & 2 -- 1 & $10.18$ & $2.00\times 10^{-4}$ & $0.7108$ & $0.0848$ & $6.2895$ & $0.0441$ & $0.077$ & $0.0086$ & $0.0582$ & $0.0095$ \\
72.7838 & HC$_3$N & 51501 & 8 -- 7 & 15.72 & $2.94\times 10^{-5}$ & $1.0714$ & $0.0053$ & $6.5523$ & $0.0038$ & $0.8467$ & $0.0037$ & $0.9656$ & $0.0064$ \\
\hline
\end{tabular}
\end{table}
\end{landscape}

\begin{landscape}
\begin{table}
  \caption*{\textbf{Table \ref*{tab:linelist}} continued }
\label{}
\centering
\begin{tabular}{lllcllllllllll}
\hline\hline
  Freq.  &  Species  & Tag & Transition   & E$_{\mathrm{up}}$  & A$_{\mathrm{ij}}$   &  FWHM  & $\delta$FWHM  &  v &   $\delta$v  &  Int &   $\delta$Int & Flux  & $\delta$Flux\\
  (GHz)        &       &      &              &  (K)  & (s$^{-1}$)  & (km s$^{-1}$) &  (km s$^{-1}$)  &  (km s$^{-1}$) &  (km s$^{-1}$)   &  (K)  &   (K)  & (K km s$^{-1}$)  & (K km s$^{-1}$)\\
\hline
111.823 & HC$_5$N & 75503 & 42 -- 41 & 115.39 & $1.51\times 10^{-4}$ & $0.6287$ & $0.5692$ & $6.5253$ & $0.1695$ & $0.017$ & $0.0139$ & $0.0114$ & $0.0139$ \\
72.8379 & H$_2$CO & 30501 & 1 0 1 -- 0 0 0 & $3.5$ & $8.15\times 10^{-6}$ & $1.37$ & $0.0049$ & $6.5274$ & $0.0023$ & $1.845$ & $0.0059$ & $2.6906$ & $0.0129$ \\
140.8395 & H$_2$CO & 30501 & 2 1 2 -- 1 1 1 & $21.92$ & $5.30\times 10^{-5}$ & $1.2424$ & $0.0175$ & $6.5256$ & $0.0074$ & $2.5873$ & $0.0314$ & $3.4217$ & $0.0636$ \\
145.6029 & H$_2$CO & 30501 & 2 0 2 -- 1 0 1 & $10.48$ & $7.81\times 10^{-5}$ & $1.094$ & $0.0155$ & $6.5035$ & $0.0065$ & $2.4974$ & $0.0305$ & $2.9081$ & $0.0544$ \\
110.8378 & D$_2$CO & 32502 & 2 1 2 -- 1 1 1 & $13.37$ & $2.58\times 10^{-5}$ & $0.7804$ & $0.0736$ & $6.3784$ & $0.0231$ & $0.1297$ & $0.0096$ & $0.1077$ & $0.0129$ \\
88.8657 & H$_2^{13}$CO & 28006 & 1 -- 0 & 4.26 & $1.98\times 10^{-5}$ & $0.8461$ & $0.034$ & $6.4276$ & $0.0249$ & $0.0815$ & $0.0028$ & $0.0734$ & $0.0039$ \\
\hline
\end{tabular}
\end{table}
\end{landscape}

\begin{landscape}
\begin{table}
  \caption*{\textbf{Table \ref*{tab:linelist}} continued }
\label{}
\centering
\begin{tabular}{lllcllllllllll}
\hline\hline
  Freq.  &  Species  & Tag & Transition   & E$_{\mathrm{up}}$  & A$_{\mathrm{ij}}$   &  FWHM  & $\delta$FWHM  &  v &   $\delta$v  &  Int &   $\delta$Int & Flux  & $\delta$Flux\\
  (GHz)        &       &      &              &  (K)  & (s$^{-1}$)  & (km s$^{-1}$) &  (km s$^{-1}$)  &  (km s$^{-1}$) &  (km s$^{-1}$)   &  (K)  &   (K)  & (K km s$^{-1}$)  & (K km s$^{-1}$)\\
\hline
110.155 & NH$_2$D & 18501 & 1 1 1 1 1 -- 1 0 1 0 0 & 21.31 & $5.49\times 10^{-6}$ & $0.7911$ & $0.0886$ & $6.1538$ & $0.0393$ & $0.0534$ & $0.005$ & $0.045$ & $0.0066$ \\
93.1716 & N$_2$H$^+$ & 29506 & 1 1 0 -- 0 1 1 & 4.47 & $3.63\times 10^{-5}$ & $1.5746$ & $0.0363$ & $5.3850$ & $0.0152$ & $1.744$ & $0.0345$ & $2.9232$ & $0.0888$ \\
93.1719 & N$_2$H$^+$ & 29506 & 1 1 2 -- 0 1 1 & 4.47 & $5.62\times 10^{-6}$ & - & - & $6.3339$ & $0.0154$ & - & - & - & -\\
93.1719 & N$_2$H$^+$ & 29506 & 1 1 2 -- 0 1 2 & 4.47 & $3.07\times 10^{-5}$ & - & - & $6.3339$ & $0.0160$ & - & - & - & -\\
93.1720 & N$_2$H$^+$ & 29506 & 1 1 1 -- 0 1 0 & 4.47 & $1.85\times 10^{-5}$ & - & - & $6.7751$ & $0.0155$ & - & - & - & -\\
93.1720 & N$_2$H$^+$ & 29506 & 1 1 1 -- 0 1 1 & 4.47 & $4.33\times 10^{-6}$ & - & - & $6.7751$ & $0.0165$ & - & - & - & -\\
\hline
\end{tabular}
\end{table}
\end{landscape}

\begin{landscape}
\begin{table}
  \caption*{\textbf{Table \ref*{tab:linelist}} continued }
\label{}
\centering
\begin{tabular}{lllcllllllllll}
\hline\hline
  Freq.  &  Species  & Tag & Transition   & E$_{\mathrm{up}}$  & A$_{\mathrm{ij}}$   &  FWHM  & $\delta$FWHM  &  v &   $\delta$v  &  Int &   $\delta$Int & Flux  & $\delta$Flux\\
  (GHz)        &       &      &              &  (K)  & (s$^{-1}$)  & (km s$^{-1}$) &  (km s$^{-1}$)  &  (km s$^{-1}$) &  (km s$^{-1}$)   &  (K)  &   (K)  & (K km s$^{-1}$)  & (K km s$^{-1}$)\\
\hline
154.2167 & N$_2$D$^+$ & 30509 & 2 2 2 -- 1 1 1 & 11.1 & $1.18\times 10^{-4}$ & - & - & $5.4529$ & $0.0134$ & $0.2343$ & $0.0082$ & $0.1936$ & $0.0147$\\
154.2168 & N$_2$D$^+$ & 30509 & 2 3 3 -- 1 2 3 & 11.1 & $1.62\times 10^{-5}$ & - & - & $5.5740$ & $0.0136$ & - & - & - & -\\
154.2169 & N$_2$D$^+$ & 30509 & 2 2 2 -- 1 1 2 & 11.1 & $4.90\times 10^{-5}$ & - & - & $5.7212$ & $0.0137$ & - & - & - & -\\
154.2170 & N$_2$D$^+$ & 30509 & 2 2 1 -- 1 1 1 & 11.1 & $3.37\times 10^{-5}$ & - & - & $6.0918$ & $0.0134$ & - & - & - & -\\
154.2171 & N$_2$D$^+$ & 30509 & 2 3 3 -- 1 2 2 & 11.1 & $1.81\times 10^{-4}$ & - & - & $6.1501$ & $0.0135$ & - & - & - & -\\
154.2171 & N$_2$D$^+$ & 30509 & 2 3 2 -- 1 2 1 & 11.1 & $1.56\times 10^{-4}$ & - & - & $6.1942$ & $0.0130$ & - & - & - & -\\
\hline
\end{tabular}
\end{table}
\end{landscape}

\begin{landscape}
\begin{table}
  \caption*{\textbf{Table \ref*{tab:linelist}} continued }
\label{}
\centering
\begin{tabular}{lllcllllllllll}
\hline\hline
  Freq.  &  Species  & Tag & Transition   & E$_{\mathrm{up}}$  & A$_{\mathrm{ij}}$   &  FWHM  & $\delta$FWHM  &  v &   $\delta$v  &  Int &   $\delta$Int & Flux  & $\delta$Flux\\
  (GHz)        &       &      &              &  (K)  & (s$^{-1}$)  & (km s$^{-1}$) &  (km s$^{-1}$)  &  (km s$^{-1}$) &  (km s$^{-1}$)   &  (K)  &   (K)  & (K km s$^{-1}$)  & (K km s$^{-1}$)\\
\hline
104.2393 & SO$_2$ & 64502 & 10 1 9 -- 10 0 10 & 54.71 & $1.12\times 10^{-5}$ & $1.4989$ & $0.8661$ & $8.2976$ & $0.3503$ & $0.0076$ & $0.0027$ & $0.0121$ & $0.0082$ \\
135.696 & SO$_2$ & 64502 & 5 1 5 -- 4 0 4 & 15.66 & $2.21\times 10^{-5}$ & $1.8259$ & $0.3714$ & $6.5288$ & $0.1554$ & $0.0313$ & $0.0055$ & $0.0608$ & $0.0163$ \\
151.3786 & SO$_2$ & 64502 & 2 2 0 -- 2 1 1 & 12.58 & $1.87\times 10^{-5}$ & $1.1768$ & $0.2715$ & $6.4535$ & $0.114$ & $0.0172$ & $0.0034$ & $0.0215$ & $0.0065$ \\
158.1997 & SO$_2$ & 64502 & 3 2 2 -- 3 1 3 & 15.34 & $2.53\times 10^{-5}$ & $1.3234$ & $0.2433$ & $6.5312$ & $0.1019$ & $0.0242$ & $0.0038$ & $0.0341$ & $0.0082$ \\
\hline
\end{tabular}
\end{table}
\end{landscape}

\section{Fits of the data}\label{app:fit_lines}

 \begin{figure*}
 \begin{center}
 \includegraphics[trim=0cm 0cm 0cm 0cm, width=0.98\textwidth]{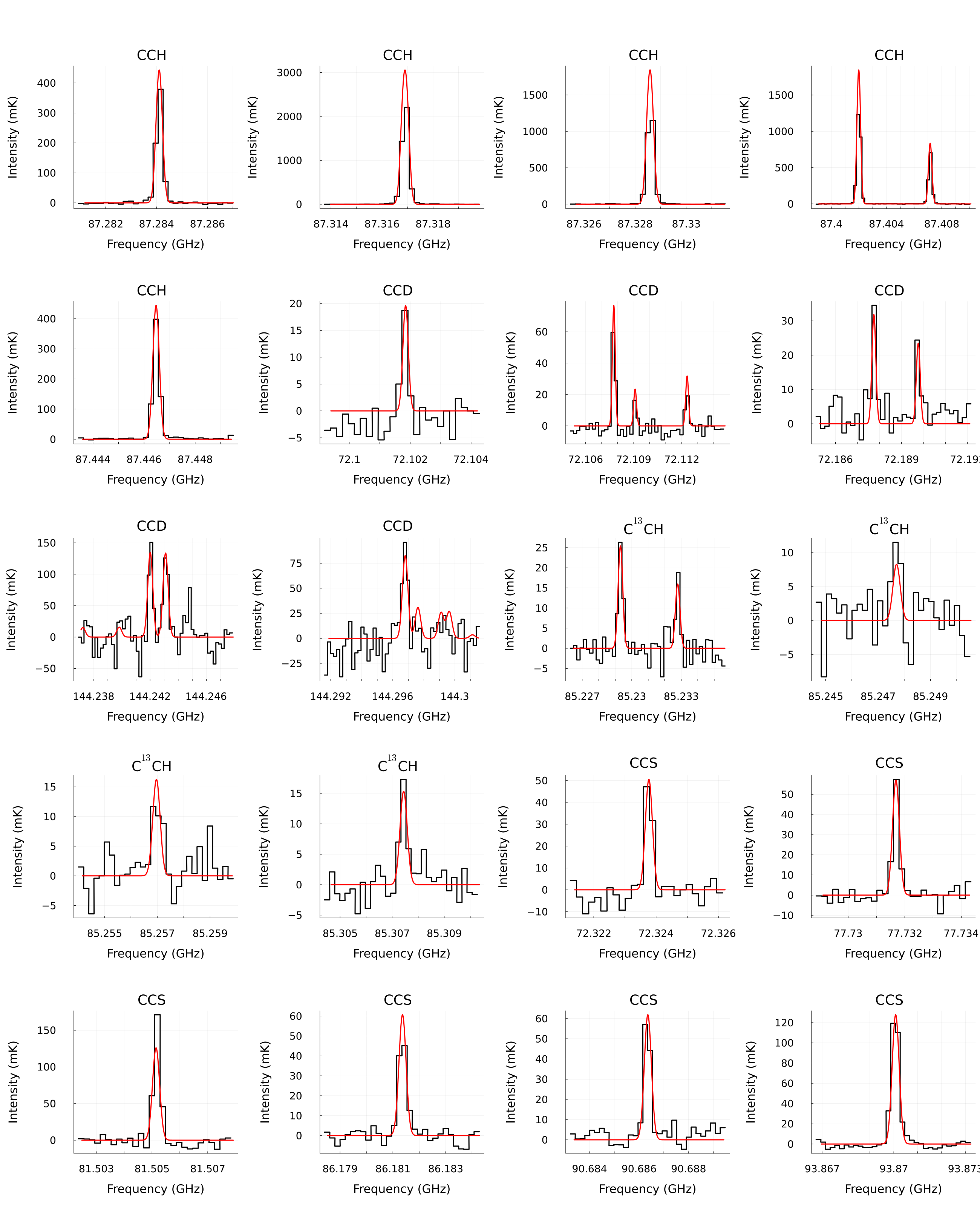}
   \caption{Observed spectrum (black histogram) and LTE radiative transfer model of the transition lines (red curves), for species listed in Table \ref{tab:texncol}.}
   \label{fig:page_1}
 \end{center}
 \end{figure*}

 \begin{figure*}
 \begin{center}
 \includegraphics[trim=0cm 0cm 0cm 0cm, width=0.98\textwidth]{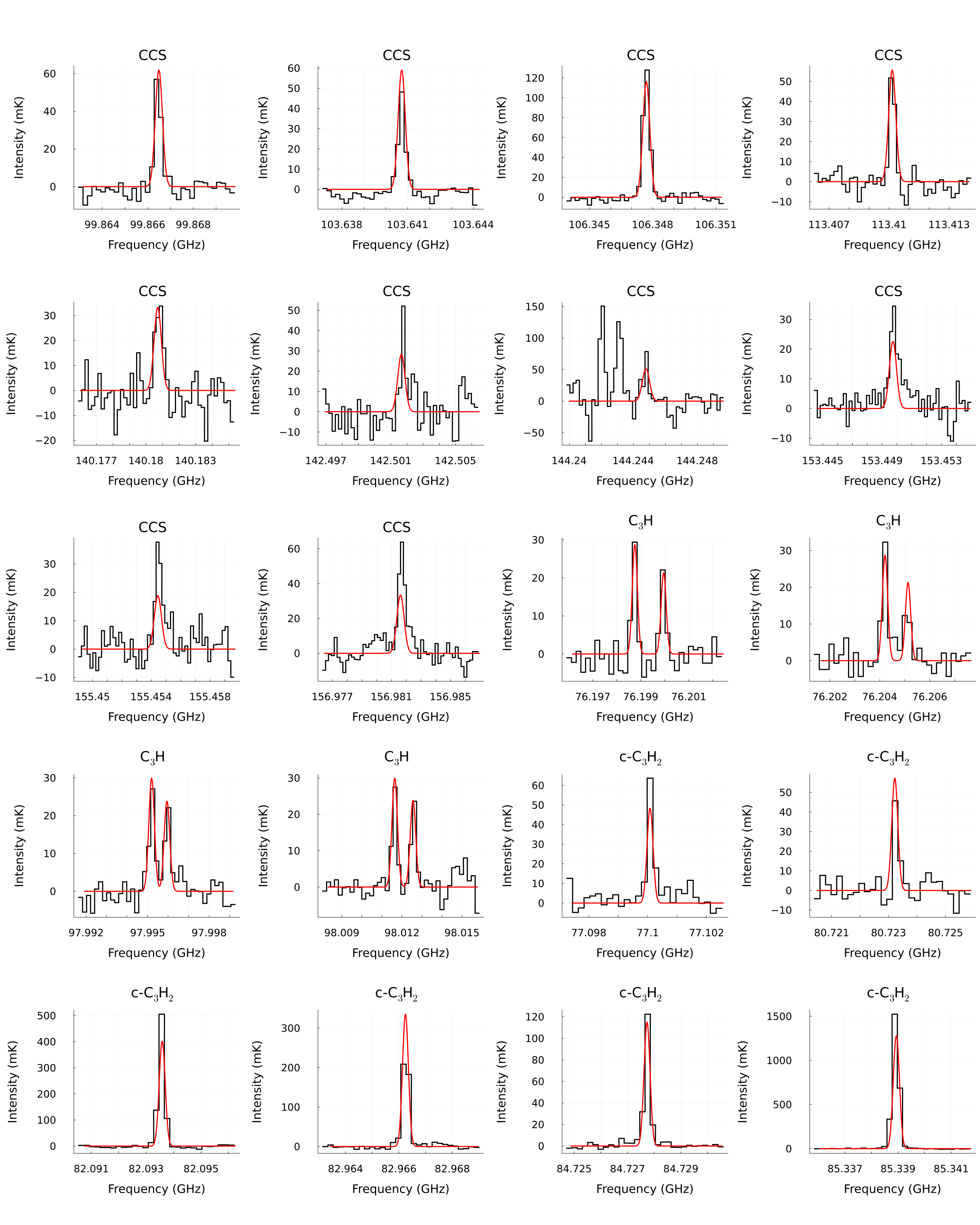}
   \caption*{\textbf{Fig. \ref*{fig:page_1}} continued.}
 \end{center}
 \end{figure*}

 \begin{figure*}
 \begin{center}
 \includegraphics[trim=0cm 0cm 0cm 0cm, width=0.98\textwidth]{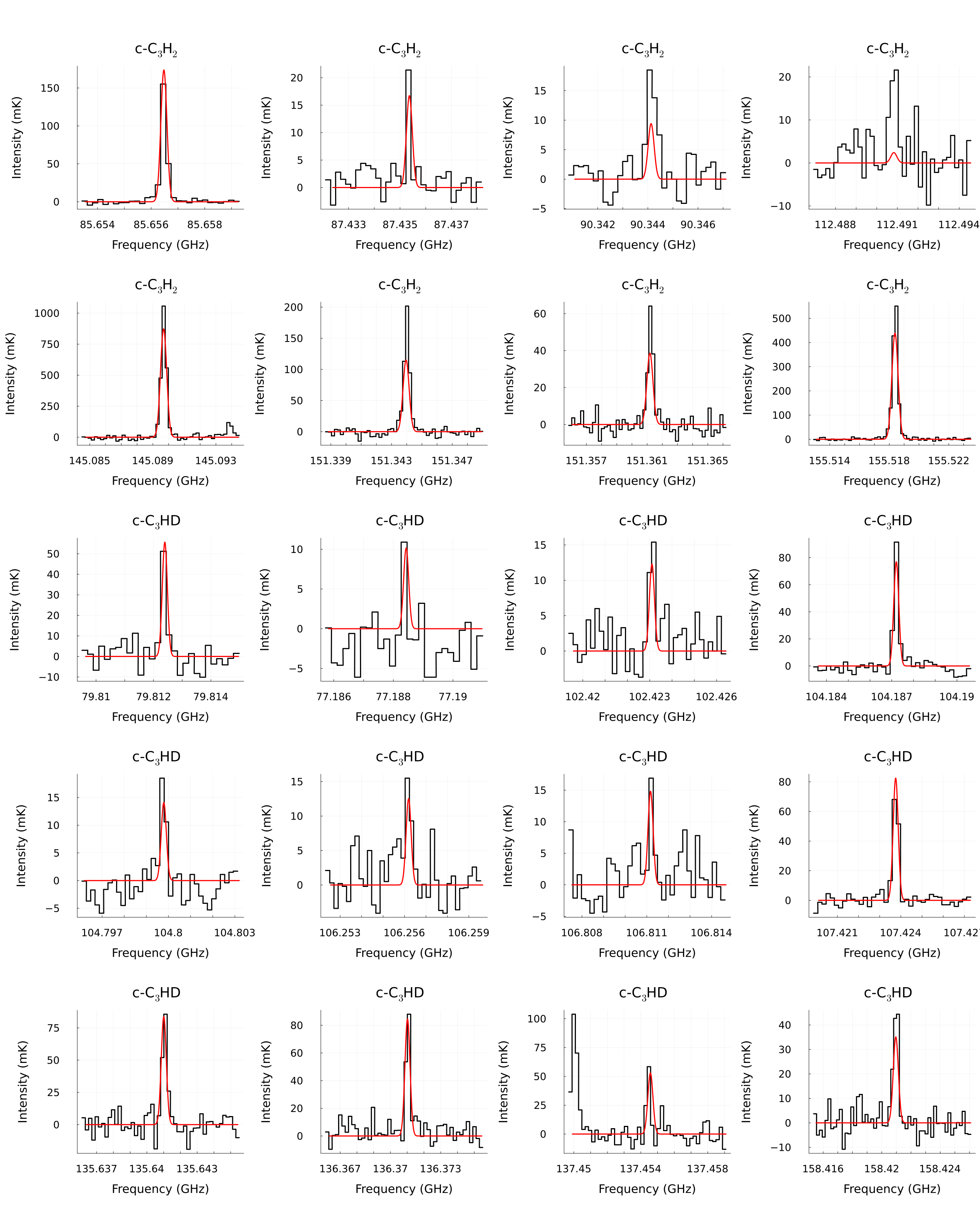}
   \caption*{\textbf{Fig. \ref*{fig:page_1}} continued.}
 \end{center}
 \end{figure*}

 \begin{figure*}
 \begin{center}
 \includegraphics[trim=0cm 0cm 0cm 0cm, width=0.98\textwidth]{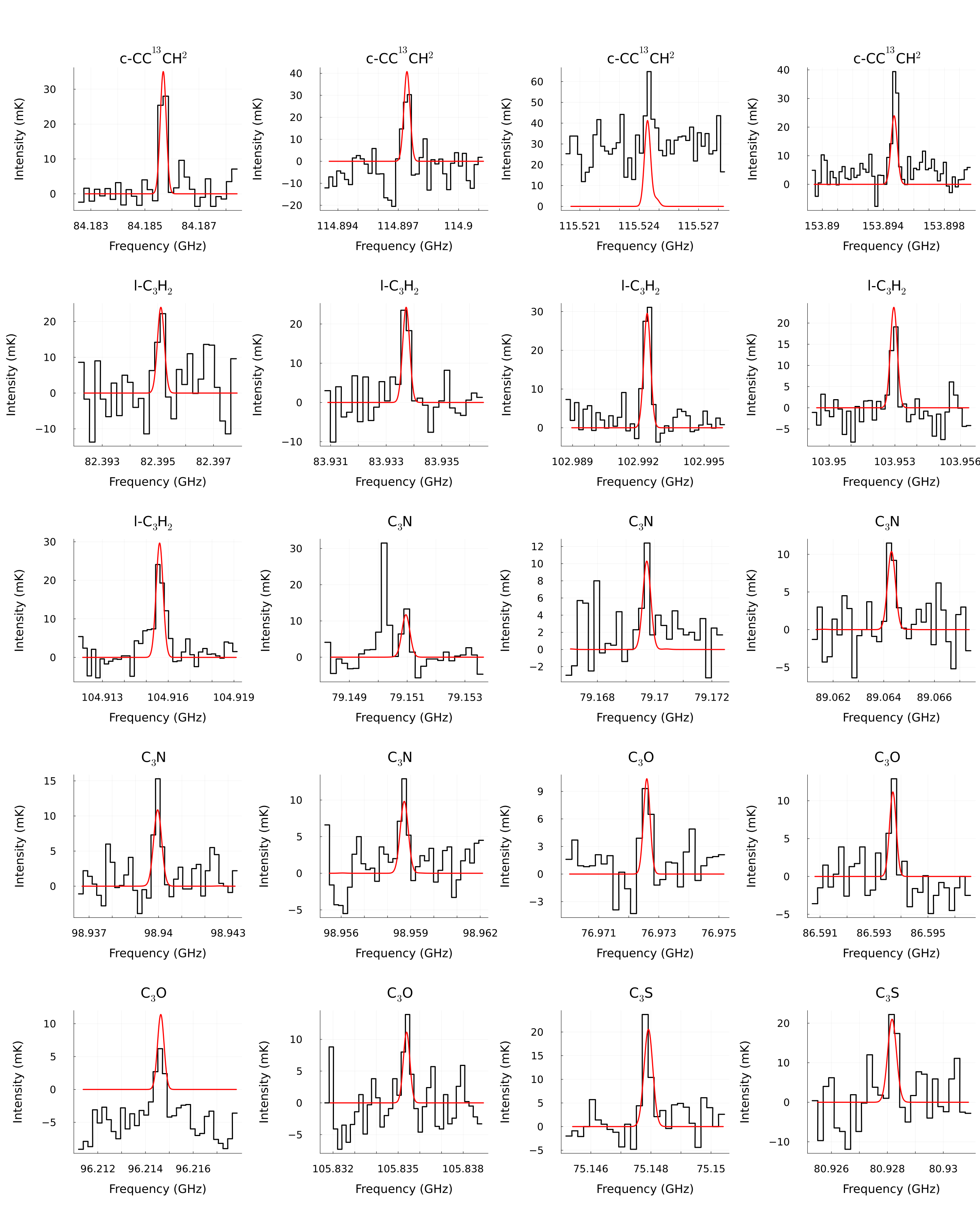}
   \caption*{\textbf{Fig. \ref*{fig:page_1}} continued.}
 \end{center}
 \end{figure*}

 \begin{figure*}
 \begin{center}
 \includegraphics[trim=0cm 0cm 0cm 0cm, width=0.98\textwidth]{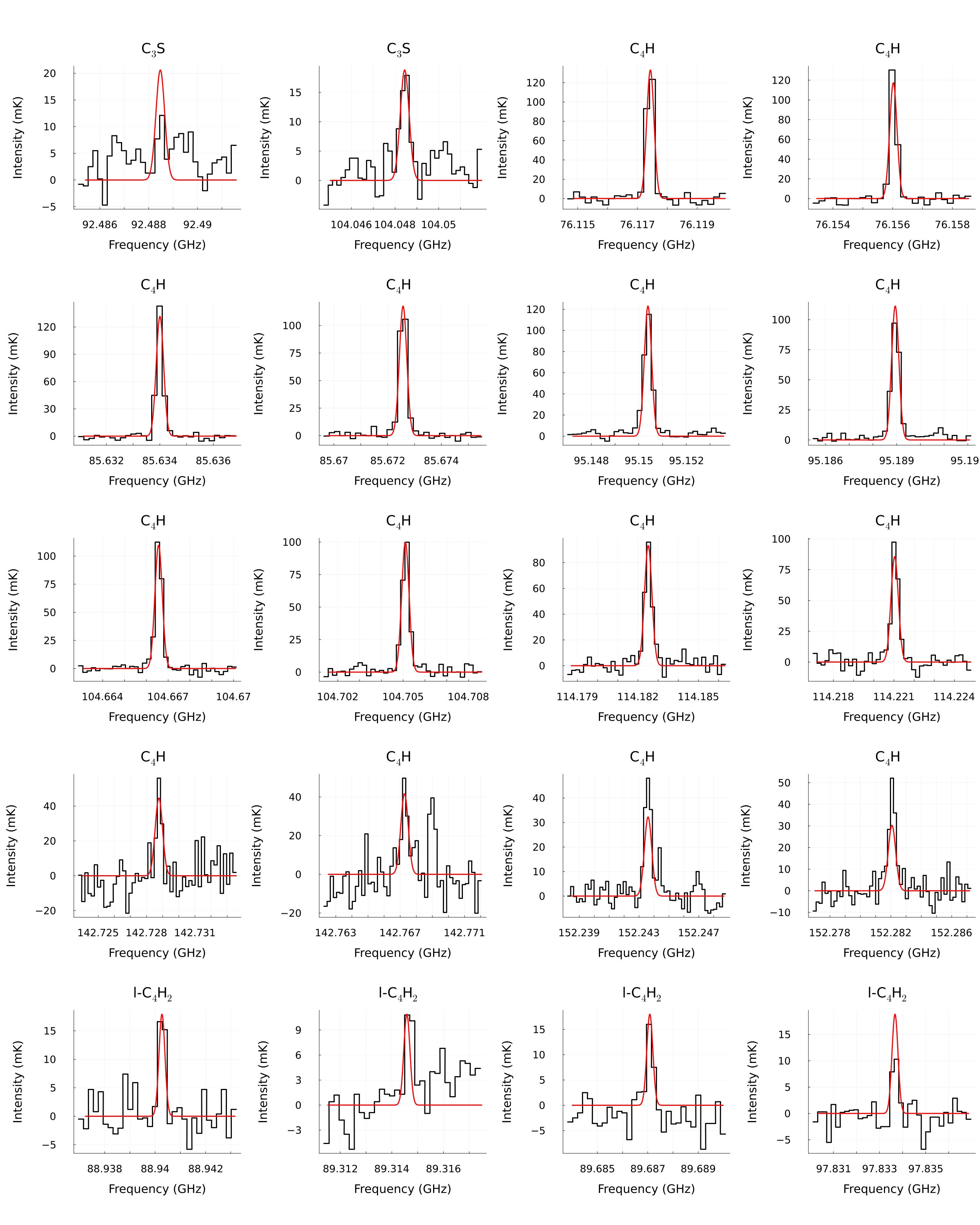}
   \caption*{\textbf{Fig. \ref*{fig:page_1}} continued.}
 \end{center}
 \end{figure*}

 \begin{figure*}
 \begin{center}
 \includegraphics[trim=0cm 0cm 0cm 0cm, width=0.98\textwidth]{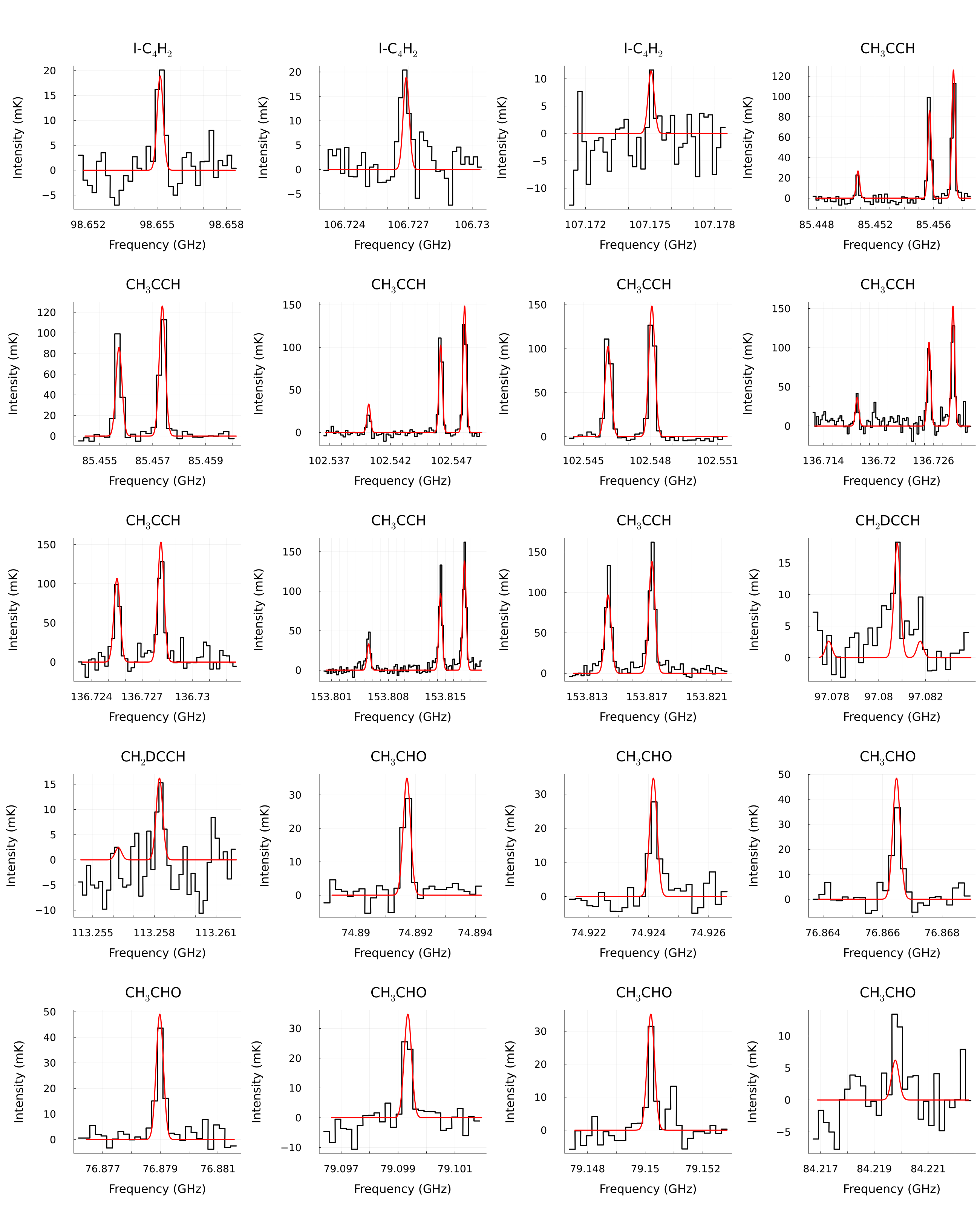}
   \caption*{\textbf{Fig. \ref*{fig:page_1}} continued.}
 \end{center}
 \end{figure*}

 \begin{figure*}
 \begin{center}
 \includegraphics[trim=0cm 0cm 0cm 0cm, width=0.98\textwidth]{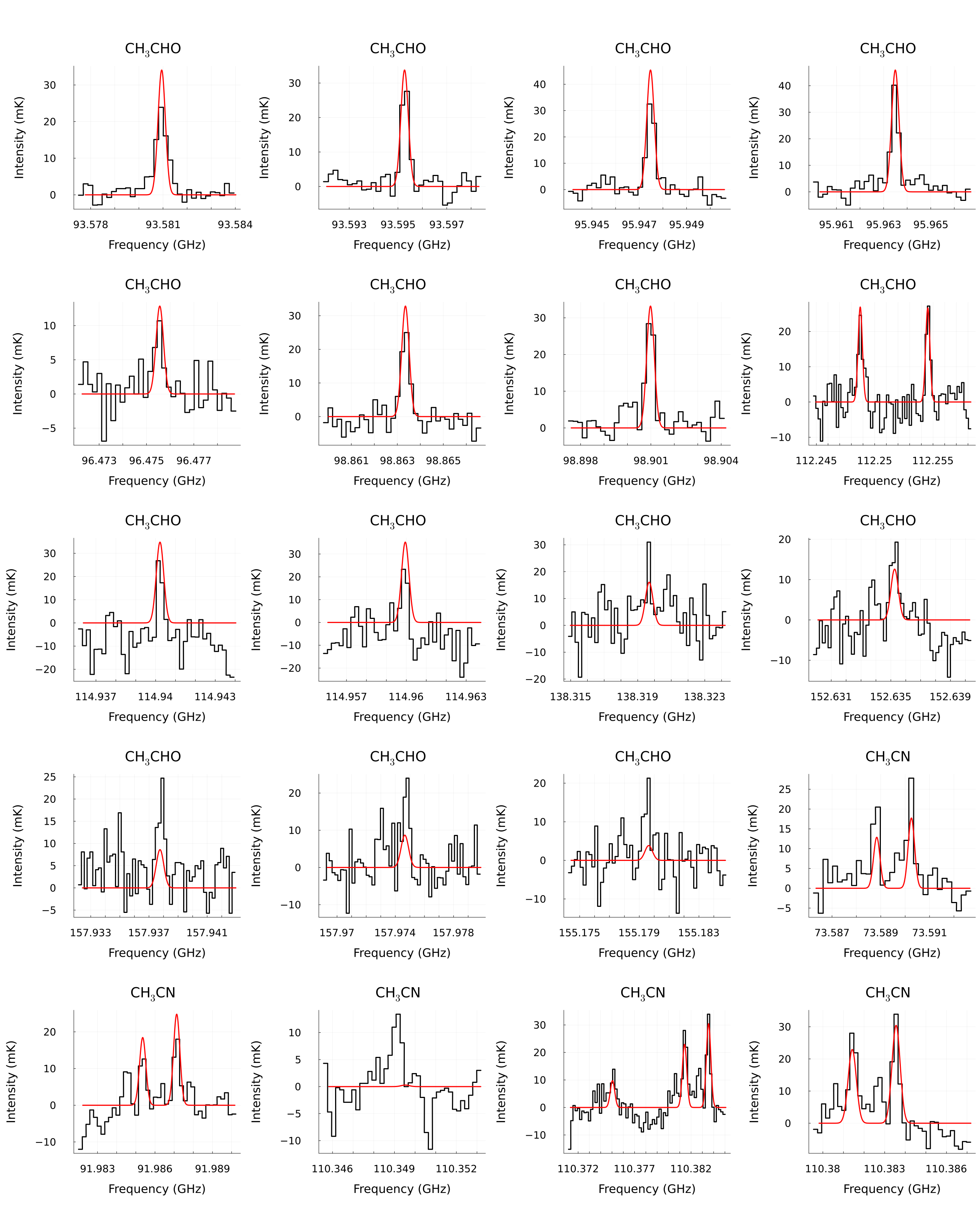}
   \caption*{\textbf{Fig. \ref*{fig:page_1}} continued.}
 \end{center}
 \end{figure*}

 \begin{figure*}
 \begin{center}
 \includegraphics[trim=0cm 0cm 0cm 0cm, width=0.98\textwidth]{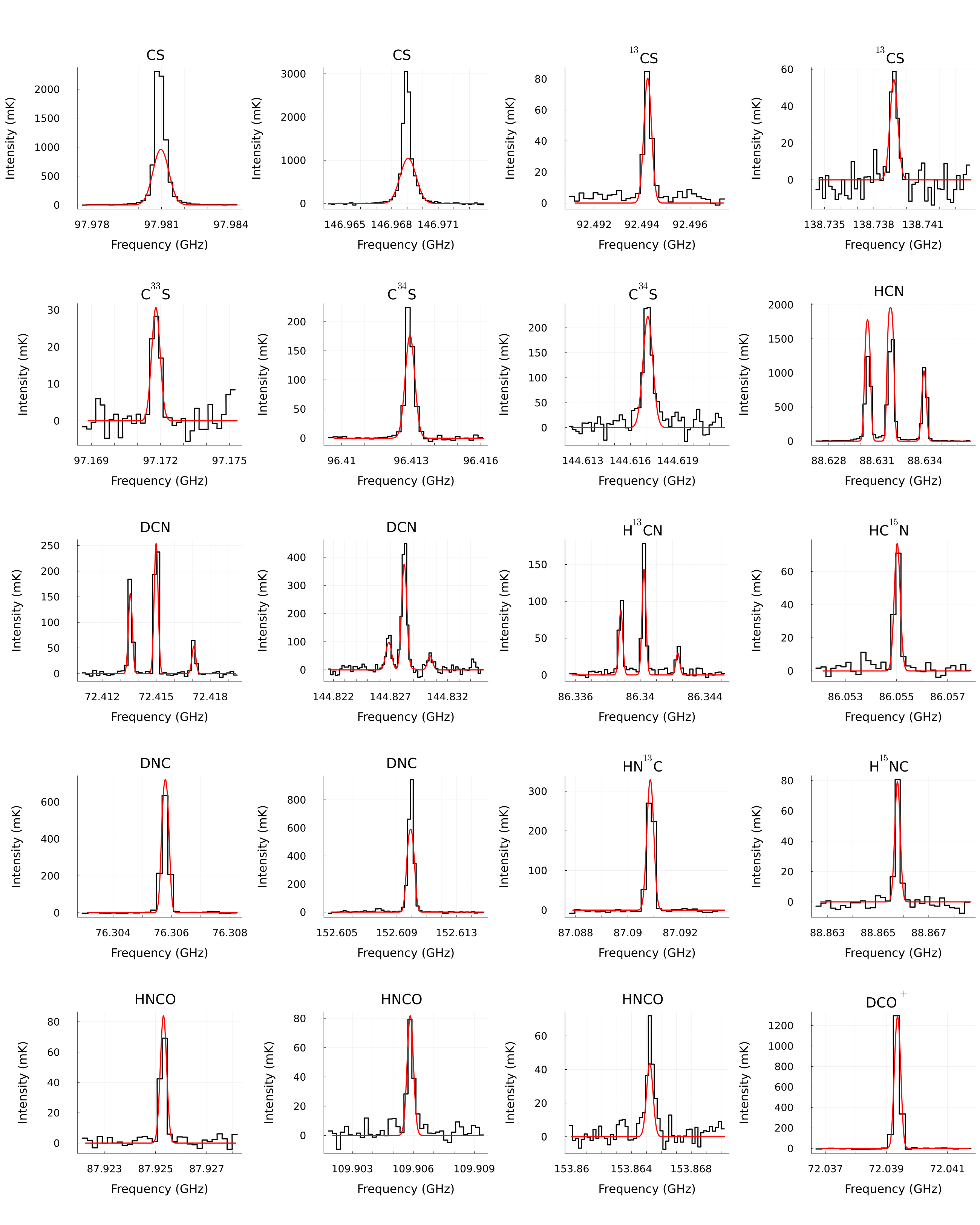}
   \caption*{\textbf{Fig. \ref*{fig:page_1}} continued.}
 \end{center}
 \end{figure*}

 \begin{figure*}
 \begin{center}
 \includegraphics[trim=0cm 0cm 0cm 0cm, width=0.98\textwidth]{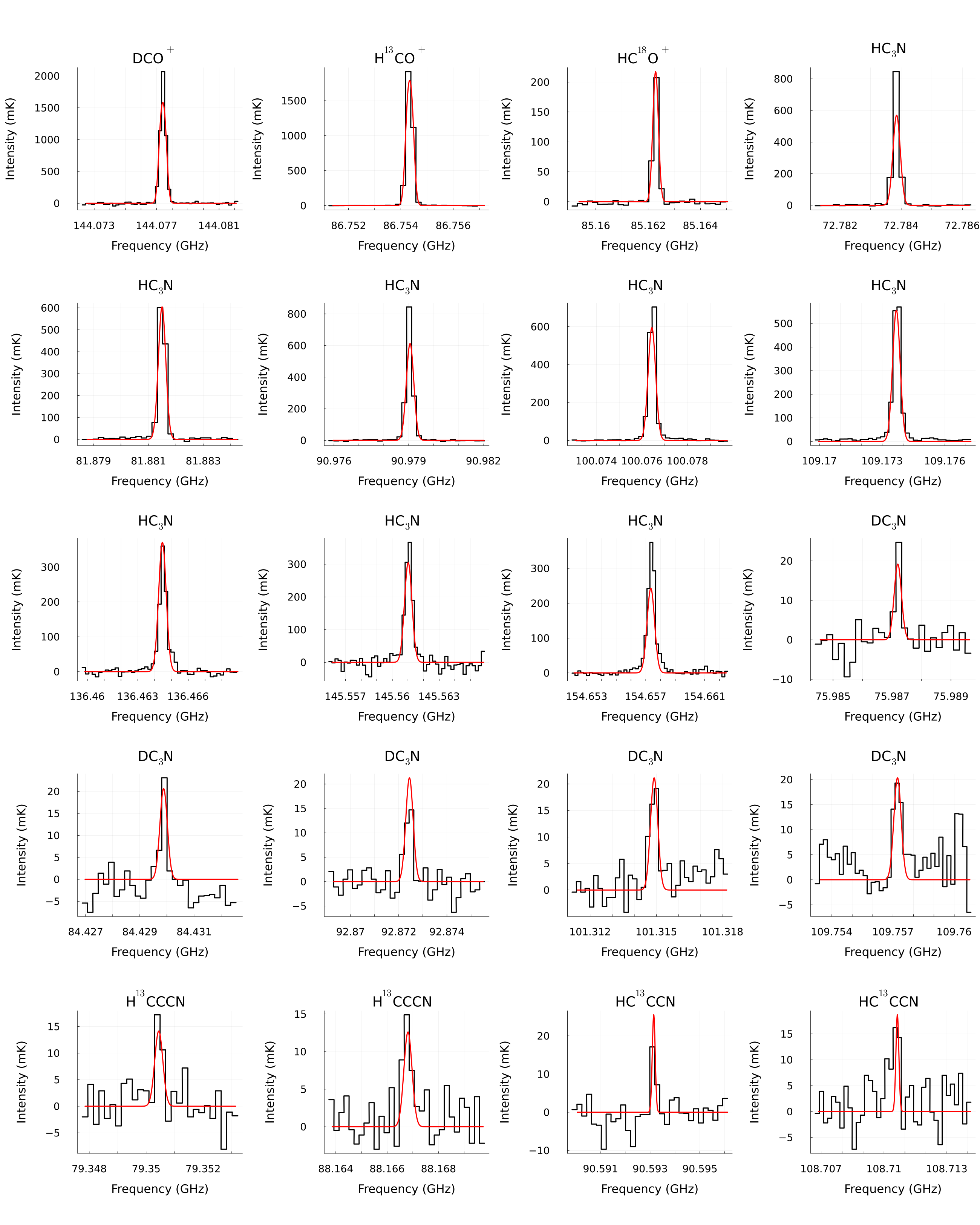}
   \caption*{\textbf{Fig. \ref*{fig:page_1}} continued.}
 \end{center}
 \end{figure*}

 \begin{figure*}
 \begin{center}
 \includegraphics[trim=0cm 0cm 0cm 0cm, width=0.98\textwidth]{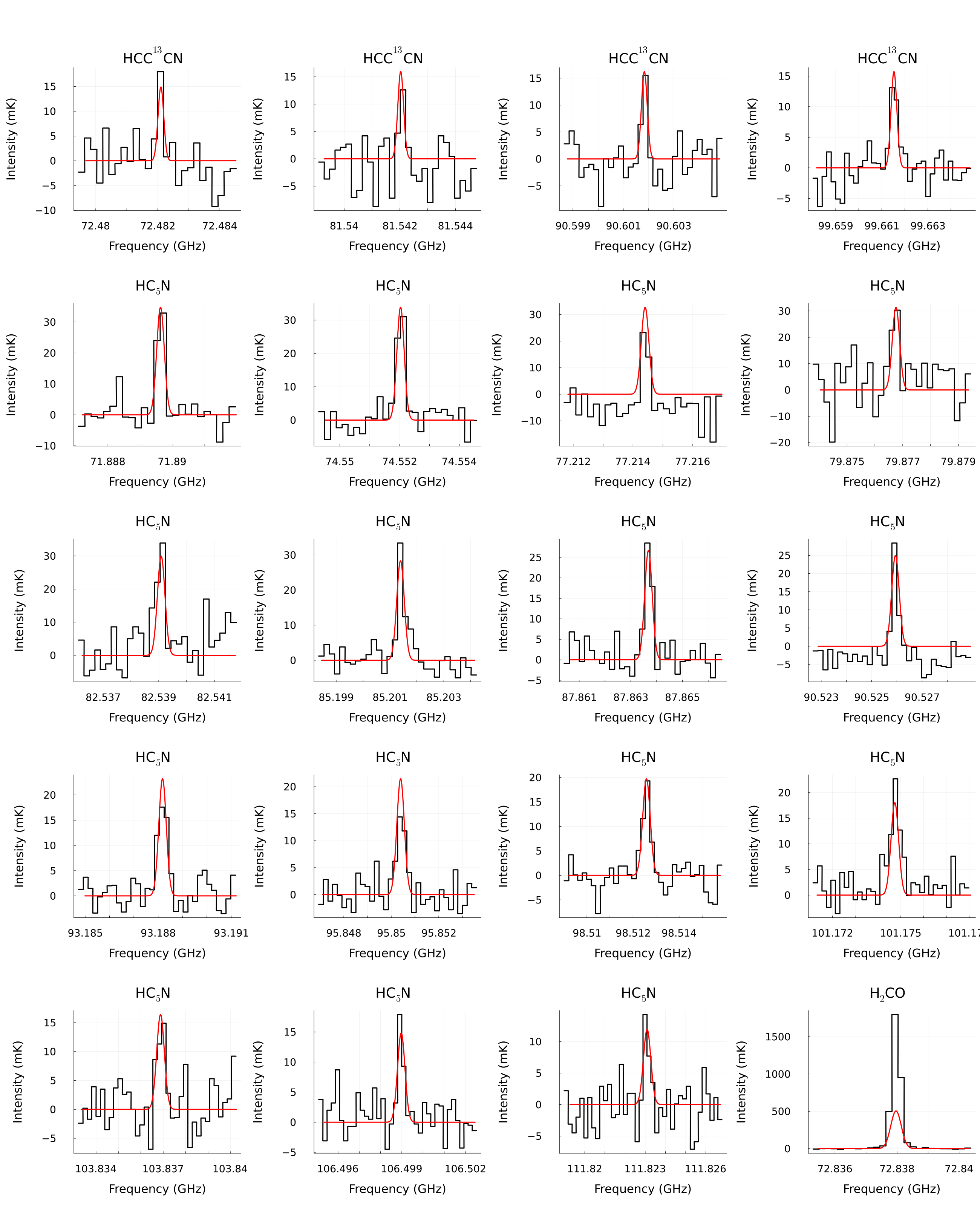}
   \caption*{\textbf{Fig. \ref*{fig:page_1}} continued.}
 \end{center}
 \end{figure*}

 \begin{figure*}
 \begin{center}
 \includegraphics[trim=0cm 0cm 0cm 0cm, width=0.98\textwidth]{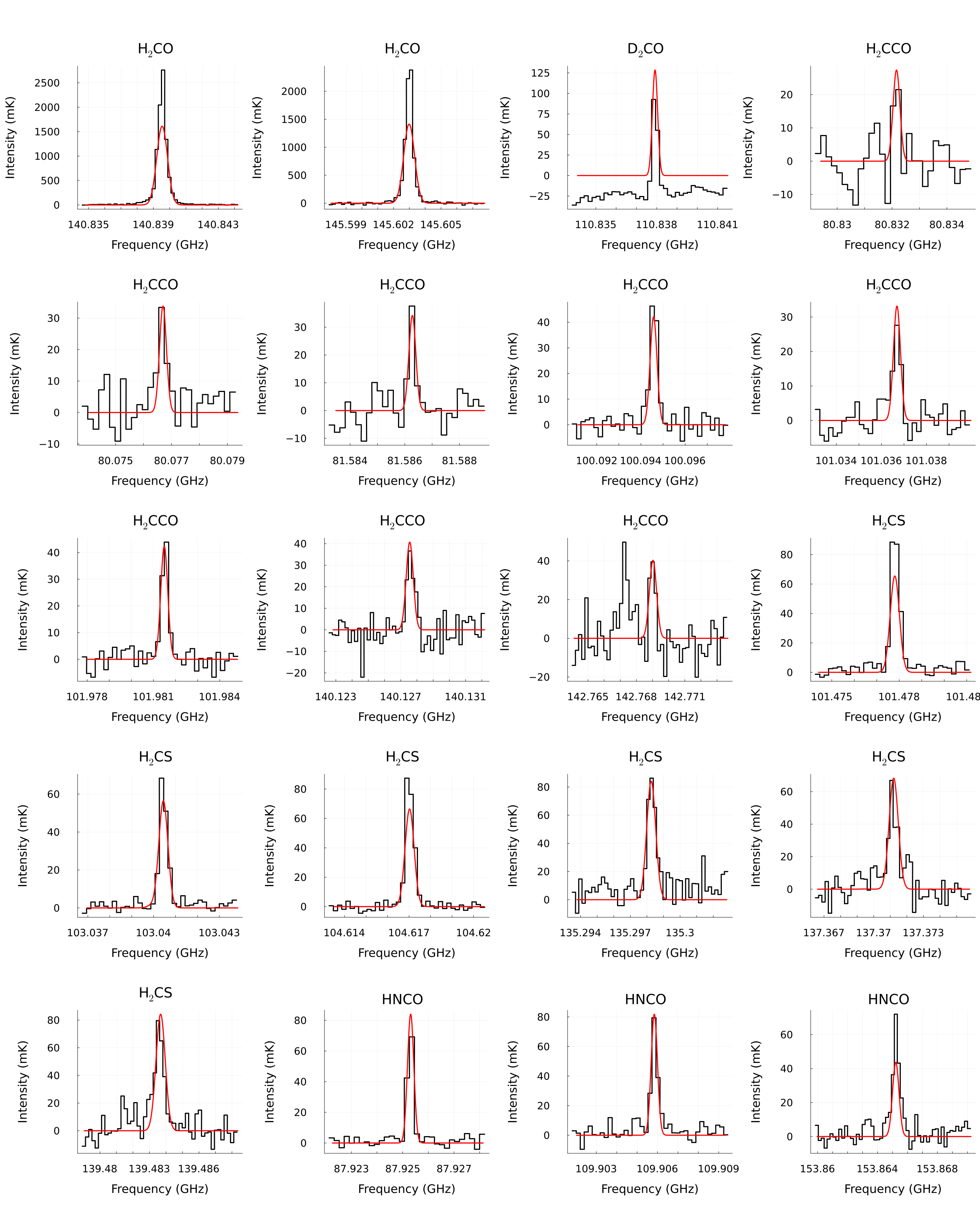}
   \caption*{\textbf{Fig. \ref*{fig:page_1}} continued.}
 \end{center}
 \end{figure*}

 \begin{figure*}
 \begin{center}
 \includegraphics[trim=0cm 0cm 0cm 0cm, width=0.98\textwidth]{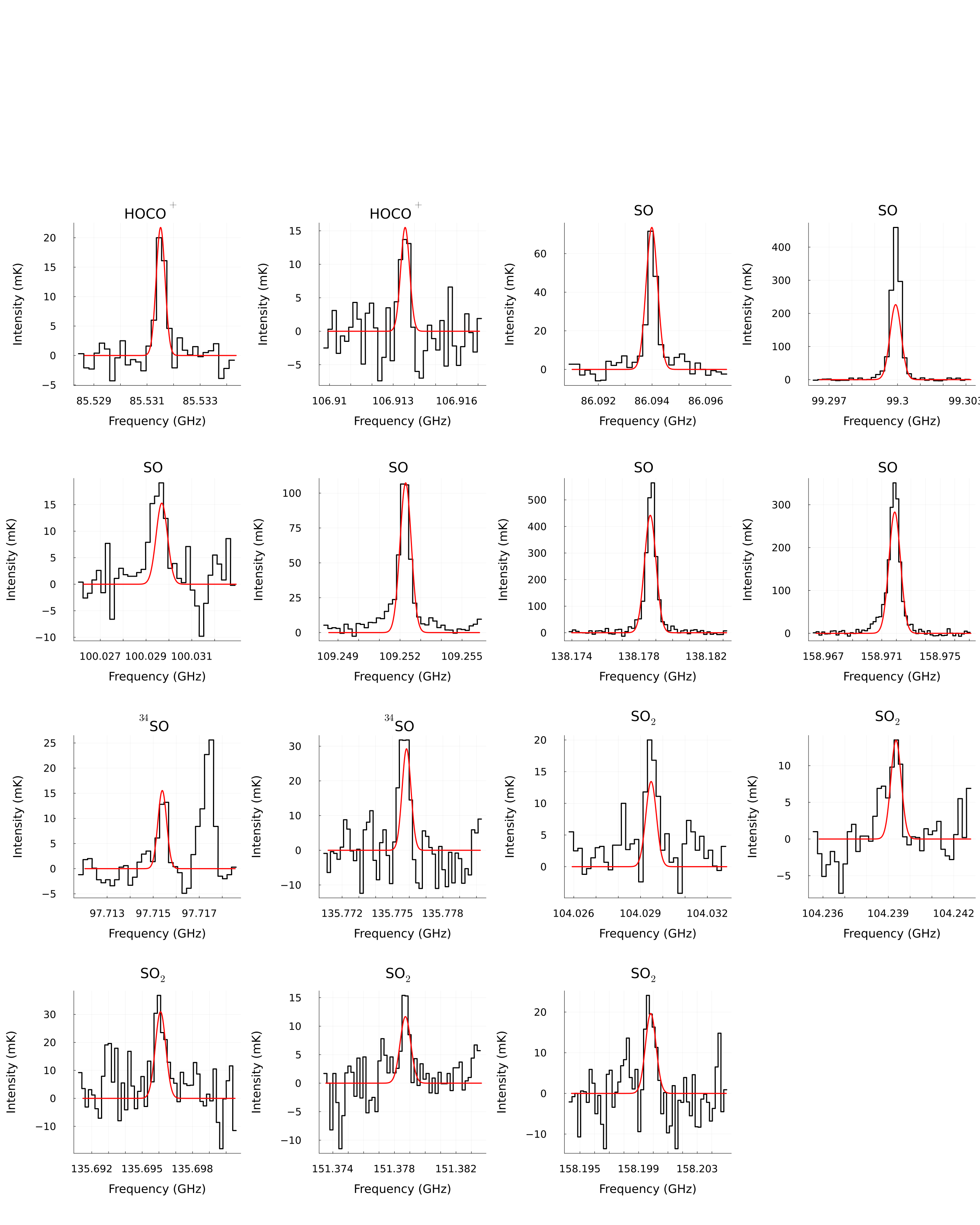}
   \caption*{\textbf{Fig. \ref*{fig:page_1}} continued.}
 \end{center}
 \end{figure*}

 \begin{figure*}
 \begin{center}
 \includegraphics[trim=0cm 0cm 0cm 0cm, width=0.98\textwidth]{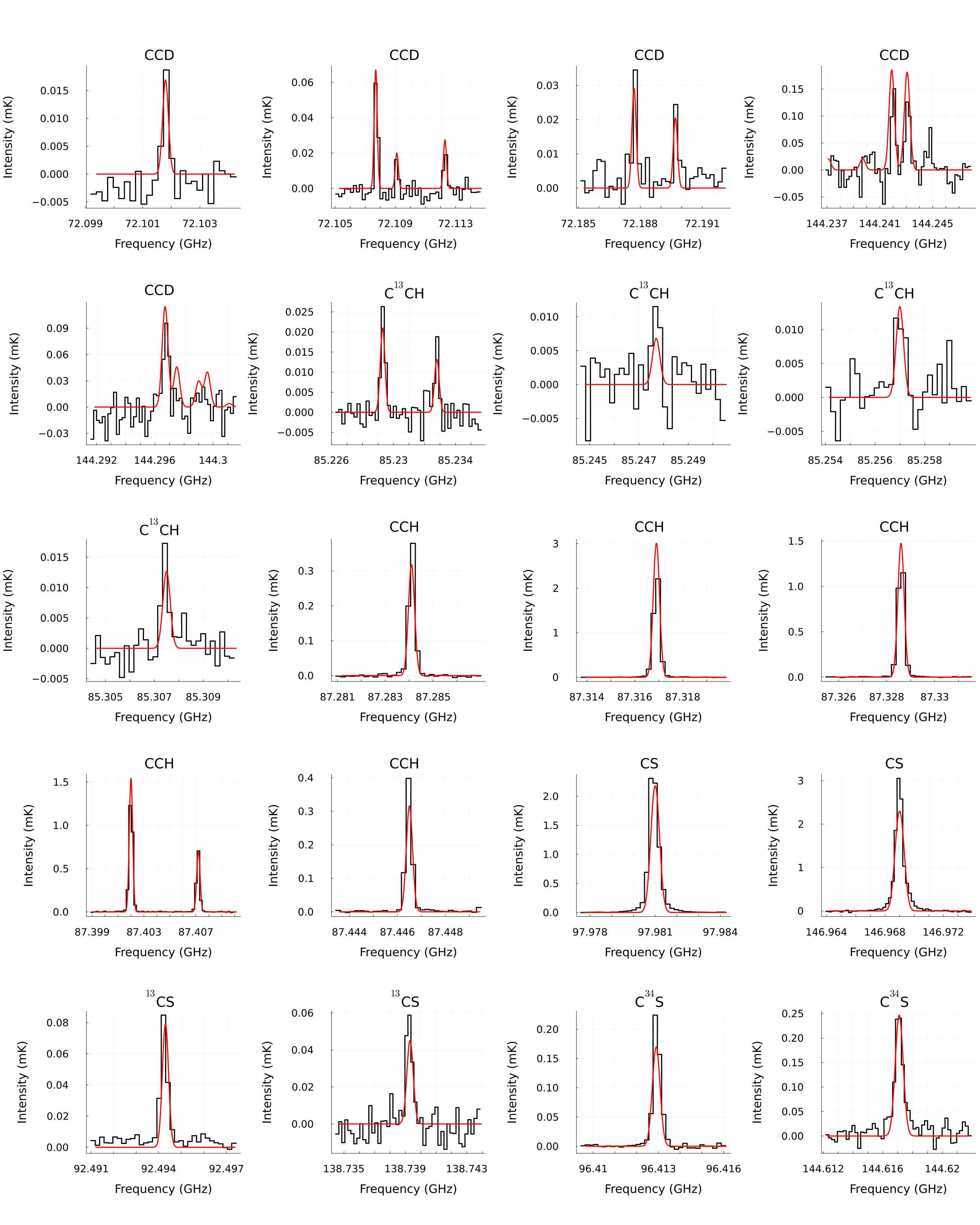}
   \caption{Same as Figure \ref{fig:page_1} for the non-LTE radiative transfer models.}
   \label{fig:nonlte_1}
 \end{center}
 \end{figure*}

 \begin{figure*}
 \begin{center}
 \includegraphics[trim=0cm 0cm 0cm 0cm, width=0.98\textwidth]{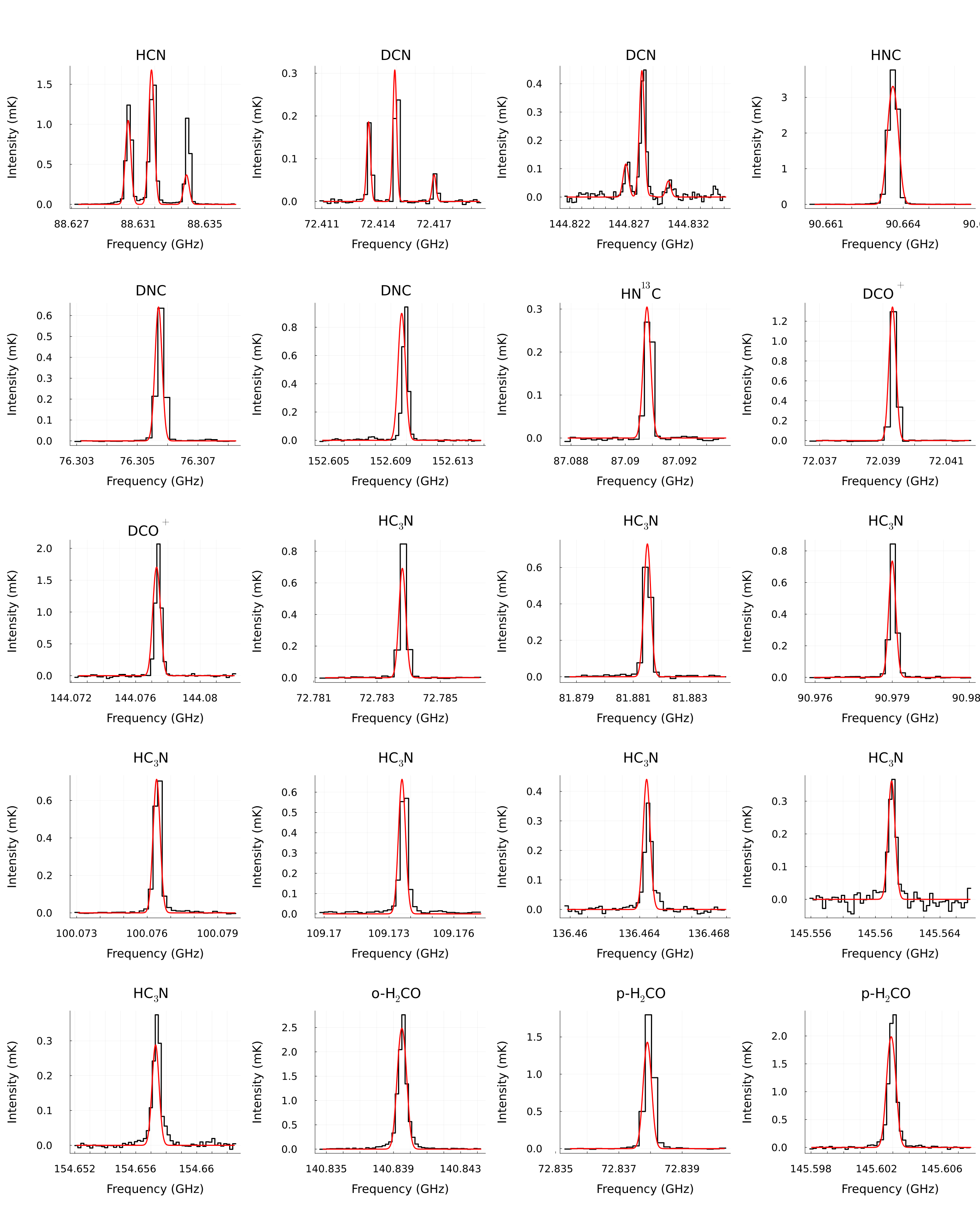}
   \caption*{\textbf{Fig. \ref*{fig:nonlte_1}} continued.}
 \end{center}
 \end{figure*}

 \begin{figure*}
 \begin{center}
 \includegraphics[trim=0cm 0cm 0cm 0cm, width=0.98\textwidth]{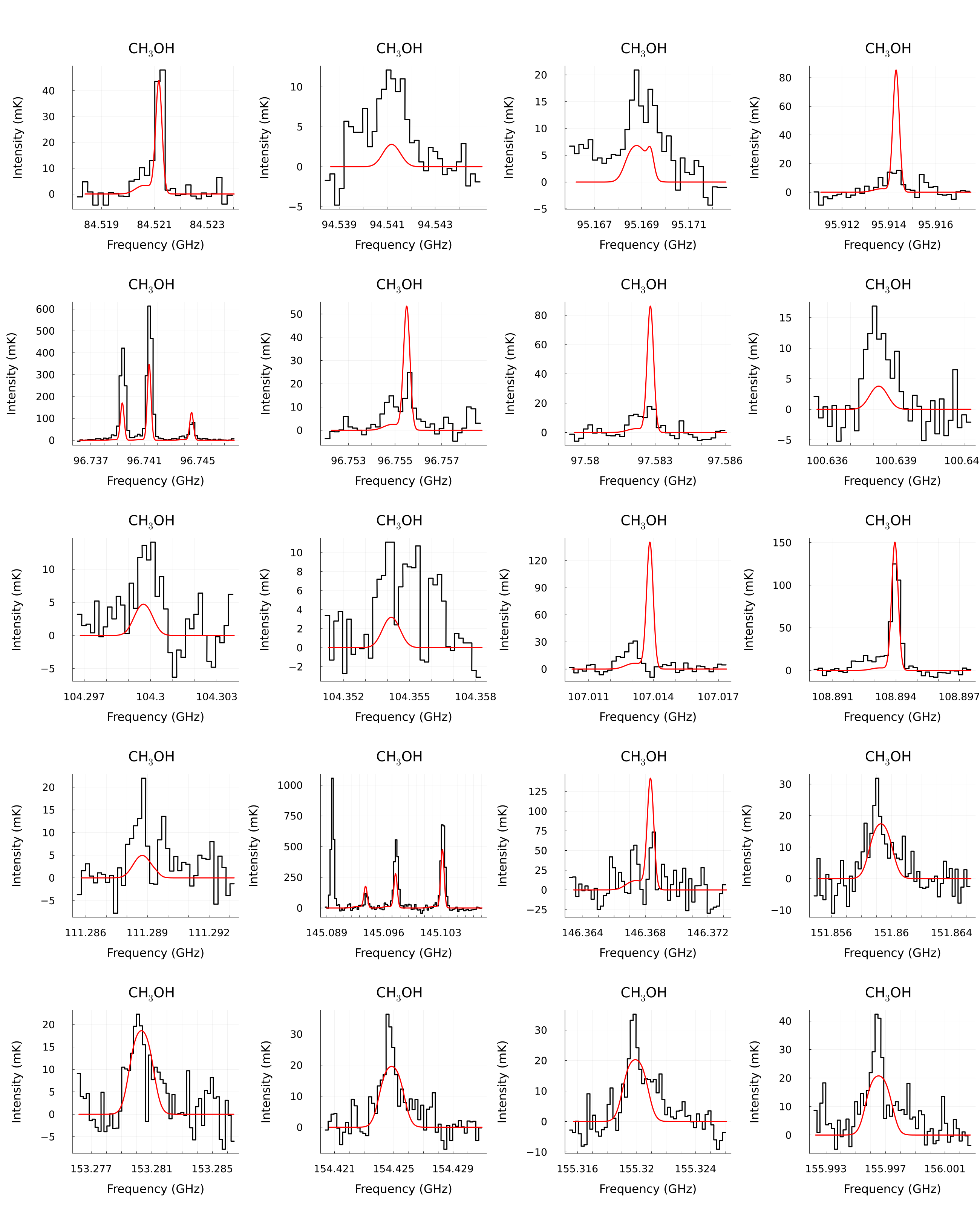}
   \caption{Reproduction of CH$_3$OH emission from its rotational diagram (see Sect. \ref{sec:methanol}).}
   \label{fig:page_meth_1}
 \end{center}
 \end{figure*}

 \begin{figure*}
 \begin{center}
 \includegraphics[trim=0cm 0cm 0cm 0cm, width=0.98\textwidth]{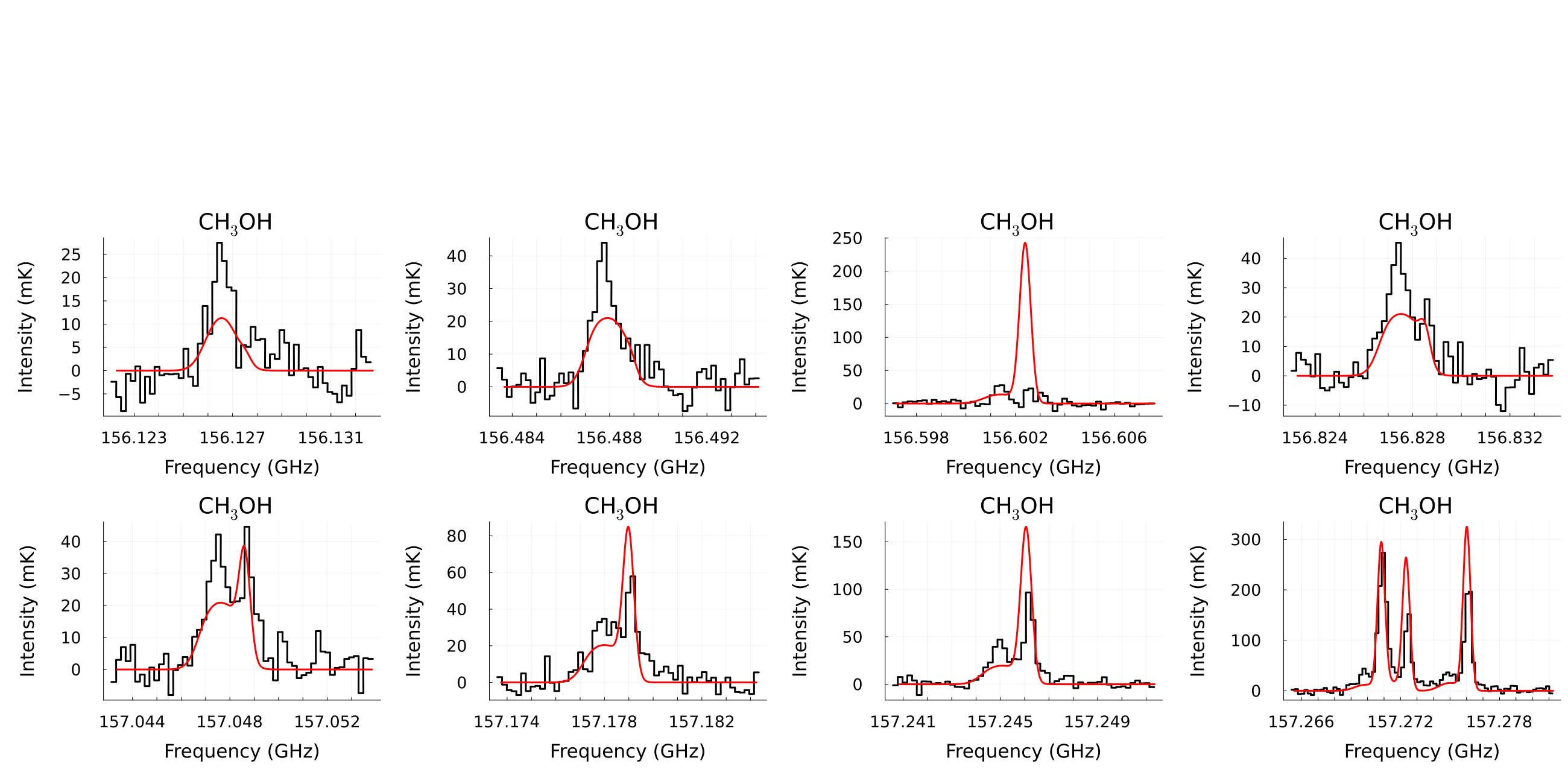}
   \caption*{\textbf{Fig. \ref*{fig:page_meth_1}} continued.}
 \end{center}
 \end{figure*}

\end{document}